\documentclass[11pt,a4paper]{article}
\pdfoutput=1
\parskip=\baselineskip

\usepackage{mathrsfs}
\usepackage[utf8]{inputenc}
\usepackage{amsmath,amsthm,amssymb} 
\usepackage{color,verbatim}
\usepackage[colorlinks=true,linkcolor=blue,citecolor=red]{hyperref}
\usepackage{amssymb}
\usepackage{hyperref}
\usepackage{graphicx}
\usepackage{subfig}
\usepackage{tikz}
\usetikzlibrary{calc,arrows,decorations.markings,shapes.arrows,patterns, decorations.pathmorphing}

\tikzset{
  ->-/.style={decoration={markings, mark=at position 0.5 with {\arrow{to}}},
              postaction={decorate}},
}

\tikzset{
  -<-/.style={decoration={markings, mark=at position 0.5 with {\arrow{to reversed}}},
              postaction={decorate}},
}

\numberwithin{equation}{section}
\newcommand{\mc}{\mathcal}
\newcommand{\mf}{\mathfrak}

\def\tilde{\widetilde}
\def\bar{\overline}

\newcommand{\eps}{\epsilon}
\newcommand{\g}{\mathfrak{g}}
\newcommand{\tr}{\triangle}

\newcommand{\til}{\widetilde}
\newcommand{\mscr}{\mathscr}

\newcommand{\br}{\overline}
\newcommand{\iso}{\cong}
\newcommand{\C}{\mathbb C}

\newcommand{\Z}{\mathbb Z}

\newcommand{\op}{\operatorname}
\newcommand{\mbf}{\mathbf}
\newcommand{\mbfn}{\mathnormal}
\newcommand{\mbb}{\mathbb}

\newcommand{\ip}[1]{\left\langle #1 \right\rangle}
\newcommand{\abs}[1]{\left| #1 \right|}

\newcommand{\R}{\mbb R}
\renewcommand{\d}{\mathrm{d}}

\newcommand{\dbar}{\br{\partial}}

\def\i{\mathsf{i}}

\def\be{\begin{equation}}
\def\ee{\end{equation}}
\newcommand{\beq}{\begin{equation}\begin{aligned}}
\newcommand{\eeq}{\end{aligned}\end{equation}}

\def\su{\mathfrak{su}}

\newtheorem{theorem}{Theorem}[section]
\newtheorem{conjecture}[theorem]{Conjecture}
\newtheorem*{conjecture-non}{Conjecture}

\newtheorem{definition}[theorem]{Definition}

\def\eqn#1{eqn.\ #1}
\def\eqns#1{eqns.\ #1}
\def\fig#1{Fig.\ #1}

\newcommand{\CP}{\mathbb{CP}}

\newcommand{\GC}{G_{\C}}
\newcommand{\Gc}{G}

\newcommand{\gC}{\g_{\C}}

\newcommand{\wbar}{\br{w}} 

\newcommand{\zbar}{\br{z}}

\newcommand{\what}{\widehat}

\renewcommand{\c}{\mathrm{c}}

\newcommand{\norm}[1]{\left\| #1 \right\|}
\newcommand{\Oo}{\mscr O}

\def\c{{\sf c}}

\def\Z{\mathbb{Z}}

\newcommand{\Lax}{{\mathscr L}} 

\numberwithin{equation}{section}
\begin{document}

\pagenumbering{Alph} 
\begin{titlepage}
\begin{flushright}

\end{flushright}
\vskip 1.5in
\begin{center}
{\bf\Large{Gauge Theory And Integrability, III}}
\vskip 0.5cm 
{Kevin Costello$^1$ and Masahito Yamazaki$^2$} \vskip 0.05in 
{\small{ 
\textit{$^1$Perimeter Institute for Theoretical Physics, }\vskip -.4cm
\textit{Waterloo, ON N2L 2Y5, Canada}
\vskip 0 cm 
\textit{$^2$Kavli Institute for the Physics and Mathematics of the Universe (WPI), }\vskip -.4cm
\textit{University of Tokyo, Kashiwa, Chiba 277-8583, Japan}
}}

\end{center}


\vskip 0.5in
\baselineskip 16pt
\begin{abstract}
We study two-dimensional integrable field theories from the viewpoint of the
four-dimensional Chern-Simons-type gauge theory introduced recently.
The integrable field theories are realized as effective theories
for the four-dimensional theory coupled with 
two-dimensional surface defects,
and we can systematically compute their Lagrangians and the Lax operators 
satisfying the zero-curvature condition.
Our construction includes many known integrable field theories, such as Gross-Neveu models,
principal chiral models with Wess-Zumino terms and symmetric-space coset sigma models.
Moreover we obtain various generalization these models in a number of different directions,
such as trigonometric/elliptic deformations, multi-defect generalizations and 
models associated with higher-genus spectral curves, many of which seem to be new.

\end{abstract}
\date{December, 2018}
\end{titlepage}
\pagenumbering{arabic} 

\tableofcontents
\newpage

\section{Introduction}\label{sec:introduction}

Recently a new approach to integrable model has been pursued \cite{Costello:2013zra,Costello:2017dso,Costello:2018gyb}
based on a four-dimensional Chern-Simons-type gauge theory proposed in \cite{Costello:2013zra}.\footnote{See \cite{Costello:2013sla,Witten:2016spx,Ashwinkumar:2018tmm,Ishtiaque:2018str,Costello:2018txb,Saidi:2018mdg,Bittleston:2019gkq,Yamazaki:2019prm,Aamand:2019evs} for recent discussion of four-dimensional Chern-Simons theory.} 
This approach has been successful in reproducing many results in integrable models
from purely field-theoretic arguments, and provides new insight into integrable models,
including conceptual explanation of the spectral parameters and the integrability of the model.

In previous two papers of the series \cite{Costello:2017dso,Costello:2018gyb} we studied integrable {\it discrete lattice models} 
from the viewpoint of four-dimensional Chern-Simons-type gauge theory proposed in \cite{Costello:2013zra}. 
In \cite{Costello:2017dso,Costello:2018gyb} we constructed discrete lattice of the statistical mechanical models from horizontal and vertical Wilson lines of the theory.
Perturbative computations of the gluon exchange diagrams reproduces the R-matrix of the the integrable model, and among other things the integrability of the resulting model
is manifest from the four-dimensional perspective.

The aim of the present paper is to engineer two-dimensional integrable {\it continuum field theories} from the same four-dimensional gauge theory, but now with two-dimensional surface defects included. The coupled 4d--2d system can be integrated out to obtain an effective two-dimensional field theory.  
As we will explain, formal properties of four-dimensional Chern-Simons theory guarantee that this effective two-dimensional theory has a Lax operator satisfying the zero-curvature condition, which therefore generates infinitely-many commuting conserved charges, at least at the classical level.

Our construction covers a wide variety of two-dimensional integrable field theories.  For example, we find Gross-Neveu model, Thirring model, and principal chiral models, Wess-Zumino-Witten (WZW) models, and sigma models whose target spaces are symmetric space cosets.  Introducing super-groups, we also find the $PSU(2,2 \, | \, 4)/ SO(4,1) \times SO(5)$ coset sigma model 
that is related to the superstring on $AdS_5 \times S^5$.\footnote{As we will see, the model we find is related to the pure-spinor formulation, not the Green-Schwarz formulation.} 
Our construction works equally for rational, trigonometric and elliptic cases,
giving a uniform construction of a very wide range of two-dimensional integrable field theories.  For example, in the trigonometric case, our construction yields exotic models such as the Fateev-Onofri-Zamolodchikov (FOZ) sausage model \cite{Fateev:1992tk}, as well as many generalizations which appear to be new.

It is natural to conjecture that \emph{all} known two-dimensional integrable field theories arise from our construction.  While we prove that a reasonable sample of integrable field theories come from our construction, we certainly do not check that all known theories arise.  A conspicuous absence is the sine-Gordon model and its cousins.  We will leave the general discussion of this conjecture for future study.

While the literature on integrable field theories is vast (too vast to be summarized here, see e.g.\ \cite{Faddeev:1987ph,Abdalla:1991vua} for classic textbooks, and also \cite{Zarembo:2017muf} for a recent pedagogical exposition which is useful for the understanding of this paper), it seems that many of the theories we discuss in this paper are new. Even for the known theories our viewpoint is rather different from the existing approaches.  The big advantage of our approach is that the existence of the Lax operator is automatic: it does not need to be guessed. 

In this paper we concentrate on classical integrability of two-dimensional theories.
In general such a classical integrability might be spoiled by quantum corrections, which will be manifested by quantum anomalies in the coupled 4d/2d system.
The quantum integrability of the our models will be discussed in a separate publication  \cite{Part4}.\footnote{See \cite{Yamazaki_Strings2018} for preview.}

The rest of this paper is organized as follows. We divide the paper into two parts, where in part \ref{part:order} we study \emph{order} surface defects. These are obtained by coupling four dimensional Chern-Simons to two-dimensional Lagrangian theories, such as free fermions, free bosons, etc.     In section \ref{sec:engineering} and section \ref{sec:general_Lax} we explain how this procedure will always give rise to an effective two-dimensional integrable field theory, with a Lax operator satisfying the zero-curvature condition.

In the sections \ref{sec:fermions}--\ref{sec:betagamma} we discuss specific examples of surface defects, namely free chiral and anti-chiral fermion (section \ref{sec:fermions}), free scalar (section \ref{sec:scalars}) and the curved $\beta-\gamma$ system (section \ref{sec:betagamma}).

Next, in part \ref{part:disorder}, we turn to the analysis of integrable field theories constructed by \emph{disorder} operators.  Disorder surface defects, in our setting, are obtained by allowing the gauge fields to have certain poles and branch cuts along the spectral parameter plane.  Poles are only allowed where the one-form $\omega$ that appears in the Lagrangian as $\int \omega \, {\rm CS}(A)$ has a zero.  We summarize analysis of disorder operators in section \ref{sec:disorder}.   The models that we engineer by using disorder operators include principal chiral models with Wess-Zumino terms (sections \ref{sec:WZW}, \ref{sec:PCM}), symmetric spaces (section \ref{sec:symmetric}),  multi-defect generalizations of the principal chiral model (the section \ref{sec:general_PCM}), two-parameter trigonometric deformation (section \ref{sec:PCM_deformed}), and generalized symmetric spaces including the integrable $AdS^5 \times S^5$ $\sigma$-model (section \ref{sec:gss}).

In section \ref{sec:higher_genus} we discuss the further extension of the construction to higher genus spectral curves.  As far as we can tell, all of these higher-genus examples are new.  These higher-genus models are $\sigma$-models whose target is an open subset of the moduli space of real-algebraic $G$-bundles on the spectral curve $C$ (which we assume is equipped with an anti-holomorphic involution).   These moduli spaces are equipped with a certain metric and closed three-form, which can be written in terms of a quantity known as the non-Abelian Sz\"ego kernel \cite{fay1992nonabelian}. 
 
We also include appendices for review and technical materials.

\part{Order Surface Defects}
\label{part:order}

\section{Engineering Two-Dimensional Field Theories}
\label{sec:engineering}

In this section we spell out the general construction of our two-dimensional theories,
and in particular explain why the resulting theory is classically integrable.

\subsection{Four-dimensional Chern-Simons Theory}

We start with the four-dimensional theory introduced in \cite{Costello:2013zra}, whose action is given by\footnote{
Compared with \cite{Costello:2017dso} we included a factor of $\hbar$ into the action,
so that the VEV (vacuum expectation value) of an observable $\mathcal{O}$ is given by the path-integral
\begin{align}
\label{path-integral}
\langle \mathcal{O} \rangle= \frac{ \displaystyle \int \mathcal{D} A \, \mathcal{O} \exp\left(\i S\right)}{ \displaystyle \int \mathcal{D} A \, \exp\left(\i S\right)} \;.
\end{align}
}
\begin{align}
S_{\rm 4d}
=\frac{1}{2\pi \hbar} \int_{\R^2 \times C} \omega \wedge  \textrm{CS}(A)
\;.
\label{eq.action}
\end{align} 

We take the four-dimensional space-time to be $\R^2 \times C$, where the $C$ is  curve with complex structure,
and is obtained by removing a finite number of (or zero) points from a closed curve $\bar{C}$.
In the discussion of integrable lattice models in \cite{Costello:2017dso}, the curve $C$ is either $\C, \C^{\times}$ or an elliptic curve $E$, where the three cases correspond to rational, trigonometric and elliptic integrable models.
For the cases of $\C, \C^{\times}$ we need to impose appropriate boundary conditions at 
the boundaries (namely $z=0$ for $\C$, and $z=0, \infty$ for $\C^{\times}$, see \cite{Costello:2017dso} for details).
We will later also discuss the cases of higher genus curves, for reasons which we will explain in section \ref{sec:higher_genus}.

The four-dimensional theory is topological along $\R^2$ and 
holomorphic along $C$, and is T-dual to the topological three-dimensional Chern-Simons theory \cite{Yamazaki:2019prm} .
We choose coordinates $x,y$ for $\R^2$ and a complex coordinate $z$ on $C$.

In the discussion of surface defect we specify a complex structure for  $\R^2$,
given by a holomorphic coordinate $w=x+ \i y$ (anti-holomorphic coordinate is denoted by $\wbar=x-\i y$). Note that for physical discussion
we often need to switch to Lorentzian signature, so that the `holomorphic' coordinate is then $w = x +  y$
and its `conjugate' is given by $\wbar=x-y$. 

For the application to integrable lattice models in \cite{Costello:2017dso,Costello:2018gyb} the topological invariance along 
the `topological plane' played a crucial role in explaining integrability of the models,
and the coordinates $x, y$ are chosen merely for computational purposes.
In this paper, however, the complex structures on the $\R^2$ are essential for the discussion of integrable field theories,
and the topological invariance on the $\R^2$ is often broken by defects.

Inside the Lagrangian, $\omega$ is a holomorphic one-form on the curve $C$,
which can be written as $\d z$ in a holomorphic coordinate $z$ on $C$.

The field $A$ is a three-component gauge field 
$A=A_w \d w+A_{\wbar} \d\wbar+ A_{\zbar} \d\zbar$
associated with gauge symmetry $G$,
and $\textrm{CS}(A)$ is the Chern-Simons three-form
\begin{equation}
\textrm{CS}(A):=\textrm{Tr}\left(A \wedge \d A +\frac{2}{3} A\wedge A\wedge A\right)
\;.
\label{eq.CS_A}
\end{equation} 

The parameter $\hbar$ in front of the action is the quantization parameter of the four-dimensional theory,
which is not integer-quantized as explained in \cite{Costello:2017dso}.
As we will see below, it also plays the role of the 
quantum parameter for the two-dimensional integrable field theories discussed in this paper.

\subsection{Two-Dimensional Theory from Surface Defects}

We next include a two-dimensional surface defect $D$ to this theory.  There are two essential classes of defects we consider:
\begin{enumerate} 
	\item \emph{Order defects}, where we introduce new degrees of freedom on the defect which are coupled to the bulk gauge theory.
	\item \emph{Disorder defects}, where the four-dimensional gauge field is required to have some singularities. 
\end{enumerate}
Order defects are simpler to define, and are studied in Part \ref{part:order} of this paper. Disorder defects, studied in Part \ref{part:disorder}, are more difficult to define, but ultimately richer: we only know how to engineer many of the most familiar integrable theories (such as the symmetric space coset models) using disorder defects.  

Let us pick up a two-dimensional theory with global symmetry $G$.
This means that the 
theory has a coupling for the form $\int J A$ between the current $J$ for the $G$-symmetry
and the background gauge field $A$.
We can then couple the defect theory to the four-dimensional theory in the bulk,
by regarding this gauge field $A$ as the components of the four-dimensional gauge field 
along the defect.
We now have a coupled 4d--2d system, which still keeps manifest Lorentz symmetry along the directions of the defect.

Note that this discussion is rather general---we can couple the four-dimensional theory for example to an arbitrary two-dimensional field theory with flavor symmetry $G$, and even those without Lagrangian descriptions. For concreteness and partly for perturbative analysis, however,
we assume that the theories on the surface defects have Lagrangian descriptions.

In this paper we choose the surface defect to be spreading along $\mathbb{R}^2$,
and located a particular point $z$ on the spectral curve $C$.

We can more generally couple several surface defects $D_{\alpha}$ ($\alpha=1, \dots n$),
where the $\alpha$-th surface defect is located at a point $z_\alpha$ of the curve $C$ (where none of the $z_\alpha$'s are on top of each other). 

As mentioned already, we assume in this paper that the two-dimensional theories on surface defects
have Lagrangian description. Let us denote the 
Lagrangian at the $\alpha$-th defect as  $\mc{L}_{\alpha}(\phi_\alpha; A_w|_{z_\alpha}, A_{\wbar}|_{z_\alpha})$,
where we in general have different Lagrangians on different defects. 
Here $\phi_\alpha$ denotes (in general, a collection of) fields localized on the $\alpha$-th defect,
and the four-dimensional bulk gauge fields $A_w, A_{\wbar}$ appear as
background gauge fields for the $G$-symmetry of the defect.
In general the Lagrangian $\mc{L}$ can contain $z$-derivative couplings for the gauge field $A_w, A_{\wbar}$.

The action for the coupled 4d--2d system is then
\begin{equation} 
S_{\rm 4d-2d}=
\frac{1}{2\pi \hbar} \int_{\R^2 \times C} \d z\, \textrm{CS}(A) + 
\sum_{\alpha=1}^n \frac{1}{\hbar} \int_{\R^2 \times z_\alpha } \mc{L}_\alpha\left(\phi_\alpha; A_{w} |_{z_\alpha}, A_{\wbar} |_{z_\alpha}\right) .  
 \end{equation}
Here we extracted the factor of $\hbar$ from the definition of the defect Lagrangian $\mc{L}_{\alpha}$.
Of course, each defect Lagrangian $\mc{L}_{\alpha}$ can have their own coupling constants hidden inside.

Let us impose specific non-singular boundary conditions for 
bulk four-dimensional gauge field $A$ at the boundaries of the curve $C$.
Since the surface defects are placed away from the boundary points, we can 
regard our theory as defined on the compactification $\bar{C}$ of the curve $C$ (recall that $C$ is a closed curve $\bar{C}$ with possibly several boundary points removed).
We can then reduce our theory along the curve $\bar{C}$, by integrating out the Kaluza-Klein (KK) modes.
We then obtain an \textit{effective} two-dimensional theory on the remaining directions, which we have taken to be $\mathbb{R}^2$.

The claim of our paper is that the resulting two-dimensional effective theory is
classically integrable: we show below that the two-dimensional theory is equipped with an infinite number of conserved currents and with a Lax operator satisfying the zero-curvature equation.

\subsection{Comments on Thermodynamic Limit}

Before proceeding further, let us make a small remark concerning the relation between
the integrable lattice models discussed in \cite{Costello:2017dso,Costello:2018gyb} and the integrable field theories discussed 
in this paper.

While there are crucial differences between integrable lattice models and integrable field theories,
one expects relations between them in the thermodynamic limit.
Namely one can start with a discrete lattice model, and then consider the limit where the lattice spacing goes to zero, such that we recover the translation symmetry and  hence a two-dimensional field theory.

\begin{figure}[htbp]
\centering\includegraphics[scale=0.38]{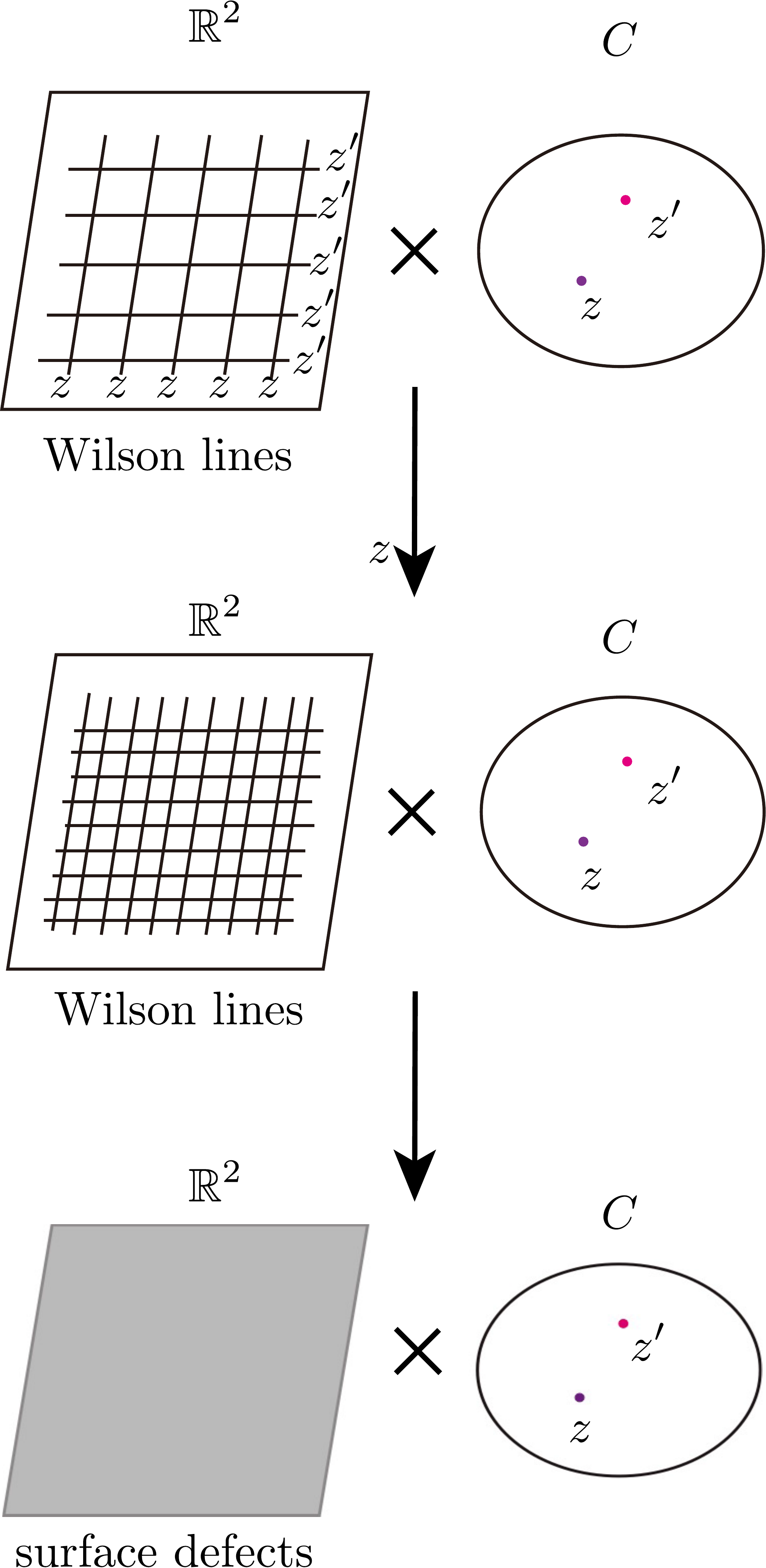}
\caption{One expects that the thermodynamic limit of an integrable lattice model
gives rise an integrable field theory. In the language of four-dimensional Chern-Simons theory, 
this is the limit where an infinite parallel Wilson lines, in both vertical and horizontal Wilson lines, fill out the two-dimensional plane,  thereby becoming a surface defect. While we do not directly take advantage of this mental picture, this is of help in understanding the relation between the present paper and the part I, II of the series \cite{Costello:2017dso,Costello:2018gyb}. Note that vertical and horizontal Wilson lines in this Figure are placed at the same point $z$ and $z'$ of the spectral curve. We can more generally inhomogeneous lattice models and Wilson lines at more than two points of the spectral curve, which will lead to the setup with multiple surface defects.}
\end{figure}

As stated in introduction, in \cite{Costello:2017dso,Costello:2018gyb} we constructed two-dimensional discrete
lattice models by considering horizontal and vertical Wilson lines along $\mathbb{R}^2$.
In this language, the thermodynamic limit corresponds to 
the limit where the spacing between Wilson lines become smaller and smaller,
so that in the limit the Wilson lines combine into a two-dimensional defect filling
the whole of $\mathbb{R}^2$. In fact, since there are two types of Wilson lines
located at points $z$ and $z'$ of the spectral curve $C$ one would expect that 
we have two surface defects, one at the point $z$ and another at $z'$.

This is one of the motivations for considering two-dimensional surface defects to the four-dimensional theory.

While it is an interesting question to consider such a limiting procedure in detail (see e.g.\ \cite{Destri:1987ug,Destri:1987ze} for related discussion),
such a limit often requires a rather careful study of the limit.
In this paper we instead choose to directly start with surface defects.
Indeed, the discussion of integrable field theories in this paper is richer than their integrable lattice model counterparts,
and it seems that some integrable models discussed in this paper (such as those with 
non-chiral defects and  those with  higher-genus spectral curves discussed in section \ref{sec:higher_genus})
might have no lattice model counterparts, at least in an obvious way.

Despite these caveats it is useful to have in mind the thermodynamic limits,
and we will indeed take advantage of this in what follows.

\subsection{Identification of Effective Two-Dimensional Theory}

Let us now come back to our effective two-dimensional theory.
Our theory will contain fields originating from surface defects $D_{\alpha}$ located at points $z_\alpha$ of the curve $C$, together with fields coming from the four-dimensional gauge theory
compactified on $\bar{C}$.

For the reasons which will become clear later, 
in the most of this paper we consider the situations where
the resulting two-dimensional theory has no gauge symmetries themselves.
This requires in particular that the two-dimensional defect theories have no gauge symmetries.

This also requires that there are no zero modes from the four-dimensional gauge field.
In the setup of integrable lattice models in \cite{Costello:2017dso}
this is indeed the case when we have a standard Yang-Baxter equation;
when we have zero modes we have a dynamical Yang-Baxter equation, 
as discussed in section 10 of \cite{Costello:2017dso}.
Assuming this,
the two-dimensional theory as originating from four-dimensional theory, after integrating out the KK modes, is trivial---a classical solution to the equation of motion is isolated, and has no moduli.

We still need to take into account fields from two-dimensional defects.
Our effective two-dimensional theory is not simply the product of the defect theories:
instead they are coupled in a non-trivial way by the exchange of four-dimensional gauge fields. 

Such couplings can be worked out by evaluating Feynman diagrams.
Here the parameter $\hbar$ will be a loop expansion parameter for the effective two-dimensional theory.  This means that
we will find a Lagrangian for a \emph{classical} two-dimensional field theory if we only include tree diagrams in our Feynman diagram expansion (this is the coefficient of $\hbar^{-1}$) .

While details of the computation depends on the choice of the theories on the surface defects, let us here make some general comments.

In principle, the tree-level Feynman diagram calculation that computes the effect of integrating out the four-dimensional gauge field can be complicated. The analysis simplifies dramatically, however, if we fix the gauge symmetry appropriately.

In our previous paper \cite{Costello:2017dso} we used the gauge 
where the four-dimensional gauge field satisfies the equation
\begin{equation} 
	\partial_{\wbar} A_w + \partial_w A_{\wbar} + g^{z \zbar} \partial_{z} A_{\zbar} =0 \;.
\end{equation}
Here $g_{z \zbar}$ is a chosen K\"ahler metric on $C$, and $g^{z \zbar}$ is its inverse.  
This is the closest analog to the Lorentz gauge.

In the limit in which the volume of $C$ becomes small (which is the situation we wish to analyze), 
the gauge fixing condition above becomes simply
\begin{equation} 
	g^{z \zbar} \partial_z A_{\zbar} = 0 \;.
	\label{gauge}
\end{equation}
This gauge is too singular to work with at the quantum level, but computations in this gauge are valid at the classical level.

In this gauge, the propagator only involves $A_{w}$ and $A_{\wbar}$ (and not $A_{\bar{z}}$).  Since the cubic interaction of the gauge theory involves all three components $A_w$, $A_{\wbar}$, $A_{\zbar}$ of the gauge field, we see that this bulk vertex can not play a role in the Feynman diagram analysis.  

\begin{figure}[htbp]
\centering\includegraphics[scale=0.35]{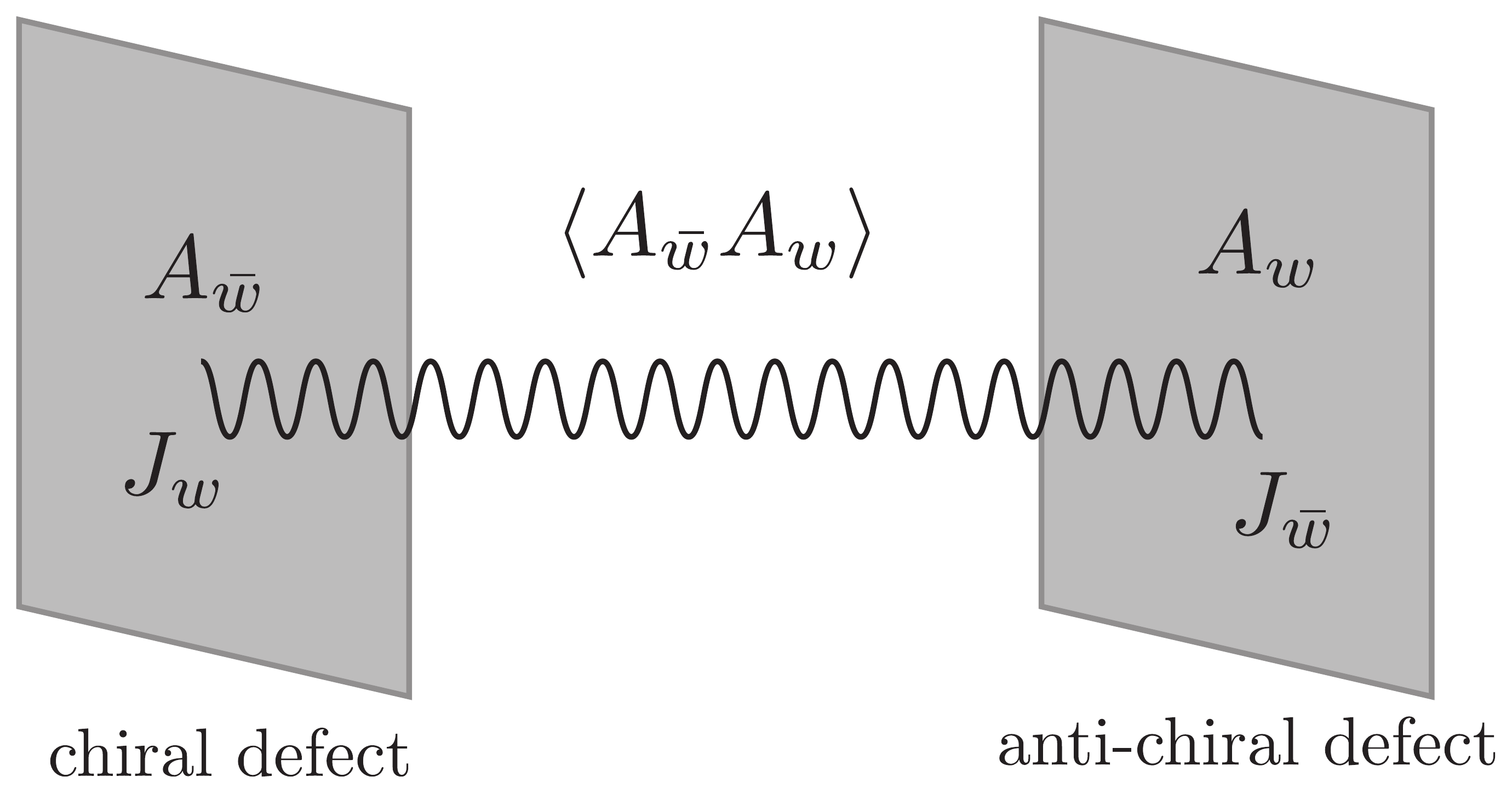}
\caption{A tree-level diagram contributing to the effective action. Since the propagator has a power of $\hbar$,
and since each of the two vertices contributes a factor of $\hbar^{-1}$, this Feynman diagram contributes with the power of 
$\hbar^{1-2}=\hbar^{-1}$, and hence should be included in the computation of the effective two-dimensional action.}
\label{fig:gluon_exchange}
\end{figure}

In the tree level computation (i.e.\ to leading order in $\hbar$), the coupling comes from the exchange of a single gluon between the two surface defects, as shown in Figure \ref{fig:gluon_exchange}.
There was relationship between the classical $r$-matrix and the propagator for the four-dimensional gauge theory \cite{Costello:2017dso}, which implies that the integral of the propagator $\langle A_w (z)A_{\wbar}(z') \rangle$
is given by the classical $r$-matrix $r(z,z')$,
which is the first non-trivial expansion of the R-matrix 
($R(z,z')=\textrm{id}+ \hbar r(z,z') + \mathcal{O}(\hbar^2)$):
\begin{equation}
\int \d\wbar \d w  \, \langle A_{a \wbar} (z)A_{bw}(z') \rangle =  r_{ab}(z,z')  \;.
\end{equation}
Here $r_{ab}(z,z')$ is the component of $r(z,z')$ in a orthonormal basis $\{ t_a\}$ of the Lie algebra 
$\mf{g}$, namely $r(z,z')=\sum_{a,b} r_{ab}(z,z') (t_a\otimes t_b)$.
Due to the translational symmetry along the curve $C$, the classical $r$-matrix $r_{ab}(z,z')$
depends only on the relative positions between $z$ and $z'$, and for this reason we will also write
$r_{ab}(z,z')=r_{ab}(z-z')$. This is to be interpreted as $r_{ab}(z,z')=r_{ab}(z/z')$ for trigonometric case with $C=\mathbb{C}^{\times}$ with the one-form $\omega=\d z/z$ in the multiplicative coordinate $z$, however we will for the most part use the additive notation for 
notational simplicity.

We have therefore learned that the Lagrangian for the effective two-dimensional theory can be written down
using the classical $r$-matrix and the fields on the surface defects. 

\begin{figure}[htbp]
\centering\includegraphics[scale=0.32]{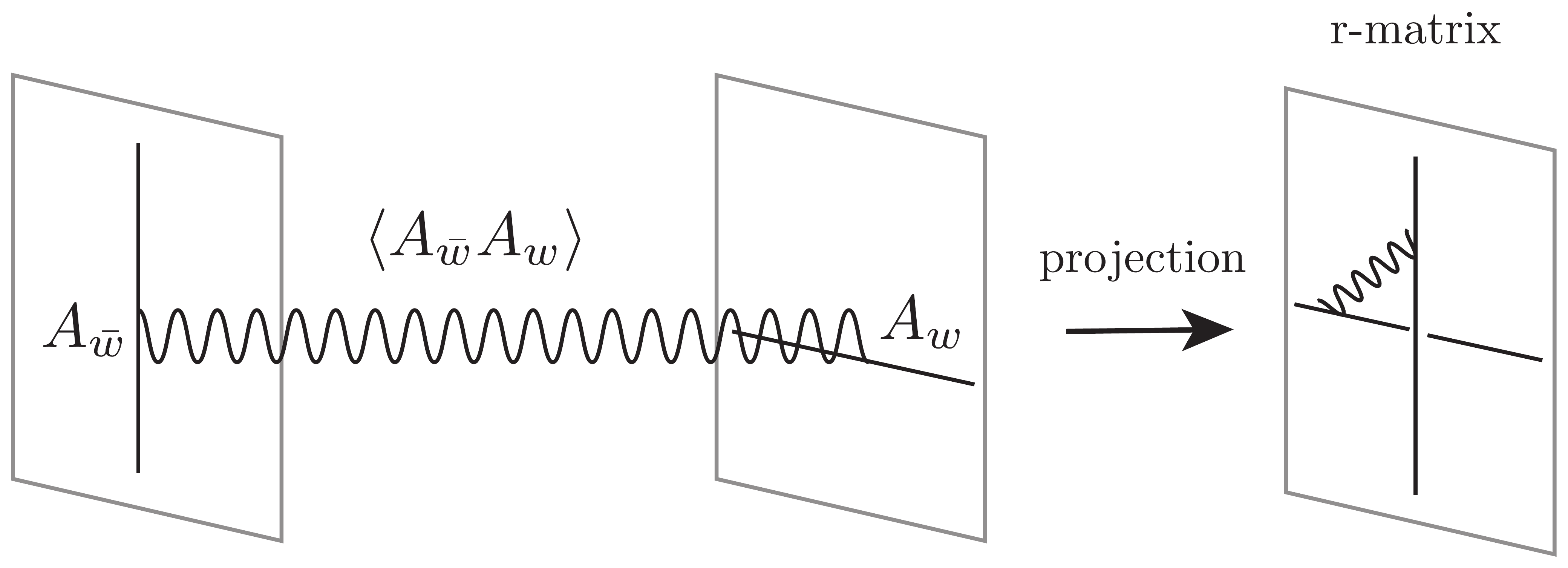}
\caption{The Feynman diagram computation of Figure \ref{fig:gluon_exchange} is the same as in the tree-level Feynman diagram
computation as in this figure, which gives the first tree-level contribution to the $R$-matrix, namely the computes the classical $r$-matrix.}
\label{fig:III_3}
\end{figure}

\begin{figure}[htbp]
\centering\includegraphics[scale=0.3]{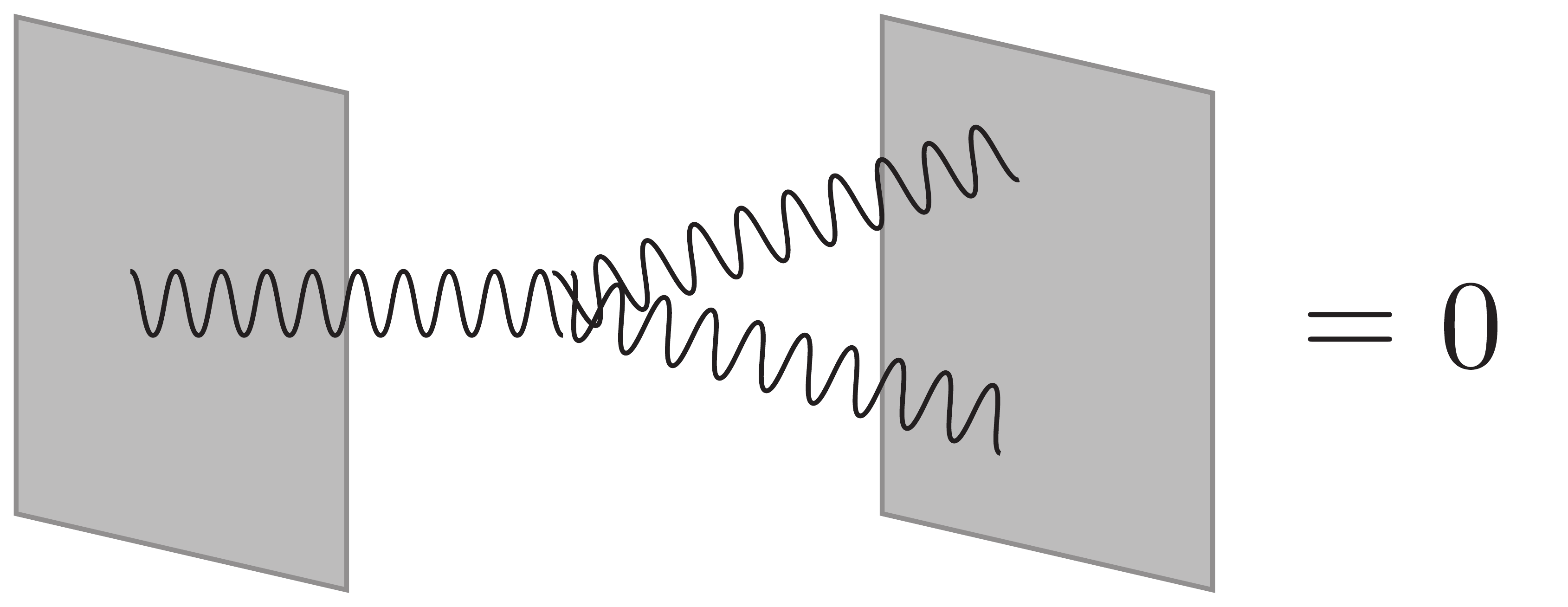}
\caption{The tree-level diagram in Figure \ref{fig:gluon_exchange} is the only non-zero Feynman diagram.
For example, the tree-level diagram in this Figure is zero, since the only vertex in the Chern-Simons theory involves all the three components $A_w, A_{\wbar}, A_{\zbar}$ of 
the gauge field, while in our gauge the propagator does not involve the $A_{\zbar}$ component.}
\label{fig:III_2}
\end{figure}

For example, suppose that we have $n_+$ chiral surface defects at $z_1,\dots,z_{n_+}$ and $n_-$ anti-chiral defects at $z_1',\dots, z_{n_-}'$,
so that the total number of defects is $n=n_++n_-$. Here a chiral (or anti-chiral) defect means that 
the coupling of the $\alpha$-th defect to the gauge field is
of the form $J_w^\alpha A_{\wbar}$ or $\bar{J}_{\bar{w}}^{\beta} A_w$ (in particular linear in the bulk gauge field $A$),
where $J_w^\alpha$, $\br{J}_{\wbar}^{\beta}$ are the currents for the $G$-action on the defects:
\begin{equation}
\begin{split}
\mc{L}_{\alpha}\left(\phi_\alpha; A_{\wbar}\right) 
&=\mc{L}_{\alpha}\left(\phi_\alpha\right) + J_w^\alpha A_{\wbar} \quad (\alpha=1, \dots n_+) \;,\\
\br{\mc{L}}_{\beta}\left(\phi_\beta; A_{w}\right) 
&=\br{\mc{L}}_{\beta}\left(\phi_\beta\right) + \bar{J}_{\wbar}^{\beta} A_{w} \quad (\beta=1, \dots n_-) \;.
\end{split}
\end{equation}
Note that the inclusion of such surface defects requires the specification of the 
complex structure (given by a  holomorphic coordinate $w$) on the topological plane.

In the following we often drop indices $w, \wbar$ from the currents $J_w^\alpha, \bar{J}_{\wbar}^{\beta}$,
to write $J^\alpha, \bar{J}^{\beta}$. We also write expand the currents with respect to an orthonormal basis $t_a$ of 
the symmetry algebra $\mathfrak{g}$, as $J^\alpha=J^\alpha_a t_a, \bar{J}^{\beta}=\bar{J}^{\beta}_b t_b$.

Integrating out the four-dimensional gauge field, we obtain 
a two-dimensional theory where the chiral and  anti-chiral theories are coupled by
the product of their currents:
\begin{equation} 
	\sum_{\alpha = 1}^{n_+} \sum_{\beta = 1}^{n_-} r_{ab}(z_\alpha- z'_\beta) J_a^\alpha \br{J}_b^\beta \;.
\end{equation}  

The action for the effective two-dimensional theory is thus given by
\begin{equation} 
S^{\rm eff}_{{\rm 2d}}=
\frac{1}{\hbar} \int_{\R^2 } \left[ 
\sum_{\alpha=1}^{n_+} \mc{L}^c_{\alpha}\left(\phi_{\alpha} \right) 
+
\sum_{\beta=1}^{n_-} \mc{L}^a_\beta\left(\phi_\beta\right) 
+
\sum_{\alpha = 1}^{n_+} \sum_{\beta = 1}^{n_-} r_{ab}(z_\alpha- z'_\beta) J_a^\alpha \br{J}_b^\beta
\right] \;.
 \end{equation}

This analysis applies equally in the rational, trigonometric, and elliptic cases,
where difference cases lead to different expressions for 
the classical $r$-matrix.
For example, for the rational case 
we have (see e.g.\ section 4 of \cite{Costello:2017dso})
\begin{equation}
r_{ab}(z,z')=\frac{c}{z-z'} \delta_{ab}
\;,
\label{rational_r}
\end{equation}
where $c=\sum_a (t_a\otimes t_a)$ is the quadratic Casimir operator.

\section{The Lax Operator and Conserved Non-Local Charges}
\label{sec:general_Lax}

\subsection{The Lax Operator in Four Dimensions}

Let us now come to the construction of the Lax operators for the two-dimensional theory introduced in the previous section.

We first construct the Lax operator in the coupled 4d--2d system. 
We will then integrate out the gauge fields along $\bar{C}$, to be left with a two-dimensional Lax operator.

The basic idea is rather simple.

Let us take $z \in \C \setminus \{z_\alpha\}_{\alpha=1}^n$ to be a point away from the surface defects. 
We will define a $\mf{g}$-valued one-form $\Lax^{4d}(z)$ on the plane $\R^2 \times \{z\}$
by the formula
\begin{equation} 
	\Lax^{4d}(z) =
	 A_{w}(z) \d w  +  A_{\wbar}(z)  \d \wbar
	\;.
	\label{eq.Lax_def}
 \end{equation}
Note that here $z$, which is one of the coordinates along the curve $C$,
is regarded as a parameter for the one-form on $\R^2 \times \{z\}$.

The zero-curvature equation for the one-form (on the plane $\R^2 \times z$)
\begin{equation}
\d\Lax^{\rm 4d}(z) +  \Lax^{\rm 4d}(z) \wedge \Lax^{\rm 4d}(z) = 0 \;,
\label{Lax_4d}
\end{equation}
then follows trivially thanks to the equations of motion for the four-dimensional gauge field $A_{\bar{z}}$:
\begin{equation}
\partial_{\wbar} A_w - \partial_{w} A_{\wbar}+ [A_{\wbar}, A_{w}]=0\;.
\end{equation}
We can therefore identify $\Lax^{4d}(z)$ as the two-dimensional Lax operator,
whose presence is one of the characterizations of classical integrability.

We now make several comments regarding this Lax operator.

Let us first note that the Lax operator defined in \eqn \eqref{eq.Lax_def} is not necessarily holomorphic in $z$,
since the gauge field $A_{w}(z), A_{\wbar}(z)$ in general have $\zbar$ dependence. This is related with the fact that the our Lax operator is gauge-dependent. In order to match the 
more standard definition of the Lax operator in the literature we can choose a gauge
$A_{\zbar}=0$, so that the equation of motion $F_{zw}=F_{z\wbar}=0$ implies
$\partial_{\zbar} A_w = \partial_{\zbar} A_{\wbar} =0$:
\begin{equation}
\Lax^{\rm 4d}(z)  \quad \textrm{is holomorphic in $z$ in the gauge} \quad  A_{\zbar}=0 \;.
\label{eq.Lax_holomorphic}
\end{equation}

Of course such a choice of gauge does not 
affect the physics, and we will see later that  what really matters is the gauge-invariant Wilson line for the gauge fields.

We can also use the BRST formalism. In this context it is natural to define
 an enhanced version of the four-dimensional Lax operator,
which contains a zero form in addition to a one-form:
\begin{equation} 
	\widehat{\Lax}^{\rm 4d}(z) 
	=
	\c +   \d w A_{w} + \d \wbar A_{\wbar} 
	\;. 
\end{equation}
Since the ghost number for $\c$ is one, if we treat an $i$-form as being of ghost number $i$,
then the total quantity $\widehat{\Lax}^{\rm 4d}(z)$ will be of ghost number one.  

Instead of asking that $\Lax^{4d}(z)$ satisfies the ordinary zero-curvature equation, we ask that it satisfies a modification of the equation which incorporates the BRST operator:
\begin{equation} 
 \d \widehat{\Lax}^{4d}(z) +\frac{1}{2} \left[ \widehat{\Lax}^{4d}(z) , \widehat{\Lax}^{4d}(z) \right]= Q_{\rm BRST} (-\c+A)=\textrm{(BRST closed)}\;.\label{equation_BRST}
\end{equation}
We can verify 
\eqn \eqref{equation_BRST} using the BRST transformation
\begin{equation}
\begin{split}
&Q_{\rm BRST}A= D_A \c \;, \\
&Q_{\rm BRST}\c=-\frac{1}{2}[\c,\c]
\;,
\end{split}
\end{equation}
together with the on-shell condition for the gauge field $F_{w \wbar}=0$, which holds away from the positions of the surface defects. 

As a side remark, it is also possible to have the equation \eqref{equation_BRST} off-shell, without imposing the equations of motion by hand. This is achieved by further enhancing the Lax operator to be
\begin{equation} 
\widehat{\Lax}^{\rm 4d}(z) =\c +   \d w A_{w} + \d \wbar A_{\wbar}   +\d w \d \wbar A_{\zbar}^\ast
\;, 
\end{equation}
where we have introduced the anti-field $A_{\zbar}^\ast$ to the field $A_{\zbar}$ as the two-form component; the BRST operator applied to the anti-field $A_{\zbar}^\ast$ includes  the equations of motion obtained by varying the field $A_{\zbar}$.  In the following 
we will for simplicity omit the anti-field component and will discuss one-shell zero-curvature conditions.

In the discussion above we have stayed away from surface defects. Since the gauge field couples to defects, there will be source terms at the defects and hence the one-form will in general have poles at the defects, as we will see.

We also note that on the boundaries of the spectral curve $C$ the Lax operator obeys the same boundary conditions on the gauge field components $A_w, A_{\wbar}$. For example, for the rational case $C=\C$ the boundary condition at infinity states that the Lax operator is zero at infinity.

Finally, in this paper we adopted the existence of the spectral-parameter dependent flat connection as a definition of classical integrability. One should keep in mind, however, that this does necessarily ensure that the whole system is integrable in the intuitive sense. An extreme example given by two-dimensional surface defects which are singlets under the four-dimensional gauge symmetry, so that the two-dimensional defects are completely decoupled from the four-dimensional theory. This example still fits into our formalism at the abstract level, and the two-dimensional degrees of freedom can obviously be non-integrable, despite the presence of the Lax operator. This does not arise, however, in the examples discussed in this paper.

\subsection{The Lax Operator in Two Dimensions}
After we integrate out the four-dimensional gauge field, the Lax operator in four dimensions gives rises to a Lax operator in the effective two-dimensional field theory, which we denote by $\Lax^{\rm 2d}(z)$.  This is a holomorphic function of $z$ with poles at the location of the surface defects.  

\begin{figure}[htbp]
\centering\includegraphics[scale=0.35]{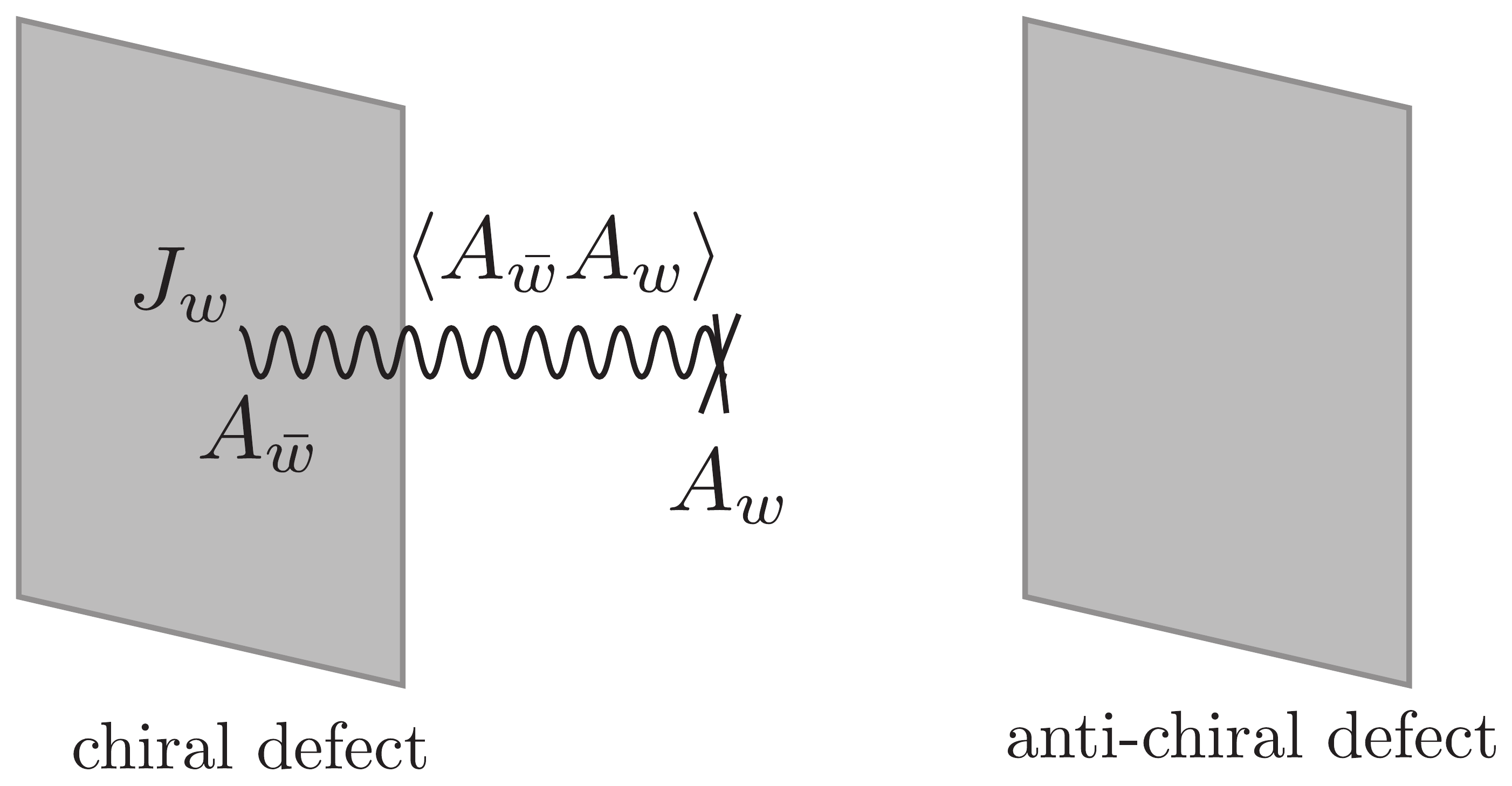}
\caption{The tree-level diagram for the computation of the Lax operator. We insert $A_w(z)$ at the position of the cross. We have a similar diagram for the 
$A_{\wbar}(z)$. The two diagrams are the only non-zero tree-level diagrams.}
\label{fig_tree_Lax}
\end{figure}

We can calculate $\Lax^{\rm 2d}(z)$ by tree-level Feynman diagrams (see Figure \ref{fig_tree_Lax}). We sum over tree-level Feynman diagrams, whose vertices are labelled by the terms in the Lagrangian coupling a surface defect to the four-dimensional gauge field; and with one special vertex where we insert the four-dimensional Lax operator.  A vertex has an incoming gluon line for every occurrence of $A_{\wbar}$, and an outgoing one for every occurrence of $A_w$, and a propagator is given by the classical $r$-matrix. We disallow diagrams where two vertices on the same surface defect are connected by a propagator. 

The fact that the four-dimensional Lax operator satisfies the flatness condition 
\eqref{equation_BRST} implies that the two-dimensional Lax operator also satisfies the flatness condition:
\begin{equation} 
 \d \Lax^{\rm 2d}(z) + \frac{1}{2}  [ \Lax^{\rm 2d}(z), \Lax^{\rm 2d}(z)]  \quad \textrm{is $Q_{\rm BRST}$-closed} \;.\label{2d_BRST} 
\end{equation}
When our set-up engineers two-dimensional field theories without gauge symmetry\footnote{If we work in the setting from which the four-dimensional gauge theory produces the dynamical Yang-Baxter equation, instead of the ordinary Yang-Baxter equation, we will find gauge symmetry in the effective two-dimensional theory.  The gauge symmetry will come from the non-trivial zero modes of the four-dimensional gauge theory. More generally, one could consider situations where the surface defects themselves had some gauge symmetry, and the four-dimensional gauge theory was coupled to flavor symmetry of these two-dimensional gauge theories.  In such a situation, one would also expect the effective two-dimensional theory to be a gauge theory.  In all these situations, one should use the BRST formalism in two dimensions as well.}, the BRST formalism is not necessary once we pass to two dimensions.  The Lax operator $\Lax^{2d}(z)$ can not have any zero-form component, because in two dimensions there are no local operators of ghost number $1$. 

 It follows that the operator-valued one-form $\Lax^{2d}(z)$ satisfies the zero-curvature equation
\begin{equation} 
 \d \Lax^{\rm 2d}(z) + \frac{1}{2}
  [ \Lax^{\rm 2d}(z), \Lax^{\rm 2d}(z)] = 0 \;,
 \label{2d_zero}
 \end{equation}
once we impose the equations of motion\footnote{If we use the  $4d$ Lax operator which includes a two-form component built from anti-fields,  then the $2d$ Lax operator will also  have such a two-form component. We then find that \eqn \eqref{2d_BRST} holds off-shell.} of the two-dimensional theory.  
Moreover, since the four-dimensional Chern-Simons theory (whose propagator we use) is holomorphic along the curve $C$, the Lax operator $\Lax^{\rm 2d}(z)$ is indeed holomorphic in the spectral parameter $z$.

As an illustration,
let us again consider the cases with $n_+$ chiral defects and $n_-$ anti-chiral defects.
We then have the one-form Lax operator to be 
\begin{equation}
\Lax^{\rm 2d}(z) =
\sum_{\alpha=1}^{n_+} J^{\alpha}(z_{\alpha}) r(z_{\alpha},z) dw +\sum_{\beta=1}^{n_-} \br{J}^{\beta}(z_{\beta}) r(z,z_{\beta})  d\wbar
\;.
\label{L2d_ca}
\end{equation}

Let us further specialize to the case $n_+=n_-=1$, with
chiral (anti-chiral) defect located at $z=1$ ($z=-1$). We moreover assume that 
we consider the rational case ($C=\mathbb{C}$), so that we can use \eqref{rational_r}.
We have
\begin{equation}
\Lax^{\rm 2d}(z) = 
\frac{ J(1)  \d w}{1-z}  +  \frac{\br{J}(-1)  \d\wbar}{z+1} 
=\frac{j+z \star j}{1-z^2}
\label{Lj}
\;,
\end{equation}
where we defined a $\g$-valued one-form $j$ (current) by 
\begin{equation}
j:= 
J(1) \d w+ \br{J}(-1) \d\wbar 
\;, 
\end{equation}
and $\star$ is the Hodge-star operator in the two-dimensional space 
so that $\star j=J(1) \d w - \br{J}(-1) \d\wbar$.\footnote{We here choose the Lorentzian signature for the two dimensions, so that we have $\d s^2=\d x^2-\d y^2$, $w=x+y, \br{w}=x-y$ and $\star (\d w)=\d w, \star(\d\br{w})=-\d \br{w}$ and $\star^2=1$. We can go back to the Euclidean signature by replacing $y$ by $\i y$, and then the spectral parameter $z$ is also rotated by a factor of $\i$.}

The relation \eqref{Lj} specifies the relation between the Lax operator $\Lax$ and the current $j$.
Under this relation, the zero-curvature equation for the Lax operator is 
translated into the two equation for the current one-form,
one the current conservation and second a flat connection equation for the current:
\begin{align}
&\d \star j =0 \;, \label{j_conserve}\\
&\d j+j\wedge j=0 \label{j_zero}\;.
\end{align}

In the literature once we often start with a current $j$ satisfying \eqns\eqref{j_conserve} and \eqref{j_zero},
and then defines the Lax operator by the relation \eqref{Lj}. For this reason one might be 
tempted to think that the current $j$ is more fundamental than the Lax operator,
and take the zero-curvature condition for the $\g$-valued current $j$ \eqref{j_zero} as the definition of 
integrability.

In our approach, however, the connection between the currents and the Lax operator is not too fundamental.  
Indeed, in the trigonometric and elliptic cases, the integrable theory will not have the full $\g$-symmetry (and hence no $\g$-current) due to boundary conditions \cite{Costello:2017dso}, but will still have a $\g$-valued Lax operator.  
Even in the rational case, the Lax operator for the most general class of models we consider, namely those associated with defects 
which is neither chiral nor anti-chiral, is not simply a linear combination of the currents (see section \ref{sec:scalars}).  

\subsection{Line Defects and the Lax operator}

The existence of the Lax operator satisfying the zero-curvature condition \eqref{2d_zero}
is often used as a definition of the integrability of the system---the utility of the Lax operator is that it allows one to define infinitely-many non-local conserved charges. Let us recall this well-known fact.

Suppose we have a two-dimensional theory with a Lax operator for the Lie algebra $\mf{g}$ satisfying the zero-curvature equation  \eqref{2d_zero}. 
Let us put the two-dimensional theory on a cylinder $\R \times S^1$ with coordinates $t,\theta$, where we choose the period of $\theta$ to be
$2\pi$.   Choose a representation $V$ of $\mf{g}$.  We can construct an operator
\begin{equation} 
W(t_0,z, V) = \op{Tr}_V P \exp \int_{\theta = 0}^{2 \pi} \Lax(z) 
\label{eq.monodromy}
 \end{equation}
as the trace in $V$ of path-ordered exponential of the Lax operator along the circle $t = t_0$ in the cylinder.

\begin{figure}[htbp]
\centering\includegraphics[scale=0.3]{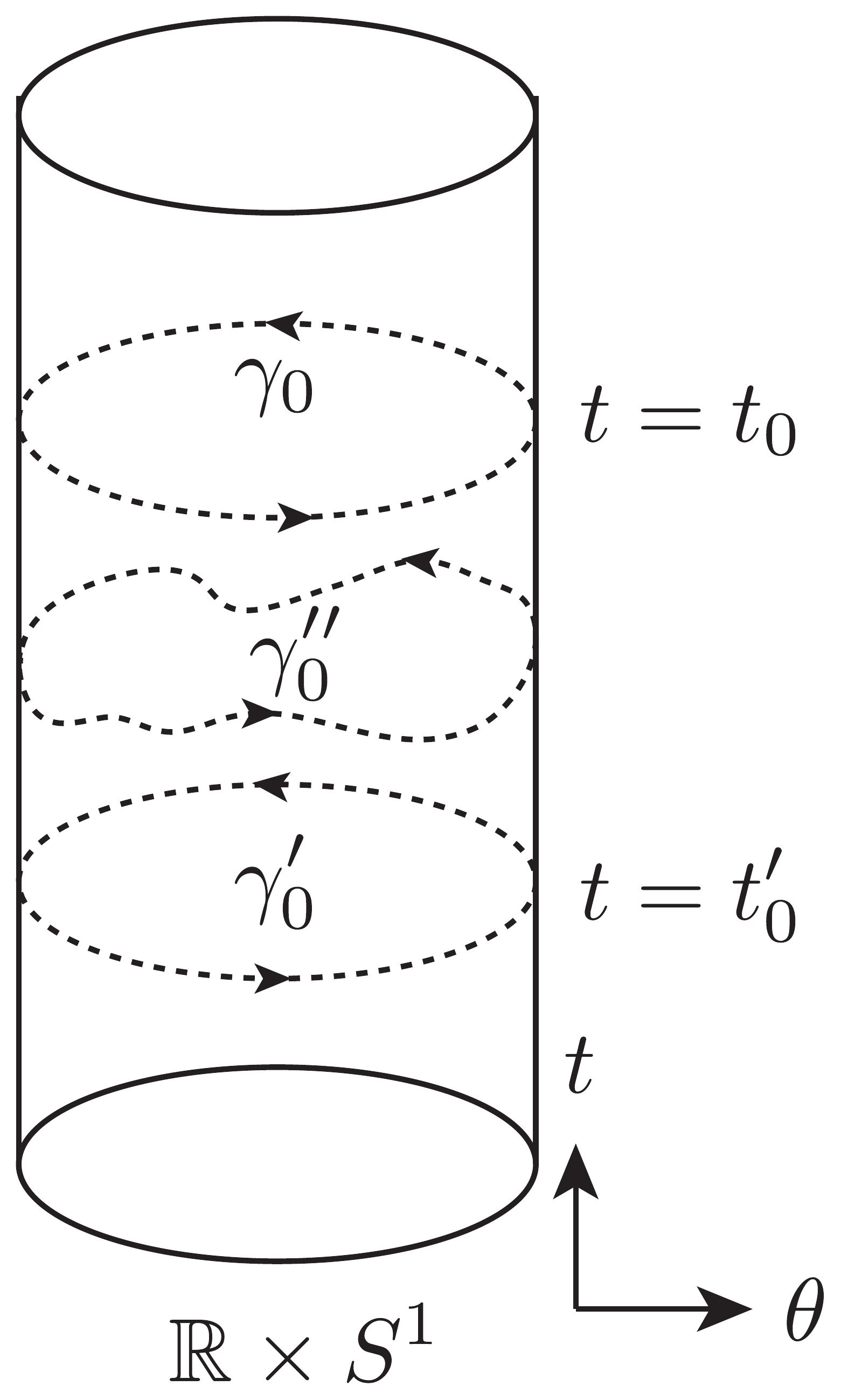}
\caption{We compactify the $\mathbb{R}^2$ into the cylinder $\mathbb{R}_t\times S^1_{\theta}$,
and consider the monodromy operator around a contour $\gamma_0$ at a fixed radial coordinate $t=t_0$ as in \eqn \eqref{eq.monodromy}. The trace of the monodromy is invariant under continuous deformation of the contour, we can deform the contour $\gamma_0$ to $\gamma_0''$ for example. }
\end{figure}

The zero-curvature equation  \eqref{2d_zero} implies immediately that the operator $W(t_0, V)$ is independent of the value of $t_0$:
\begin{equation} 
\partial_{t_0} W(t_0,z, V) = 0 \;. 
 \end{equation}
 That is, $W(t_0,z,V)$ is conserved.    By expanding in $z$, say at around $z=\infty$, we find that the Lax operator provides an infinite number of conserved (non-local) charges acting on the Hilbert space of the theory on a circle:
 \begin{equation}
 W(t_0,z, V) = \exp\left( \sum_{n=0}^{\infty} \frac{Q_n}{z^n} \right) \;.
 \end{equation}

In case one is interested in the case where the effective two-dimensional theory has a gauge symmetry,
then one should use the 
zero-curvature equation with the BRST operator included \eqref{equation_BRST}. This equation implies that the operator $\widehat{W}(t_0,z, V)$
defined by
\begin{equation} 
\widehat{W}(t_0,z, V) = \op{Tr}_V P \exp \int_{\theta = 0}^{2 \pi} \widehat{\Lax}(z) 
 \end{equation}
 satisfies
\begin{equation} 
\partial_{t_0}  \d t \, \widehat{W}(t_0,z, V)  \quad \textrm{is $Q_{\rm BRST}$-closed}  \;. 
 \end{equation}
Expanding the operator $\widehat{W}$ as
\begin{equation}
\widehat{W}(t_0,z, V) =w_0 + w_1 \d t \;,
\end{equation}
we find that 
 the quantity $w_0$, which is $Q$-closed and hence defines an element of the BRST cohomology,
is conserved up to a BRST exact term $Q_{\rm BRST} w_1$. This is what one wishes to show.

\subsubsection{Back to Four Dimensions}

We now have engineered a two-dimensional integrable theory, with its Lax operator, from surface defects in a four-dimensional gauge theory.  However,  one might get the impression that the discussion above is ad-hoc.
Where do these conserved non-local charges really come from physically? In our narrative many things are more transparent if we go back to the coupled 4d--2d system before integrating out KK modes.

As we have explained, 
the two-dimensional Lax operator $\Lax^{\rm 2d}(z)$ is obtained form the four-dimensional Lax operator $\Lax^{\rm 4d}(z)$ by integrating out the gauge field along the curve $C$.  Instead of performing the path-ordered exponential after integrating out the four-dimensional gauge field, we can do it before we integrate out the gauge field: the answer will be the  same.  

The path-ordered exponential of the four-dimensional Lax operator yields, rather tautologically, the Wilson line of the four-dimensional gauge theory, at some point $z$ in the holomorphic plane and on some path in the topological plane:
\begin{equation} 
W(z, V) = \op{Tr}_V P \exp \int_{\gamma \times \{ z \}} A  \;,
\end{equation}
where $\gamma$ is a winding cycle around the cylinder.

We have concluded that the non-local conserved currents built from the Lax operator have a very straightforward interpretation in terms of our construction. They are obtained by simply inserting a Wilson line into the four-dimensional gauge theory, at some value of $z$ distinct from the location of the surface defects, and then integrating out the four-dimensional gauge field.  The fact that the four-dimensional theory is topological in the $w,\wbar$ plane (away from the locations of the surface defects in general) implies that moving the Wilson line in this plane has no effect. This is implies that the corresponding non-local current in two dimensions is conserved. Simply put, the effective two-dimensional theory is integrable simply because there exists topological Wilson lines in four dimensions!

Summarizing, in the four-dimensional perspective we have integrability simply because there are Wilson lines in the four-dimensional theory.
This perspective will be of great use when we discuss quantum integrability in the companion paper \cite{Part4}.

\subsection{Spectral Curves}
The Wilson line extracts gauge-invariant information from the monodromy operator
\begin{equation} 
\Omega(z) = P \exp \int_{\gamma \times \{ z \}} A \;.
\label{eq.Omega}
\end{equation}
by computing a trace in some representation.

There are, however, other gauge-invariant combination of the monodromy.
We can consider the 
spectral curve
\begin{equation} 
\tilde{C}_V: \quad 
\left\{ (\lambda, z)\, \Big| \, \lambda\in \mathbb{C}, \quad z\in C, \quad
\det{}_{V} (\lambda \, \textrm{Id}-\Omega(z) ) =0 \right\}\;,
\end{equation}
for a representation $V$. This is a branched cover over the base spectral curve $C$.
For example, for $G=SL_N$ we can choose $V$ to be an $N$-dimensional vector representation,
so that $\tilde{C}_V$ is an $N$-fold branched cover of $C$.

This spectral curve is also the spectral curve for the group-valued Hitchin system (also called the
multiplicative Hitchin system, see e.g.\ \cite{Elliott:2018yqm} and references therein).
This is because the moduli spaces of classical equations of motion of the four-dimensional Chern-Simons theory define the
group-valued Hitchin system on the spectral curve $C$.

On the spectral curve the $A_{w}$ and $A_{\wbar}$ components 
combine into a complex scalar field (the Higgs field).
The remaining $A_{\zbar}$ component gives rise to 
the holomorphic monodromy operator $\Omega(z)$ as in \eqn \eqref{eq.Omega}, 
and hence defines a holomorphic section of the group-valued
adjoint bundle. Note that $\Omega$ is group ($G$)-valued, as opposed to Lie-algebra ($\mf{g}$)-valued in the standard Hitchin system.

The moduli space of the group-valued Hitchin system is infinite-dimensional
if we allow arbitrary poles, but is truncated to finite-dimensional
once we specify the singularity structures appropriately.
In the rest of this paper we will discuss many such boundary conditions.

\section{Coupling with Free Fermions}\label{sec:fermions}

\subsection{Two-dimensional Lagrangian}

A simple example of chiral and and anti-chiral defect theories 
is provided by a free chiral or anti-chiral fermion, living in a real representation $R$ of the gauge group $G$.  The chiral fermion is a field
$\psi \in \Omega^{1/2,0}(\C,R)$ at $\C \times z_0$.  The action functional is 
\begin{equation} 	
S_{z_0}=\frac{1}{\hbar} \int_{\C \times z_0} \ip{\psi, \dbar_A \psi} \;,
\label{psi_chiral}
\end{equation}
where $\dbar_A=\partial_{\wbar} + A_{\wbar}$ and 
$\langle - ,-\rangle$ is a $G$-invariant symmetric pairing
in the representation $R$.
This action only depends on the $A_{\wbar}$ component of the four-dimensional gauge field.

Similarly, at $z_1$ we can introduce an anti-chiral fermion $\br{\psi} \in \Omega^{0,1/2}(\C, R)$ in the representation $R$, with action
\begin{equation} 
	S_{z_1}=\frac{1}{\hbar} \int_{\C \times z_1}\ip{ \br{\psi}, \partial_A \br{\psi}} \;,
	\label{psi_antichiral}
\end{equation}
where $\partial_A=\partial_{w} + A_{w}$.
This action only depends on the $A_w$ component of the four-dimensional gauge field. 

In the following we will sometimes omit the brackets $\langle -, - \rangle$ to write 
for example $\psi \dbar_A \psi$ and $\br{\psi} \partial_A \br{\psi}$ instead of $\ip{\psi , \dbar_A \psi}$ and $\ip{\br{\psi}, \partial_A \br{\psi}}$, as is often done in the physics literature.

We can include $n$ chiral fermions $\psi^{(\alpha)}$ at $z_\alpha$ ($\alpha=1, \dots n$)
and $m$ anti-chiral fermions $\br{\psi}^{(\beta)}$ at $z'_{\beta}$ ($\beta=1, \dots m$).
The action for the full coupled 4d--2d system is, including the factors of $\hbar$, is
\begin{equation} 
S_{\rm 4d-2d}=\frac{1}{\hbar}\left[
\frac{1}{2 \pi} \int \d z \, {\rm CS}(A) + \sum_{\alpha=1}^n  	\int_{\C \times z_\alpha} \psi^{(\alpha)}  \dbar_A \psi^{(\alpha)}  + 
\sum_{\beta=1}^m  \int_{\C \times z'_\beta} \br{\psi}^{(\beta)} \partial_A \br{\psi}^{(\beta)}
\right]
 \;.
\label{fermionic_coupled} 
 \end{equation}

To simplify the notation, let us focus below on the simple situation where we have inserted two surface defects into our four-dimensional gauge theory, one given by chiral fermions and one by anti-chiral fermions in the same real representation $R$ of $\g$ (in the previous notation we consider the case $n_-=n_-=1$, and we drop the indices $(\alpha)$ and $(\beta)$); generalization to the case of multiple chiral/anti-chiral defects is straightforward.

Let us choose an orthonormal basis of the real representation $R$ where the fermions live, and let $(t_a)_{ij}$ denote the matrix of the action of the Lie algebra basis element $t_a$.  Then the currents as given from \eqref{psi_chiral} and \eqref{psi_antichiral} are given by
\begin{equation} 
\begin{split}
	J_a(\psi) &=\ip{\psi, t_a \psi} = \psi t_a \psi \;, \\
	\br{J}_a(\br{\psi}) &=\ip{\br{\psi}, t_{a} \br{\psi}} = \br{\psi} t_{a} \br{\psi}   \;,
\end{split}	
\end{equation}
where $\psi t_a \psi  =\psi_i (t_a)_{ij}\psi_j$ if we explicitly write the $i,j,\dots$ indices for the weight space
of the representation.

Integrating out the four-dimensional gauge field gives a classical coupling $r_{ab}(z_0 - z_1) J_a \br{J}_b$. 
The full Lagrangian 
for the effective two-dimensional theory is
\begin{equation}	
	S_{\rm 2d}^{\rm eff}=\frac{1}{\hbar}\left[
	\int \psi \dbar \psi +
	\int \br{\psi} \partial \br{\psi} + 
	 r_{ab}(z_0 - z_1) \int J_a(\psi) \br{J}_b(\br{\psi}) 
	\right]\;. 
	\label{2d_psi}
\end{equation}

\subsection{The Lax Operator}

We have seen in section \ref{sec:general_Lax} that a formal analysis of the four-dimensional set-up guarantees that the Lax operator satisfies the zero-curvature equation. In this subsection we verify as a consistency check the zero-curvature equation more directly from the classical Yang-Baxter equation.

The Lax operator is (recall \eqn \eqref{L2d_ca})
\begin{equation} 
	\Lax_a(z) =
	r_{ab}(z_0- z) J_b(\psi) \d w + r_{ab}(z-z_1) \br{J}_b(\br{\psi}) \d \wbar 
	\;. 
\end{equation}
This is a one-form valued in $\g$. Note that we have
$(t_a)_{ij}=-(t_a)_{ji}$ since the matrices $(t_a)_{ij}$ define a real representation of $\mf{g}$, 
 
The equations of motion for the fields $\psi,\br{\psi}$, as derived from the action \eqref{2d_psi} are given by
\begin{equation} 
\begin{split}
	-\partial_{\wbar} \psi_i +
	r_{ab}(z_0 - z_1) (t_a)_{ij} \psi_j \br{J}_b(\br{\psi}) &= 0 \;,\\ 
	-\partial_w \br{\psi}_i + 
	r_{ab}(z_0 - z_1) J_a(\psi) (t_b)_{ij} \br{\psi}_j &=  0 \;.
\end{split}	  
\end{equation}
From this we see that
\begin{align} 
	\partial_w \Lax_{a,\wbar} (z) &=  2 r_{ab}(z-z_1) (t_b)_{ij} (\partial_w \br{\psi}_i) \br{\psi}_j  \nonumber  \\
	&= 2  r_{ab}(z-z_1 ) (t_b)_{ij} (t_d)_{ik}  r_{cd}(z_0 - z_1)  J_c \br{\psi}_k \br{\psi}_j \;. 
\label{delL_psipsi}
 \end{align} 
In \eqn \eqref{delL_psipsi} the presence of $\br{\psi}_k \br{\psi}_j$ means that the the expression is anti-symmetric in the $j,k$ indices.  
Taking advantage of this fact, and also the commutation relation $[t_b, t_d]=f_{bde} t_e$, we obtain
\begin{align} 
	\partial_w \Lax_{a,\wbar} (z) 
	&=- r_{ab}(z-z_1) [t_d, t_b]_{kj}  r_{cd}(z_0 - z_1)  J_c \br{\psi}_k \br{\psi}_j \;,\nonumber\\	
	&=-r_{ab}(z-z_1)  f_{dbe} (t_e)_{kj}  r_{cd}(z_0 - z_1)  J_c \br{\psi}_k \br{\psi}_j \;, \nonumber\\
 	&=  r_{ab}(z-z_1) r_{cd}(z_0 - z_1) f_{bde}  \br{J}_e J_c \;. 
 \end{align}
 Similarly,
 \begin{equation} 
 \partial_{\wbar} \Lax_{a,w}(z) = r_{ab}(z_0 - z) r_{cd}(z_0 - z_1) f_{bce}  \br{J}_d J_e \;. 
  \end{equation}
Finally, 
  \begin{equation} 
  [\Lax_{w}(z), \Lax_{\wbar}(z)]_a =r_{be}(z_0-z) r_{cd}(z-z_1) f_{bca} \br{J}_d J_e \;. 
   \end{equation}
 The zero-curvature equation  
 \begin{equation} 
\partial_w \Lax_{a,\wbar}(z) - \partial_{\wbar} \Lax_{a,w}(z) +  [\Lax_w(z), \Lax_{\wbar}(z)]_a = 0 \;,
\end{equation}
reduces to
\begin{equation}
r_{ab}(z_{21}) r_{ec}(z_{01}) f_{bcd}
+ r_{ab}(z_{02})r_{cd}(z_{01})  f_{bce}  +  r_{be}(z_{02}) r_{cd}(z_{21}) f_{bca} =0 \;,
\end{equation}   
where we wrote down the coefficient of $\br{J}_d J_e$,
we introduced $z=z_2$ as well as a shorthand notation $z_{ij}=z_i-z_j$.
Let us contract the indices $a,d, e$ with an element $t^d\otimes t^a\otimes t^e$.
We obtain, after some help with the unitarity constraint for the classical $r$-matrix $r_{ab}(-z) = - r_{ba}(z)$, 
\begin{equation}
\begin{split}
& \left[r_{ba}(z_{12})(  t^b\otimes t^a\otimes 1) ,   r_{ce}(z_{10}) ( t^c \otimes 1 \otimes t^e)\right] \\
& \quad + \left[  r_{dc}(z_{10}) (t^d \otimes 1\otimes t^c) , r_{ab}(z_{20}) (1\otimes t^a\otimes t^b )  \right] \\
& \quad +\left[   r_{dc}(z_{12}) (t^d\otimes t^c\otimes 1) ,r_{be}(z_{20}) (1\otimes t^b\otimes t^e)  \right]=0 \;.
\end{split}
\end{equation}  
or rather
\begin{equation}
\left[  r_{12}(z_{12}) ,r_{13}(z_{10})  \right] 
 + \left[  r_{13}(z_{10}), r_{23}(z_{02})    \right] 
  +\left[   r_{12}(z_{12})  , r_{23}(z_{02})    \right]=0 \;,
\end{equation}  
where e.g.\ $r_{12}$ acts on the first and the second components of the tensor product: $r_{12}(z)=r_{ab}(z)(  t^a\otimes \otimes t^b \otimes 1)$. This is nothing but the classical Yang-Baxter equation.

\subsection{Rational Cases} 

Integrating out the gauge field $A$ couples the chiral and anti-chiral fermions. 
To describe this, let us choose an orthonormal basis $t_a$ of the real representation $R$, and of the Lie algebra $\g$. 
Integrating out the gauge field leaves the action 
\begin{equation} 
S_{\rm 2d}^{\rm eff}=\frac{1}{\hbar}\left[\int_{\R^2} \psi \dbar \psi + 	\int_{\R^2} \br{\psi} \partial \br{\psi} + \frac{1}{z_0 - z_1}\int_{\R^2} (\psi t_a \psi) ( \br{\psi} t_{a}\br{\psi})\right]\;. \label{fermion_-ction} 
\end{equation}

There are two simple special cases of this construction which lead to familiar models.  If $\g = \mf{so}_n$ and $R$ is the vector representation, we find the Gross-Neveu model \cite{Gross:1974jv}.  If $g = \mf{sl}_n$ and $R$ is the sum of the fundamental anti-fundamental representation, we find the (massless) Thirring model \cite{Thirring}.  

There are clearly many other models of this type, including the obvious generalization where the chiral and anti-chiral fermions live in different real representations. As we will see in \cite{Part4}, not all configurations can be realized consistently at the quantum level.  In order for a certain anomaly to cancel, we require that the real representations for the chiral and anti-chiral fermions have the same Dynkin index.  

As a consistency check, we note that the equations of motion implied by \eqn \eqref{fermion_-ction} are the same as those of the four-dimensional gauge theory in the presence of the surface defects.
In the gauge $A_{\br{z}}=0$, the four-dimensional equations of motion for the gauge field are given by
\begin{equation}
\begin{split}
\partial_{\br{z}} A_{\br{w}} &= \pi \br{J}^a \delta_{z_0} \;, \quad
\partial_{\br{z}} A_{w} = -\pi J^a \delta_{z_1} \;, \quad
F_{w \br{w}}=0 \;.
\end{split}
\end{equation}
These equations can be solved by
\begin{equation} 
A^a_w=-\frac{1}{2(z - z_0)} J^a  \;, \quad
A^a_{\br{w}}=\frac{1}{2(z - z_1)} \br{J}^a \;.
 \end{equation}
We then insert this value of $A$ into the original Lagrangian \eqref{fermionic_coupled} to obtain the Lagrangian \eqref{fermion_-ction}. 

\subsection{Trigonometric and Elliptic Cases} \label{subsec:fermion_trig_ell}
Let us specialize to the case when the fermions are in the fundamental representations of $\mf{g} = \mf{sl}_2$ and write down the the Lagrangians coming from our construction in the trigonometric and elliptic cases.  We will need to recall formulae for the classical $r$-matrix in these two cases.  

For the trigonometric case, the classical $r$ matrix is
(see section 9.4 of \cite{Costello:2017dso}, where we chose $\tilde{h}=0$)
\begin{equation} 
	r(z_1,z_2) = \frac{1}{1 - z_1/z_2} e \otimes f - \frac{1}{1 - z_2 / z_1} f \otimes e + \frac{z_2 + z_1}{4 (z_2 - z_1)} h \otimes h \;,
	\label{trig_r}
\end{equation}
where we use the basis $e,f,h$ for $\mf{sl}_2$, with $[e,f] = h$, $[h,e] = 2 e $, $[h,f] = -2f$. Note that in the rational limit 
this reduces to the rational classical $r$ matrix (recall \ref{rational_r})
\begin{equation} 
	r(z_1-z_2) = \frac{e \otimes f + f \otimes e + \frac{1}{2} h \otimes h }{z_1- z_2} \;,
	\label{rational_r_sl2}
\end{equation}
 
For elliptic case we consider a rigid holomorphic $PGL_2$ bundle on the elliptic curve so that we have no four-dimensional zero modes along $C$.
It will be convenient to write the $r$-matrices in terms of the Pauli matrix basis of $\mf{sl}_2$, given by $t_1,t_2,t_3$ which satisfy $[t_1,t_2 ] = t_3$ and cyclic permutations.  These are related to the basis $e,f,h$ used above by relations $t_1 = \i h/2$, $t_2 = (e-f)/2$, $t_3 = \i (e+f)/2$.  

Let us define three doubly quasi-periodic functions of a complex variable by
\begin{equation} 
\begin{split}
	w_1(z) &= \sum_{n,m \in \Z} (-1)^n \frac{1}{z + m + n \tau} \;,\\
        w_2(z) &= \sum_{n,m \in \Z} (-1)^m \frac{1}{z + m + n \tau} \;,\\
	w_3(z) &= \sum_{n,m \in \Z} (-1)^{n+m} \frac{1}{z + m + n \tau} \;.	
\end{split}	 
\label{w_def}
\end{equation}
Each function is uniquely determined by the quasi-periodicity properties they satisfy with respect to the transformations $z \mapsto z + 1$, $z \mapsto z + \tau$, and by the fact that they have a simple pole on the lattice points $n + m \tau$ with residue $1$ and no other poles (These functions can also be written as ratios of $\theta$-functions).

Then the elliptic $r$-matrix is (see section 10.3 of \cite{Costello:2017dso})
\begin{equation} 
	r(z_1,z_2) = (t_1  \otimes t_1) w_1(z_1 - z_2) +( t_2 \otimes t_2) w_2 (z_1  - z_2) + (t_3 \otimes t_3) w_3 (z_1 - z_2) \;.  
\end{equation}

Let us calculate the Lagrangian we find in the trigonometric and elliptic cases when our chiral and anti-chiral surface defects are given by a free fermion valued in the adjoint representation of $\mf{sl}_2$ (or equivalently the vector representation of $\mf{so}_3$).  We will place the chiral and anti-chiral defects at $z_0$ and $z_1$.  We will write the Lagrangian in the basis of the fermions given by the $e,f,h$ basis of  $\mf{sl}_2$: $\psi=\psi_e e+\psi_f f+\psi_h h$. We will choose the normalization of the invariant form $\langle e, f \rangle=1, \langle h, h \rangle
=2$.

In the trigonometric case, we find that the two-dimensional action is\footnote{
Note, for example, $\psi e \psi=\langle \psi, e \psi \rangle=
\langle h, e f\rangle \psi_h \psi_f +\langle f, e h \rangle \psi_f \psi_h
=2 \psi_h \psi_f$ and $\psi h \psi=\langle \psi, h \psi \rangle=
\langle e, h f \rangle \psi_e \psi_f +\langle f, h e \rangle \psi_f \psi_e
=2 \psi_f \psi_e$.
} 
\begin{equation} 
\begin{split}
S^{\rm eff}_{\rm 2d}=\frac{1}{\hbar} & \left[ 2 \int \psi_e \dbar \psi_f  + 2 \int \psi_h \dbar \psi_h + 2 \int \br{\psi}_e \partial \br{\psi}_f + 2 \int \br{\psi}_h \partial \br{\psi}_h  \right. \\
	&\qquad+ \frac{4}{1 - z_1 / z_2}   \int \psi_f \psi_h \br{\psi}_f \br{\psi}_h + \frac{4}{1 - z_2 / z_1}  \int \psi_e \psi_h \br{\psi}_e \br{\psi}_h  \\
	&\qquad +\left.    \frac{z_2 + z_1}{z_2 - z_1}  \int \psi_e \psi_f \br{\psi}_e \br{\psi}_f  \right]\;.  
\end{split}	
\end{equation}

In the elliptic case, using the basis $t_{i}$ ($i=1,2,3$) as before (i.e. $\psi=\sum_{i=1}^3 \psi_i t_i$) with the normalization of the invariant form given by $\langle t_i, t_j \rangle=-\delta_{ij}/2$, we find 
\begin{equation} 
\begin{split}
S^{\rm eff}_{\rm 2d}=\frac{1}{\hbar}\Bigg[ &-\frac{1}{2}	\sum_{i=1}^3 \int \psi_i \dbar \psi_i -\frac{1}{2} \sum_{i=1}^3 \int \br{\psi}_i \partial \br{\psi}_i    \\
& \qquad + \frac{1}{4} \sum_{i,j,k,l,m=1}^3 w_i(z_0 - z_1) \eps_{ijk} \eps_{ilm}\int\psi_j \psi_k \br{\psi}_l \br{\psi}_m  \Bigg]\;.
\end{split}	
\end{equation}
This model is generally not unitary.

\section{Coupling with Free Scalars}\label{sec:scalars}

Let us next discuss an example of a defect which is neither chiral nor anti-chiral: a free scalar field\footnote{As we will see in \cite{Part4}, there is a quantum anomaly in the gauge theory coupled to a free scalar. This example is mostly of interest at the classical level. It is possible to modify this example to make it anomaly free by replacing the free scalar by certain $\sigma$-models on homogeneous spaces.} theory living in some real representation $R$ of the gauge group $G$.  Such a field theory has fields $\phi_i$, as $i$ ranges over an orthonormal basis of $R$, with action
\begin{equation} 
\begin{split}
	&	\int_{\R^2 \times z} \partial_{A} \phi \dbar_A \phi \\
	&= \int_{\R^2 \times z} \partial_w \phi_i \partial_{\wbar}  \phi_i + A_{w,a} t_{aij} \phi_{j} \partial_{\wbar} \phi_i  
	+ \partial_w \phi_i A_{\wbar,a} t_{aij} \phi_j + A_{\wbar,a} A_{w,b} t_{aij} \phi_j t_{bik} \phi_k \;.
\end{split}	
\end{equation}
Here $t_{aij}$ is the matrix for the action of the Lie algebra element $t_a \in \g$  on the representation $R$.

A surface defect of this form is coupled both to $A_w$ and $A_{\wbar}$, and the Lagrangian describing the coupling has a term quadratic in $A$.  

Let us consider the Feynman diagram analysis for integrating out the gauge field $A$ at tree level, in the presence of some chiral and anti-chiral defects and a single free scalar defect.   We find that there are extra diagrams that can contribute compared to the previous situation.  There are diagrams with a single propagator, which connects a chiral defect with an anti-chiral defect, a chiral defect with the free scalar defect, or an anti-chiral defect with the free scalar defect.  There are also diagrams with two propagators, where one propagator connects a chiral theory with the free scalar defect, and the other connects an anti-chiral theory with a free scalar defect.   

\begin{figure}[htbp]
\centering\includegraphics[scale=0.23]{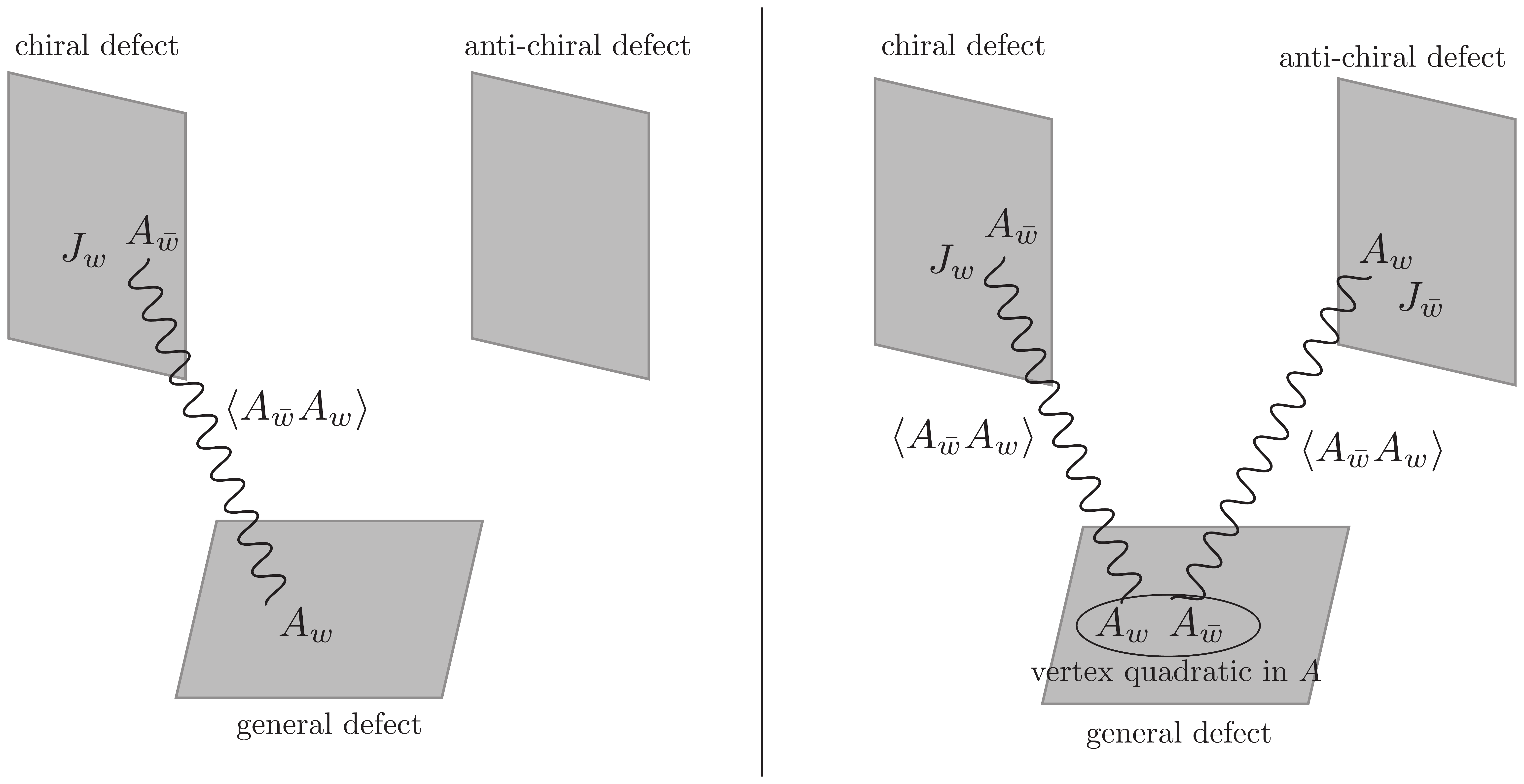}
\caption{Once we have a defect which is neither chiral nor anti-chiral, we have many more tree-level diagrams contributing to the 
two-dimensional effective action.}
\label{fig_tree_chiral}
\end{figure}

\begin{figure}[htbp]
\centering\includegraphics[scale=0.35]{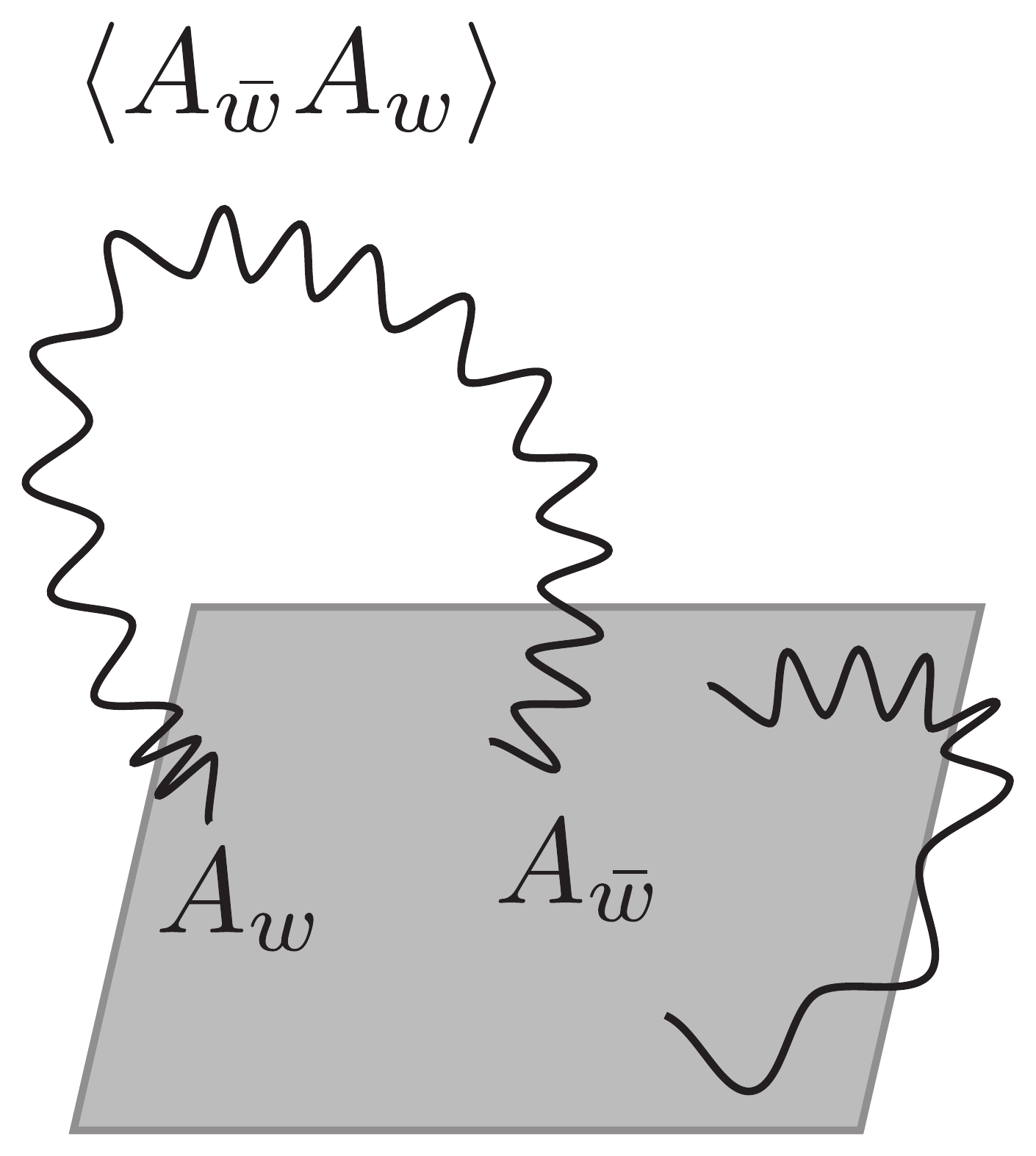}
\caption{The propagator(s) connecting the non-chiral defect to itself.}
\label{fig_tree_self}
\end{figure}

In principle, one can have diagrams where there are several copies of the bivalent vertex associated to the free scalar defect,  connected by a propagator.  Such diagrams have a UV divergence, as they involve evaluating the propagator $P(z,z')$ at $z = z'$.   The solution to this is to regularize the propagator, and then analyze what happens when we take the regulator away.  One natural regulator to use is based on the Feynman parametrization: we can write 
\begin{equation} 
\frac{1}{z} = \int_{t = 0}^{\infty} \zbar \,  e^{-\frac{ z \zbar}{t}} \frac{\d t}{t^2}  \;.
 \end{equation}
A UV regulator is defined by performing the integral over the region where $t \ge \eps$.  If we do this, and insert the regulated propagator into the calculation, we find that propagators that link a surface defect to itself yield zero.  This is because the regulated propagator involves $\zbar - \zbar'$, and becomes zero if we set $\zbar = \zbar'$.  

With this regulator, we find that we should only sum over trees where no two vertices associated to the same surface defect are connected by a propagator.  

Let us write out explicitly the result of this calculation in the case when there is one chiral defect at $z_0$, one anti-chiral defect at $z_1$, and one free scalar field theory at $z_2$.  Let $J_a$, $\br{J}_a$ be the currents for the chiral and anti-chiral theories. 
The result of this Feynman diagram calculation is that the chiral, anti-chiral, and free scalar theories are coupled by the Lagrangian
\begin{equation} 
\begin{split}
	S^{\rm eff}_{\rm 2d}&=\int r_{ab}(z_0, z_1) J_a \br{J}_b + r_{ab}(z_0, z_2) J_a  t_{bij} \phi_{j} \partial_{\wbar} \phi_i  
	+ r_{ab}(z_2, z_1) t_{aij}  \partial_{w}   \br{J}_b \\
	&\quad \quad + r_{ab}(z_0,z_2) r_{cd}(z_2,z_1) J_a \br{J}_d t_{bij} \phi_j t_{cik} \phi_k \;.  
\end{split}	
\end{equation}
This result holds in the rational, trigonometric, or elliptic cases, where $r_{ab}(z,z')$ is the corresponding classical $r$-matrix. The generalization to multiple surface defects is straightforward.

\section{\texorpdfstring{Coupling with Curved $\beta-\gamma$ Systems}{Coupling with Curved beta-gamma Systems}}
\label{sec:betagamma}

\subsection{\texorpdfstring{Engineering $\sigma$-Models}{Engineering sigma-Models}}

Some of the most interesting integrable field theories are given by non-linear $\sigma$-models to manifolds with particular metrics. One can ask if these can be engineered from our perspective. In this section we will explain how to engineer a particular class of non-linear $\sigma$-models, where the target is a K\"ahler manifold.  Other models, such as non-K\"ahler symmetric space models, will be discussed later in section \ref{sec:symmetric}.  

The surface defects we introduce to engineer this class of models are curved $\beta-\gamma$ systems and their complex conjugates. 

Let $X$ be a complex manifold with a holomorphic $G$-action.   At $z_0$ we will introduce fields $\gamma : \C \to X$ and $\beta \in \Omega^{1,0}(\C, \gamma^\ast T^\ast X)$ (where $T^\ast X$ is the holomorphic cotangent bundle of $X$).    The action is written succinctly as 
\begin{equation} 
S_{\beta-\gamma} = \int \beta\, \dbar_A \gamma \;. 
 \end{equation}
We can make this more explicit. Let $\rho : \g \to \op{Vect}(X)$ be the Lie algebra homomorphism from $\g$ to the Lie algebra of holomorphic vector fields on $X$, which defines the infinitesimal $G$-action on $X$.  Let us choose local holomorphic coordinates $u_1,\dots,u_n$ on $X$, and a basis $t_a$ of the Lie algebra $\g$.   In these coordinates
\begin{equation} 
\rho(t_a) = \sum \rho_{a,i} (u) \partial_{u_i}  \;,
 \end{equation}
where $\rho_{a,i}(u)$ are some holomorphic functions of the variables $u_j$.  Then the expanded form of the action functional is 
\begin{equation} 
\int_{w \in \C} \beta^i \partial_{\wbar} \gamma_i  \d \wbar + A_{a,\wbar} \beta^i \rho_{a,i}(\gamma) \d \wbar \;. 
 \end{equation}

Let us consider introducing a complex-conjugate $\br{\beta}-\br{\gamma}$ system at $z_1$, where $\br{\gamma}$ is a map to $\br{X}$ (which is $X$ with the opposite complex structure) and $\br{\beta}$ is an element of $\Omega^{0,1}(\C, \br{\gamma}^\ast T^\ast \br{X})$. The action function is $\int \br{\beta} \partial_A \br{\gamma}$, which can be expanded out as  
\begin{equation} 
 \int_{w \in \C} \br{\beta}^i \partial_{w}\br{ \gamma}_i  \d w + A_{a,w} \br{\beta}^i \br{\rho}_{a,i}(\br{\gamma}) \d \wbar \;. 
 \end{equation}
Integrating out the gauge field leaves the theory whose fields are $\beta^i, \gamma_i,\br{\beta}^i,\br{\gamma}_i$ with the action 
\begin{equation} 
\int  \beta^i \partial_{\wbar} \gamma_i  \d \wbar +  \br{\beta}^i \partial_{w}\br{ \gamma}_i  \d w + \frac{1}{z_0 - z_1} \sum \beta^i \br{\beta}^j \rho_{a,i} (\gamma) \br{\rho}_{a,j}(\br{\gamma}) \;.\label{action_betagamma} 
\end{equation}
It is known in the literature that, under a certain non-degeneracy hypothesis (which we will elaborate momentarily around \eqn \eqref{eq.non-degeneracy}), this action is equivalent to the action given by an ordinary bosonic $\sigma$-model with target $X$, equipped with a certain K\"ahler metric.  This follows from a general feature of the $\sigma$-model on a K\"ahler manifold: in the large volume limit, the $\sigma$-model on a K\"ahler manifold decomposes as a product of a $\beta-\gamma$ system and its complex conjugate $\br{\beta}-\br{\gamma}$-system. Moving away from the large volume limit introduces a $\beta-\br{\beta}$ interaction of the type we have written down.  In order to be self-contained we will present a derivation of this fact in appendix \ref{app:interlude}.

Let us consider the situation where at $z_0$ we couple the $\beta-\gamma$ system on a complex manifold $X$ with a $G$-action, and at $z_1$ we couple the complex conjugate of this. We find that the action \eqref{action_betagamma} is that of a $\sigma$-model on $X$, as long as the tensor 
\begin{equation} 
   g^{i \br{j}}(u,\br{u}) = \sum_a \rho^{a,i} (u) \br{\rho}^{a,\br{j}}(\br{u}) 
   \label{eq.non-degeneracy}
 \end{equation}
is invertible. The metric tensor $g_{i \br{j}}$ is the inverse of this tensor $g^{i \br{j}}$.  

Our construction includes a rather large class of $\sigma$ models with K\"ahler target space.
While there is huge literature on sigma models with symmetric space target spaces (which we will study with different methods in section \ref{sec:symmetric}), there seems to be little work on more general non-symmetric cosets,
which are included  in our construction. See \cite{Bykov:2014efa,Bykov:2016rdv} for recent works in this direction.

\subsection{\texorpdfstring{$\CP^1$ $\sigma$-Model}{CP(1) Sigma-Model}}

The simplest example for the construction above is when $X = \CP^1$. There is a local coordinate $u$ in which the holomorphic action of the Lie algebra of $\su(2)$ is given by the vector fields $\sigma_1 =  \i u \partial_u, \sigma_2 = \tfrac{1}{2}(-\partial_u - u^2 \partial_u), \sigma_3 =\tfrac{1}{2}(\i \partial_u - \i u^2 \partial_u)$.   Since the Casimir of $\mf{sl}_2$ is, in terms of the Pauli matrices $\sigma_i$, 
\begin{equation} 
\sum \sigma_i \otimes \sigma_i  \;,
 \end{equation}
we find that the inverse metric is
\begin{equation} 
\left( - u \br{u} + \tfrac{1}{2} u^2 + \tfrac{1}{2}  \br{u}^2 \right) \partial_u \partial_{\br{u}} \;. 
 \end{equation}

In this example, we should not integrate over the contour $u = \br{\br{u}}$, as this will not give rise to a non-degenerate inverse metric.  Instead,  we first perform the change of coordinates $\br{u} \mapsto - \br{u}^{-1}$ (recalling that both $u$ and $\br{u}$ take values in $\CP^1$).  Then the inverse metric becomes
\begin{equation} 
 \left(  u \br{u}^{-1} + \tfrac{1}{2} u^2 + \tfrac{1}{2}  \br{u}^{-2} \right)\br{u}^{2}  \partial_u \partial_{\br{u}} = \tfrac{1}{2}(1 + u \br{u} )^2 \partial_u \partial_{\br{u}} \;. 
 \end{equation}
Therefore the metric is the Fubini-Study metric
\begin{equation} 
\d s^2=2 \frac{1}{(1 + u \br{u})^{2}} \d u \d \br{u} \;. 
 \end{equation}
 
The $\CP^1$ sigma model in this setup has a topological term $\int \phi^{\ast} \omega$, as in \eqn \eqref{eq.action_sigma} in appendix \ref{app:interlude}. This has an overall factor $1/\hbar$,
and since the K\"ahler form $\omega$ is given by
\begin{equation}
 \omega = \i \frac{\d z \d \br{z}} {1+\abs{z}^2} \in 2\pi H^1(\CP^1, \mathbb{Z}) \;,
\end{equation}
the value of the $\theta$ angle to be  
\begin{equation}
\theta=\frac{2\pi}{\hbar} \;.
\end{equation}

The topological term is known to dramatically affect the IR physics---for example, $\CP^1$ is 
gapped for $\theta=0$, while gapless for $\theta=\pi$ \cite{Haldane:1982rj,Haldane:1983ru}. While the theory with special values $\theta=0$ \cite{Zamolodchikov:1978xm} and $\theta=\pi$ \cite{Zamolodchikov:1992zr} are known to be integrable, one expects that this is not the case for a general value of $\theta$. Despite such a rich phase-space structure as a function of $\theta$,
the analysis in this paper (as well as in the previous two papers in the series \cite{Costello:2017dso,Costello:2018gyb}) is
restricted to the perturbative analysis of the theory, and hence is insensitive to the 
precise values of the $\theta$-angle. One expects that this issues will be fully addressed in a 
proper non-perturbative treatment of the theory, which involves the specification of the 
integration cycle of the complexified path-integral (see also later discussion around \eqn \eqref{eq.level},
which comments on similar issues for the WZW model).

Later in section \ref{sec:symmetric} we will discuss sigma models with symmetric space targets, which include the $\CP^1$ model as a special example. We find that there is no topological theta term in this 
latter realization.\footnote{A sigma model whose target space is topologically $\CP^1$ can also be constructed from $\mathbb{T}^2$-dimensional reduction of four-dimensional pure $SU(N)$ Yang-Mills theory,
and this gives a rather different four-dimensional realization of the $\CP^1$-model \cite{Yamazaki:2017ulc} (albeit in with a non-standard metric).}

\subsection{\texorpdfstring{Trigonometric and Elliptic Deformations of the $\CP^1$ $\sigma$-Model}{Trigonometric and Elliptic Deformations of the CP(1) Sigma-Model}}

The analysis above goes through with very few changes in the trigonometric and elliptic cases.   In the trigonometric  case, the classical $r$-matrix is written most conveniently in the $e,f,h$ basis of $\mf{sl}_2$.  These act on $\CP^1$ by the vector fields $\i \partial_u,\i u^2 \partial_u, -2 u \partial_u$.   The trigonometric $r$-matrix \eqref{trig_r} gives rises, after the transformation $\br{u} \mapsto- \br{u}^{-1}$, to the inverse metric    
\begin{equation} 
	\frac{1}{1 - z_0/z_1} \partial_u \partial_{\br{u}} - \frac{1}{1 - z_1 / z_0} u^2 \br{u}^2 \partial_u \partial_{\br{u}} + \frac{z_1 + z_0}{z_1 - z_0}  u \br{u} \partial_u \partial_{\br{u}} \;. 
\end{equation}
We find the metric is 
\begin{equation} 
	\d s^2=(z_1 - z_0) \left( z_1 + z_0 \abs{u}^4 +  (z_1 + z_0) \abs{u}^2   \right)^{-1} \d u \d \br{u} 	 \;.
	\label{trigonometric_metric} 
\end{equation}
This is a one-parameter family of metrics deforming the Fubini-Study metric, preserving $O(2)$ symmetry of $\CP^1$.
Note that this metric reduces to the Fubini-Study metric (up to an overall constant factor) in the limit $z_0=0$,
where we have an enhanced $O(3)$ symmetry.

In the elliptic case, the metric is calculated most conveniently by using the vector fields  $\sigma_1 =  \i u \partial_u$, $\sigma_2 = \tfrac{1}{2}(-\partial_u - u^2 \partial_u)$, $\sigma_3 =\tfrac{1}{2}(\i \partial_u - \i u^2 \partial_u)$.   After performing the transformation $\br{u} \mapsto - \br{u}^{-1}$, the inverse metric is
\begin{equation} 
	w_1(z_{01})u \br{u} \partial_u \partial_{\br{u}} - w_2 (z_{01})\frac{1}{4} (-  \br{u}^2 - u^2 - 1 - u^2 \br{u}^2      )  \partial_u \partial_{\br{u}}   - w_3(z_{01})\frac{1}{4} \left(  \br{u}^2 +  u^2 - u^2 \br{u}^2 - 1    \right)  \;,
\end{equation}
where we introduced shorthand notation $z_{01}=:z_0-z_1$.
The metric is therefore
\begin{equation} 
	\d s^2=\frac{4 \d u \d \br{u}}{ w_1(z_{01}) \frac{\abs{u}^2}{4} + w_2(z_{01}) (\abs{u}^4 + u^2 + \br{u}^2 +1 ) + w_3(z_{01})(\abs{u}^4 - u^2 - \br{u}^2 + 1   )   } \;.
\label{elliptic_metric}	 
\end{equation}
This is a two-parameter deformation of the round metric on $S^2$, parametrized by  and the elliptic modulus $\tau$, 
in which the $U(1)$ isometry has  been broken. The parameters are $z_0 - z_1$ and the modular parameter $\tau$ of the elliptic curve, which appears in the definition of the elliptic functions $w_1,w_2,w_3$ in \eqn \eqref{w_def}. Note however that this family of metrics is complex-analytic: the coefficient of $\d u \d \br{u}$ is generally complex, not real.  

One can still define the $\sigma$-model path integral in this situation, at least in perturbation theory. 

\subsection{\texorpdfstring{Trigonometric $\CP^1$ Model and FOZ Sausage Model}{Trigonometric CP(1) model and FOZ Sausage Model}}
\label{subsec:sausage}
We introduced above a one-parameter deformation of the $\CP^1$ $\sigma$-model by applying our construction in the trigonometric (in contrast with the rational case).  In general, as we will see later, if we built a two-dimensional field theory by inserting surface defects in the trigonometric version of our four-dimensional gauge theory, we will always find an integrable  model with the quantum loop group as a symmetry.  

Fateev, Onofri and Zamolodchikov (FOZ) \cite{Fateev:1992tk} introduced a one-parameter deformation of the round metric on $S^2$ (see also \cite{King,Rosenau}).  They argued that the $\sigma$-model with this target is integrable, and that the scattering of particles in this model is related to the trigonometric $R$-matrix for the group $SU(2)$.   

It is natural to suspect that our integrable model deformation of the $\CP^1$ $\sigma$-model is the same as the one constructed by FOZ.
It turns out that this is the case, up to a certain coordinate transformation.

To see this, let us star with the metric \eqref{trigonometric_metric} we derived previously, and we set $s = z_0 / z_1$, and use polar coordinate $u = r e^{i \theta}$.  In these coordinates our metric is
\begin{equation} 
\d s^2=(1-s)  \frac{ ( r^{-2} \d r^2 +  \d \theta^2)}{r^{-2} + s r^2 + 1+s  } \;. 
 \end{equation}
If we perform the change of coordinates $x =  \log r$, $s = e^{-2t}$, the metric becomes
\begin{equation} 
\d s^2=\frac{1-e^{-2t}}{ e^{-2x} +  e^{2 x - 2 t} + 1+e^{-2t} } (\d x^2 +   \d \theta^2) \;.   
 \end{equation}
Sending $x \mapsto x - \tfrac{1}{2} t$ gives us a conformally flat metric 
in cylindrical coordinates $x, \theta$:
\begin{equation} 
\d s^2= \frac{e^t - e^{-t}}{  e^t + e^{-t} + e^{-2x } + e^{2x} } (\d x^2 + \d \theta^2)  
=\left(  \op{coth}(t)  + \frac{ \op{cosh}(2x) }{\op{sinh}(t)}  \right)^{-1} (\d x^2 + \d \theta^2) \;.
\end{equation}
This is the FOZ sausage metric \cite{Fateev:1992tk}.

FOZ identified the deformation parameter $t$ with the 
flow under the renormalization group (RG). In particular 
FOZ metric satisfies the Ricci-flow equation\footnote{For a compact two-dimensional surface, ancient solutions (i.e. solutions defined in the time $t\in (-\infty, T)$) which becomes spherical at time $t=T$ has been classified in \cite{Hamilton}. Their results states that such an ancient solution is either a family of contracting round sphere, or the two-dimensional sausage discussed above. Note that we expect our elliptic solution will not extend to the infinite past $t=-\infty$.}
\begin{equation}
	\frac{\d g_{ij}}{\d t} = -\frac{1}{2\pi} R_{ij} 	\;,
	\label{Ricci_flow}
\end{equation}
which is the one-loop approximation to the RG equation. 
Here the deformation parameter $t$
identified as the logarithm of the energy scale,
and $t\to -\infty$ ($t\to \infty$), namely $s\to \infty$ ($s\to 0$) corresponds to 
the UV limit (the IR limit). In particular in the IR limit the metric reduces to the Fubini-Study metric,
where we have an enhanced $O(3)$ isometry.

There is a slightly subtle consequence of the analysis above. 
In the transition from our metric into the FOZ metric we used a time (i.e. $t$)-dependent coordinate transformation,
under which the Ricci flow is not invariant. Since we have performed such a change of coordinates to match the trigonometric deformation engineered from our construction to the sausage metric, we deduce that our family of metrics does \emph{not} satisfy the Ricci flow equation.  It does so only after the time-dependent change of coordinates. 

As described before, we can engineer a two-dimensional theory from our four dimensional gauge theory with chiral and anti-chiral surface defects with locations $z_0$ and $z_1$.
We will show in the forthcoming paper \cite{Part4} that at the quantum level, renormalization group flow of the theory
 is obtained by moving the locations $z_0, z_1$ of the surface defects.  In the trigonometric case, if we let $s = z_0 / z_1$, we find that the renormalization group flow is the flow associated to the vector field $s \partial_s$. 

The fact that one-parameter family of metrics we engineer in \eqn\eqref{trigonometric_metric}  does not satisfy the Ricci flow equation does not contradict the statement that it gives a semiclassical description of an RG flow trajectory.  When we state that a family of Lagrangians represents an RG flow trajectory, we allow for the possibility that a change of coordinates on the field space which depends on the RG flow ``time'' has been performed.

In addition to the trigonometric metrics,
one might expect that the elliptic metric \eqref{elliptic_metric} also satisfies the Ricci flow equation,
again until suitable definition of the time coordinate. It would be interesting to verify this conjecture.

\subsection{Examples from Flag Varieties}

Let us consider possible generalizations of the $\CP^1$ example to more general flag varieties. 

The data we need for the construction is a complex manifold $X$ with an action of the complex Lie group $\GC$.  We also need to have a real slice $Y \subset X \times \br{X}$ which imposes the reality conditions on our fields. (It will often be convenient, although not strictly necessary, to choose $Y$ so that it is invariant under the action of some real form $G_{\R}$ of $\GC$). 

There is a homomorphism
\begin{equation} 
\gC \otimes \gC \to H^0_{\dbar} (X, T^{1,0}X) \otimes H^0_{\dbar}(\br{X}, T^{1,0} \br{X}) \to \Gamma(Y, T Y \otimes T Y) \;. 
 \end{equation}
In order for us to find an example that is equivalent to an actual $\sigma$-model, we need the image of the Casimir element $c \in \gC \otimes \gC$ to be a section of the bundle $T Y \otimes T Y$ on $Y$ which defines a non-degenerate symmetric pairing on the cotangent bundle of $Y$.  Then we can invert this tensor to define a metric on $Y$, and we end up with the $\sigma$-model with this metric. 

For this to work, it is necessary that the $\GC$-action on $X$ is transitive. The tangent bundle of $Y$ is the sum of the restriction to $Y$ of the $(1,0)$ tangent bundle of $X$ with the $(1,0)$ tangent bundle of $\br{X}$.  The putative inverse metric we have is built from the vector fields defining the $\GC$-action on $X$ and $\br{X}$.  For this tensor to be invertible, it is clearly necessary that the $\GC$-action is transitive in a neighborhood of $Y$.

What other conditions are necessary?  For one thing, the $\beta-\gamma$ system on $X$ must be anomaly-free.   If we aim to produce a $\sigma$-model on a compact manifold by this method, we need $X$ to be compact.  For a complex simple Lie group $\GC$, there is exactly one compact homogeneous space on which the $\beta-\gamma$ system is anomaly free: this is the flag variety $\GC / B$ where $ B \subset \GC$ is a Borel subgroup.  Equivalently, if $\Gc$ is the compact form of $\GC$, the flag variety is $\Gc / T$, the quotient of $\Gc$ by the maximal torus.  

Some of these conditions  can be relaxed, leading to generalizations of the construction. For example, as we will see later, there are interesting examples that can be constructed even if $X$ is not compact, which will lead to a $\sigma$-model on a product of a compact manifold with $\R$.   

The flag variety $\GC/B$ is the natural generalization of the $\CP^1$ $\sigma$-model.  In this case, the inverse metric arising from our construction is the image of the quadratic Casimir in 
\begin{equation} 
\gC \otimes \gC = H^0(\GC/B, T \GC / B) \otimes H^0(\GC/B, T \GC / B) \;. 
 \end{equation}
To find a real cycle on which this defines a non-degenerate metric, let us fix the compact real form $\Gc$ of $\GC$ and the corresponding compact real form $\g_{\R}$ of the Lie algebra.  

The variety $\GC/B$ is the variety of Borel subalgebras $\mf{b} \subset \gC$.  The cycle we choose is the set of pairs $(\mf{b},\mf{b}')$ of Borel subalgebras such that
\begin{equation} 
	\br{\mf{b}} = \mf{b}'  \;,
\end{equation}
 where we use our chosen real form of $\gC$ to define complex conjugation.

Since we have chosen the compact real\footnote{We could choose a non-compact real form and find a non-compact $\sigma$-model. In this case, the reality condition $\br{\mf{b}} = \mf{b}'$ needs to be supplemented by the condition that $\mf{b}$, $\br{\mf{b}}$ intersect transversely, i.e.\ span the whole Lie algebra $\g$.}   form, $\mf{b}$ contains the Lie algebra of some unique maximal torus $T \subset G$ of the compact group.  The Borel subalgebras $\mf{b}$ and $\br{\mf{b}}$ intersect transversely and the intersection is the complexification of $\op{Lie} T$.  

These conditions mean that we can decompose $\gC$ as $\gC = \mf{n} \oplus \mf{h} \oplus \mf{n}'$, where $\mf{n} \subset \mf{b}$, $\mf{n}' \subset \mf{b}'$, $\mf{h} = \mf{b} \cap \mf{b}'$ is a Cartan, and $\mf{n}, \mf{n}'$ are complex conjugate.

Next, let us check that the section of the square of the real tangent bundle on $\GC/B$ arising from this construction is invertible, and so defines a metric.  The tangent space to $\GC/B$ at a point corresponding to $\mf{b} \subset \gC$ is the quotient $\gC/\mf{b}$, and the map from $\gC$ to this tangent space is given by projecting onto the quotient.  Therefore the complexified tangent space to our real cycle in $\GC /B \times \GC / B$ is, at a point $(\mf{b}, \mf{b}')$, given by $\gC/\mf{b} \oplus \gC/\mf{b}'$.  The section of the square of the tangent bundle to our real cycle comes from the image of the quadratic Casimir under the map
\begin{equation} 
\gC \otimes \gC \to \gC/\mf{b} \otimes \gC/\mf{b}' \;. 
 \end{equation}
Because we assume that $\mf{b}$ and $\mf{b}'$ intersect transversely, the tensor given by the quadratic Casimir is non-degenerate.   

In this example we have seen how the $\sigma$-model on the full flag variety $\GC / B$ arises as an instance of our construction.  It is clear that this example generalizes readily to the trigonometric and elliptic cases. In the trigonometric case, if we place our surface operators at $z_0,z_1$ and let $s = z_0/z_1$ we find a one-parameter family of K\"ahler metrics on $\GC/B$, which has as isometries the maximal torus $T$ in the compact group $\Gc$.  

Let us formulate the following conjecture.

\begin{conjecture-non}
We can identify the trigonometric integrable deformation of the sigma model with the RG flow,
under a suitable parameter identifications.
More mathematically, after performing an $s$-dependent change of coordinates and letting $t = \log s$, the family of metrics $\GC / B$ satisfies the Ricci flow equation \eqref{Ricci_flow}.  
\end{conjecture-non}

We leave it an open question to prove/disprove this conjecture.

\subsection{Lax Operator for Flag Varieties}

Let us next discuss the Lax operator. We consider the rational case,
so that the classical $r$-matrix is given by \eqref{rational_r}.
From \eqref{L2d_ca} we find that the 
Lax operator is given by
\begin{equation}
\Lax(z)=\frac{ J}{z_0-z}
+\frac{\br{J} }{z-z_1} \;. 
\label{Lax_current}
\end{equation}
where $J$ and $\br{J}$
are the currents for the $G$-action.

We can convert this expression into that in the second-order formalism, 
by integrating out the fields $\beta$ and $\br{\beta}$. Since we expect that the Noether current will
stay the same in either first-order or second-order formulation, we expect that 
the same expression \eqref{Lax_current} will still be correct 
in the second-order formalism. We therefore obtain
\begin{equation}
\Lax(z)=\frac{ J(z)}{z_0-z}
+\frac{\br{J}(z) }{z-z_1} \;.
\end{equation}

As we discussed previously in \eqn \eqref{Lj}, when we choose $z_0=-1, z_1=1$
we obtain the formula which is often found in the literature (see e.g.\ \cite{Zarembo:2017muf}):
\begin{equation}
\Lax(z)= \frac{j+z\star j}{z^2-1} \;.
\end{equation}
Here $j$, the current for the $G$-symmetry, can be obtained from the Lagrangian as 
$j=g K g^{-1}$, where we have decomposed
$g^{-1} \d g=A+K$, with $A\in \mathfrak{h}$ and $K \in \mathfrak{n}$
according to the decomposition $\mathfrak{g}=\mathfrak{h}\oplus \mathfrak{n}$.
By a gauge transformation $\Lax(z)\to g^{-1} dg+g^{-1} \Lax(z) g$, this becomes
\begin{equation}
\Lax(z)=A +\frac{z^2+1}{z^2-1} K -\frac{2z}{z^2-1} \star K \;,
\end{equation}
which is another expression often found in the literature \cite{Zarembo:2017muf}.

\subsection{\texorpdfstring{$\sigma$-Models on Spheres}{Sigma-Models on Spheres}}\label{subsec.spheres}

Our construction does not give, at least not in any obvious way, the $\sigma$-model on $\CP^{n}$ for $n >1$.  This is because $\CP^n$ has non-trivial first Pontryagin class for $n > 1$, so that the $\beta-\gamma$ system is anomalous \cite{GMS,Witten:2005px,Nekrasov:2005wg}.  We view this as a reassuring point:  the $\CP^n$ $\sigma$-model with $n>1$ is not integrable at the quantum level \cite{Abdalla:1982yd}. 

We can, however, engineer the $\sigma$-models on spheres $S^{2n-1}$, which are $S^1$ bundles over $\CP^n$. The construction is a small variant of the one presented above. 

We take our gauge group to be $G =GL(n,\C)$, and introduces at $z_0$ a free $\beta-\gamma$ system valued in the fundamental representation. At $z_1$ we introduce a free $\br{\beta}-\br{\gamma}$ system valued in the anti-fundamental representation.  The action functional for the full system is   
\begin{equation} 
S_{\rm 4d-2d}=	\frac{1}{\hbar}\left[
\frac{1}{2 \pi} \int_{\R^2 \times \C} \d z\, {\rm CS}(A)  + \int_{\R^2 \times z_0} \beta^i \dbar_A \gamma_i  + \int_{\R^2 \times z_1} \br{\beta}_i \dbar_A \br{\gamma}^i \right]\;. 
 \end{equation}
Integrating out the gauge field $A$ leaves us with the two-dimensional theory with the classical action.
\begin{equation} 
S_{\rm 2d}^{\rm eff}=	
\frac{1}{\hbar}\left[
 \int_{\R^2 \times z_0} \beta^i \dbar_A \gamma_i  + \int_{\R^2 \times z_1} \br{\beta}_i \dbar_A \br{\gamma}^i + \frac{1}{ z_0 - z_1} \int \beta^i \gamma_j \br{\beta}_i \br{\gamma}^j 
 \right]\;. 
\end{equation}
If we give $\C^n$ coordinates $u_i, \br{u}_i$, and work on the open subset where $u_i \neq 0$, then this action is that for the $\sigma$-model on $\C^n \setminus 0$ with the metric 
\begin{equation} 
	\frac{z_0 - z_1}{\norm{u}^2} \sum \d u_i \d \br{u}_i \;. 
\end{equation}
With this metric, there is an isometry
\begin{equation} 
	\C^n \setminus \{ 0  \} \iso S^{2n-1} \times \R \;. 
\end{equation}
We see that this construction has engineered the $\sigma$-model on $S^{2n-1}$, together with a single free boson. 

\subsection{\texorpdfstring{Trigonometric Deformation of the $S^3$ Model}{Trigonometric Deformation of the S(3) Model}}

As usual, this example generalizes readily to the trigonometric and elliptic settings.  Let us write down the trigonometric and elliptic deformations of the metric on $S^3$. 

The vector fields on $\C^2\backslash \{0\}$ associated to the basis elements $e,f,h$ of $\mf{sl}_2$ act on the fundamental representation, with basis $u_i$, by $u_1 \partial_{u_2}$, $u_2 \partial_{u_1}$, $u_1 \partial_{u_1} - u_2 \partial_{u_2}$ and on the anti-fundamental representation with basis $\br{u}_i$ by $-\br{u}_2 \partial_{\br{u}_1}$, $-\br{u}_1 \partial_{\br{u}_2}$, $-\br{u}_1 \partial_{\br{u}_1} + \br{u}_2 \partial_{\br{u}_2}$.
For our purposes we need to consider $\mf{gl}_2$, and we thus consider 
$u_1 \partial_{u_1} + u_2 \partial_{u_2}$ and $-\br{u}_1 \partial_{\br{u}_1} - \br{u}_2 \partial_{\br{u}_2}$
for an addition $\mf{u}_1$ factor.

Let us now start with the rational case.
Using formula \eqref{rational_r_sl2} for the rational $r$-matrix, we find that the inverse metric on $\C^2 \setminus 0$ is
\begin{equation}\label{eq.deldel}
\begin{split} 
&u_1 \br{u}_1\partial_{u_2} \partial_{\br{u}_2} +u_2 \br{u}_2 \partial_{u_1} \partial_{\br{u_1}} +\frac{1}{2} (u_1 \partial_{u_1} - u_2 \partial_{u_2}) (\br{u}_1 \partial_{\br{u}_1} - \br{u}_2 \partial_{\br{u}_2} ) \\
& \qquad+\frac{1}{2} (u_1 \partial_{u_1} + u_2 \partial_{u_2}) (\br{u}_1 \partial_{\br{u}_1} + \br{u}_2 \partial_{\br{u}_2} ) 
= (|u_1|^2 +|u_2|^2) (\partial_{u_1} \partial_{\br{u}_1}+\partial_{u_2} \partial_{\br{u}_2})\;.  
 \end{split}
\end{equation}
We can invert this to obtain the metric
\begin{equation} 
\d s^2=\frac{1}{ \abs{u_1}^2+\abs{u_2}^2 } 
 \left( \d u_1  \d \br{u}_1 +  \d u_2 \d \br{u}_2  \right)
\;,
\end{equation}
On the slice $|u_1|^2+|u_2|^2=1$ gives the canonical metric on $S^3$,
while the radial direction, corresponding to the overall size $\abs{u_1}^2+\abs{u_2}^2$, gives an extra non-compact direction $\mathbb{R}$.

Let us now come to the trigonometric case.
Using formula \eqref{trig_r} for the trigonometric $r$-matrix, we find that $\mathfrak{sl}_2$ part of the inverse metric on $\C^2 \setminus 0$ is
\begin{equation} 
\frac{1}{1 - s} u_1 \br{u}_1\partial_{u_2} \partial_{\br{u}_2} - \frac{1}{1 - s^{-1}} u_2 \br{u}_2 \partial_{u_1} \partial_{\br{u_1}} \,  + \frac{1+s}{4(1-s)} (u_1 \partial_{u_1} - u_2 \partial_{u_2}) (\br{u}_1 \partial_{\br{u}_1} - \br{u}_2 \partial_{\br{u}_2} ) \;.  
 \end{equation}
where $s = z_0 / z_1$.  Let us rescale $u_1, \br{u}_1$ by $s^{1/8}$ and $u_2, \br{u}_2$ by $s^{-1/8}$,
and write $s=e^{2\xi}$, so that we have, after rescaling by $2\sinh \xi$,
\begin{equation} 
 (u_1 \br{u}_1\partial_{u_2} \partial_{\br{u}_2} +u_2 \br{u}_2 \partial_{u_1} \partial_{\br{u_1}} )+\frac{\cosh \xi}{2 } (u_1 \partial_{u_1} - u_2 \partial_{u_2}) (\br{u}_1 \partial_{\br{u}_1} - \br{u}_2 \partial_{\br{u}_2} )  
 \;. 
 \end{equation}
 This expression for $\xi=0$ (rational case) reduces to what we had in the $\mathfrak{sl}_2$ part of 
\eqn \eqref{eq.deldel}, and hence it is now clear how to add the $\mf{u}_1$ part, which should be independent of the deformation parameter $\xi$:
\begin{equation} 
\begin{split}
& (u_1 \br{u}_1\partial_{u_2} \partial_{\br{u}_2} +u_2 \br{u}_2 \partial_{u_1} \partial_{\br{u_1}} )+\frac{\cosh \xi}{2 } (u_1 \partial_{u_1} - u_2 \partial_{u_2}) (\br{u}_1 \partial_{\br{u}_1} - \br{u}_2 \partial_{\br{u}_2} )  \\
 & \qquad+ \frac{1}{2}(u_1 \partial_{u_1} - u_2 \partial_{u_2}) (\br{u}_1 \partial_{\br{u}_1} - \br{u}_2 \partial_{\br{u}_2} )  
 \;. 
 \end{split}
 \end{equation}

To invert this we must invert the matrix
\begin{equation}
	\begin{pmatrix}
 	\abs{u_2}^2 + \frac{\cosh \xi+1}{2}  \abs{u_1}^2 & \frac{-\cosh \xi+1}{2} \br{u}_1 u_2   \\
	 \frac{-\cosh \xi+1}{2} u_1 \br{u}_2 & \abs{u_1}^2 + \frac{\cosh \xi+1}{2} \abs{u_2}^2
	\end{pmatrix} \;,
\end{equation}
giving 
\begin{equation} 
\frac{1 }{\frac{\cosh \xi+1}{2}  ( \abs{u_1}^2 +  \abs{u_2}^2 )^2 }  
\begin{pmatrix}
	\abs{u_1}^2 + \frac{\cosh \xi+1}{2} \abs{u_2}^2 & \frac{\cosh \xi-1}{2} \br{u}_1 u_2   \\
	\frac{\cosh \xi-1}{2} u_1 \br{u}_2 &\abs{u_2}^2 + \frac{\cosh \xi+1}{2}  \abs{u_1}^2  
\end{pmatrix} \;,
 \end{equation}
and thus the metric is, after some cleaning, given by
\begin{equation} 
\begin{split}
\d s^2
&=\frac{ ( \abs{u_1}^2 +  \abs{u_2}^2 )(\d u_1  \d \br{u}_1 +  \d u_2 \d \br{u}_2) 
+ \left(\tanh^2 \frac{\xi}{2} \right)  \abs{u_1 \d \br{u}_1 -u_2 \d \br{u}_2}^2  }{ ( \abs{u_1}^2 +  \abs{u_2}^2 )^2 }    \;.
\end{split}
\end{equation}

As in the rational case, this metric is a cone with a non-compact radial $\mathbb{R}$-direction along the overall size direction $\abs{u_1}^2+\abs{u_2}^2$. When restricted to the slice $|u_1|^2+|u_2|^2=1$ we can rewrite
\begin{equation}  \label{eq.S3_sausage}
\begin{split}
\d s^2   =(\d u_1  \d \br{u}_1 +  \d u_2 \d \br{u}_2) 
+ \left(\tanh^2 \frac{\xi}{2} \right)  \abs{u_1 \d \br{u}_1 -u_2 \d \br{u}_2}^2  \;.
\end{split}
\end{equation}
This metric coincides with the Fateev three-dimensional sausage solution \cite{Fateev:1996ea} 
(see appendix \ref{app:3d_sausage} for more details). In \cite{Fateev:1996ea} it was also shown that the 
metric satisfies the Ricci flow equation, where the 
time for the Ricci flow is related with the deformation parameter $\xi$ by a suitable coordinate transformation.

We can repeat the procedure above for the elliptic case, to derive a more general deformation of the 
metric of the three-sphere, which could solve the Ricci flow equation upon appropriate coordinate transformations. We leave this exercise for future work.

\newpage

\part{Disorder Surface Defects}
\label{part:disorder}

\section{Introduction to Disorder Operators}
\label{sec:disorder}

The models we have engineered so far miss some of the most basic integrable models, such as the principal chiral model and symmetric space coset models. To build these, we need to use \emph{disorder operators} in four-dimensional Chern-Simons theory.  Disorder operators are given by specifying that the four-dimensional gauge field $A$ has some poles at points on the $z$-plane.    In order for the Chern-Simons action to be gauge invariant, these poles must occur at points where the one-form $\omega$ has zeroes.

In this section we will summarize our construction of integrable field theories from disorder operators. Because general Riemann surfaces equipped with a meromorphic one-form play such a prominent role in our story, we will start with a review of the geometry of such surfaces. 

\subsection{The Geometry of Translation Surfaces}

The general construction takes as input a Riemann surface $C$  with a holomorphic one-form $\omega$ which may have simple zeroes, simple poles, and second-order poles. Such a surface $C$ has a natural metric  $g = \omega \br{\omega}$. This metric is always flat away from the zeroes and poles of $\omega$, as there exists a local coordinate in which $\omega = \d z$.

In this metric, the first-order poles in $\omega$ correspond to infinite flat cylinders, where the metric looks like $g \sim \frac{\d z \d \br{z}}{ \abs{z}^2} $.  The second-order poles correspond to patches where the metric looks like a neighborhood of $\infty$ in the flat plane $\d x^2 + \d y^2$ (there are subleading corrections to this description when there is a non-zero residue at a second-order pole).  The poles of the one-form $\omega$ should be thought of as points at infinity on the Riemann surface $C$. 

Near the simple zeroes of $\omega$, the metric takes the form $4 u^2 \d \theta^2 + \d u^2$, giving a conical singularity.  The angle around the conical singularity is $4 \pi$. 

Conversely, we can construct $(C,\omega)$ from a flat Riemannian surface with singularities and asymptotic boundaries as above. The surface must be oriented and equipped with a covariant constant vector (the real part of the inverse to $\omega$), defined away from the singular points, which is nowhere vanishing. The singular points are, as above, cone singularities with angle $4 \pi$. The existence of a covariant constant vector means that the Levi-Civita connection on $C$ has no monodromy.

Surfaces like this have been extensively studied in the mathematics literature, in the context of billiards dynamical systems, where they are called \emph{translation surfaces}. We recommend  the beautiful survey \cite{zorich2006flat}.  

If the one-form $C$ has no poles, so that the surface is closed, then the surface can always be obtained from a planar polygon by identifying pairs of parallel edges (of the same length).  The fact that we only identify parallel edges guarantees that we have a covariant constant vector. This procedure will glue several vertices of the polygon to a single conical singularity. The angle at the conical singularity is the sum of the interior angles of the corresponding vertices of the polygon.  The angle around the cone is not necessarily $4 \pi$, but we will only consider surfaces built by this gluing procedure where each cone angle is $4 \pi$. 

We are interested in  more general surfaces, where the one-form has poles.  These are build not just from gluing polygons, but from several kinds of unbounded regions.
\begin{enumerate} 
	\item A simple pole in the one-form $\omega$ gives us a semi-infinite cylinder. To attach a semi-infinite cylinder, we should consider a semi-infinite strip, as in \fig\ref{fig:cylinder_plane}\subref{fig:cylinder}, which has a polygonal boundary at one end.  The opposite sides of the strip are identified, and the edges of the boundary polygon are identified with other edges of the same length and angle.  The width and angle of the strip encode the residue of the one-form. 
	\item A double pole with no residue gives a region in the surface which is isometric to a plane.  To attach a region like this, we consider the \emph{exterior} of a planar polygon. An example of such a surface is given in figure \fig\ref{fig:cylinder_plane}\subref{fig:plane1}. 

		(In this case, we allow the degenerate polygon with two vertices and two overlapping edges).
	\item A one-form with a double pole and non-zero residue is obtained by attaching a region which is the plane with a semi-infinite strip with polygonal boundary removed, as in figure \fig\ref{fig:cylinder_plane}\subref{fig:plane2}.  The opposite sides of the strip are identified, and the the vector which identifies the opposite sides of the strip is the residue of the one-form.  
\end{enumerate}

\begin{figure}
	\subfloat[A semi-infinite strip with sides identified and polygonal boundary, corresponding to a first-order pole]{
		\label{fig:cylinder}	
	\begin{tikzpicture}[scale=(0.55)]

	\node at (-1.5,1.5) {$\cdots$}; 
	\node at (-1.5,-1.5) {$\cdots$};
	\coordinate (A) at (3,1.5);
	\coordinate (B) at (3,-1.5);
	\coordinate (C) at (-1,1.5);
	\coordinate (D) at (-1,-1.5);
	\coordinate (E) at (3.5,0.5);
	
	\filldraw[pattern=north west lines](A)--(C)--(D)--(B)--(E) --(A);
	\draw[color=white,thick](A)--(C)--(D)--(B)--(E)--(A);
	\draw[dashed, ->-] (A) --(C);
	\draw (A) -- (E) --(B);
	\draw[dashed, ->-](B)--(D);
\end{tikzpicture}}
\hfill 
	\subfloat[A planar region which is the exterior of a polygon, corresponding to a second-order pole with no residue]{\label{fig:plane1}
		\begin{tikzpicture}[scale=0.55]
	\filldraw[ pattern=north west lines] (1,0) circle (3);  
	\draw[color=white,thick] (1,0) circle (3); 
	\coordinate (A) at (0,0);
	\coordinate (B) at (30:1);
	\coordinate (C) at ($(B) + (-30:1)$);
	\coordinate (E) at (-40:1);
	\coordinate (F) at ($(E) + (0:1)$);
	\filldraw[color=white] (A) --(B) --(C) --(F) --(E) --(A);
	\draw (A) --(B); 
	\draw (B) --(C);  
	\draw (C) --(F); 

	\draw (A) --(E);  
	\draw (E) --(F);  
	
\end{tikzpicture}}
\hfill
	\subfloat[A planar region corresponding to a one-form with a second order pole and non-zero residue]{\label{fig:plane2}		\begin{tikzpicture}[scale=0.55]

\filldraw[ pattern=north west lines] (0,0) circle (3);  
\draw[color=white,thick] (0,0) circle (3);
	
	\coordinate (A) at (-2.95,0.5);
	\coordinate (B) at (-2.95,-0.5);
	\coordinate (C) at (-1,0.5);
	\coordinate (D) at (-1,-0.5);
	\coordinate (E) at (-0.5,0);
	\filldraw[color=white](-3,0.5)--(C)--(E)--(D)--(-3,-0.5)--(-3,0.5);
	
	\draw[dashed, ->-] (A) --(C) ; 
	\draw(C)--(E) ; 
	\draw(E)--(D) ; 
	\draw[dashed, ->-](B)--(D); 

		\end{tikzpicture}}
		\caption{}
                \label{fig:cylinder_plane}	
\end{figure}

\subsection{Specification of Integrable Field Theories}

To specify the integrable field theory, we need to give boundary conditions for the theory at the first and second order poles, and specify a disorder operator which lives at the zeroes of the one-form. For the simplest constructions, the boundary conditions and disorder operators are chosen among the following possibilities:
\begin{enumerate} 
	\item At a second order pole, we will always have \emph{Dirichlet} boundary conditions, where all components of the gauge field, and the gauge transformations, vanish. This gives us a model with a global $G$-symmetry.  This boundary condition is topological in the $w-\br{w}$ plane. 
	\item At a first order pole, we have two possibilities.
		\begin{itemize}
			\item We can introduce \emph{chiral Dirichlet} boundary conditions, where $A_{\wbar} = 0$ and gauge transformations vanish.  These will lead to a model with chiral Kac-Moody currents.  We will denote a chiral Dirichlet boundary condition at a first order pole $z = z_0$ by $\mc{D}(z_0)$.  
			\item We can have \emph{anti-chiral Dirichlet} boundary conditions, where $A_w = 0$ and gauge transformations vanish.  These will give us anti-chiral Kac-Moody currents.  This boundary condition at $z = z_0$ is denoted $\br{\mc{D}}(z_0)$. 	
		\end{itemize}
	\item At a first order zero of the one-form, we have two possibilities:
		\begin{itemize}
			\item $A_w$ has a pole, leading to a \emph{chiral disorder operator}. A chiral disorder operator at $z = z_0$ will be denoted $\mc{P}(z_0)$. 
			\item $A_{\wbar}$ has a pole, leading to an \emph{anti-chiral disorder operator}. This will be denoted $\br{\mc{P}}(z_0)$. 
		\end{itemize}
\end{enumerate}
Suppose $C$ is our Riemann surface with one-form $\omega$. Suppose we break the collection of first-order poles into three groups $p_i$, $p'_j$,  at which we place the boundary conditions $\mc{D}(p_i)$ and $\br{\mc{D}}(p_j')$. Suppose we also divide the collection of first-order zeroes into two groups $q_i$, $q_j'$, at which we place the disorder operators $\mc{P}(q_i)$, $\br{\mc{P}}(q_j')$.   Compactifying four-dimensional Chern-Simons theory on $C$ with these disorder and boundary conditions will give rise to an integrable field theory
\begin{equation} 
	\mbf{IFT}_{G,k} (C,\omega,\mc{D}(p_i), \br{\mc{D}}(p_j'),  \mc{P}(q_i), \br{\mc{P}}(q_j')) \;,
\end{equation}
which we will often abbreviate to $\mbf{IFT}_{G,k}(C,\omega)$, assuming we have labelled the poles and zeroes by chiral and anti-chiral boundary conditions and disorder operators.

This depends on the group $G$ and the level $k$ of four-dimensional Chern-Simons theory, which may be an arbitrary complex number because we are considering analytically continued theories.  We do not include the Dirichlet boundary condition at a second-order pole into the notation, as this is the only boundary condition we ever consider at such poles.

These theories are automatically integrable, for the reasons we have mentioned before: the expectation value of the four-dimensional gauge field gives rise to a Lax operator, whose spectral parameter\footnote{There is a subtlety with this statement when $C$ is not of genus $0$: the Lax operator will only be holomorphic as a function of the spectral parameter up to a gauge transformation. The trace of the path-ordered exponential of the Lax matrix is the more physical observable, and this will be holomorphic.}  lives in $C$. 

The poles of the Lax operator\footnote{In the higher genus case, of its path ordered exponential} $\Lax$ reflect the features of the disorder operators we have chosen. A chiral disorder operator $\mc{P}(q_i)$ at $z = q_i$, then $\Lax^w(z)$ will have a first-order pole at $z = q_i$, and an anti-chiral disorder operator will give a pole in $\Lax^{\wbar}(z)$. 

Our goal for the rest of the paper is to analyze these integrable field theories, and certain elaborations. The main results are the following.

	 In section \ref{sec:WZW} we study the theory 
		\begin{equation}
\mbf{IFT}_{G,k} (\CP^1, \d z / z, \mc{D}(0), \br{\mc{D}}(\infty) )  \;.
		\end{equation}
		We show that this is the conformal WZW model on the group $G$, with Lagrangian
\begin{equation} 
 \frac{k}{8\pi} \int_{\R^2} \op{Tr} (j \wedge \star j) -
   \frac{k}{12\pi}  \int_{\R^2 \times \R_{\ge 0} }  \op{Tr} (\what{j} \wedge \what{j}\wedge \what{j}) 
   \;.
\end{equation}
The coefficient of the Wess-Zumino term is precisely the correct value of the Wess-Zumino term to make the model conformal.
	
In section \ref{sec:PCM}, we study the theory
		\begin{equation} 
			\mbf{IFT}_{G,k} \left(\CP^1, \frac{(z-z_0)(z-z_1)}{z^2} \d z, \mc{P}(z_0), \br{\mc{P}}(z_1)     \right) \;,
		\end{equation}
where the one-form has two second-order poles, at $0$ and $\infty$, and two first order zeroes, at $z_0,z_1$.  We show that this model gives the principal chiral model(i.e. the $\sigma$-model on the group manifold) with Lagrangian
\begin{equation} 
	\frac{k(z_1 - z_0)}{8\pi} \int_{\R^2} \op{Tr} (j \wedge \star j) -
	\frac{k(z_1 + z_0)}{12\pi}  \int_{\R^2 \times \R_{\ge 0} }  \op{Tr} (\what{j} \wedge \what{j}\wedge \what{j}) 
   \;,
\end{equation}
The flat surface corresponding to this one-form is obtained by gluing two planes with semi-infinite strips removed.

In section \ref{sec:general_PCM} we introduce a generalization of the principal chiral model, where we have a one-form $\omega$ on $\CP^1$, of the form
\begin{equation} 
	\omega = \frac{ \prod_{i = 1}^{n-1} (z-q_i^w)(z-q_i^{\wbar})}{\prod_{j = 1}^n (z-p_j)^2} \d z  \;,
\end{equation}
which has $n$ second order poles at $z = p_k$ and $2n-2$ first order zeroes. The zeroes are divided into two groups of $n-1$: we put chiral disorder operators $\mc{P}(q_i^w)$ at  $z = q_i^w$, and anti-chiral disorder operators $\mc{P}(q_i^{\wbar})$ at $z = q_i^{\wbar}$.  At the second order pole, we place, as always, a Dirichlet boundary condition on all fields. 

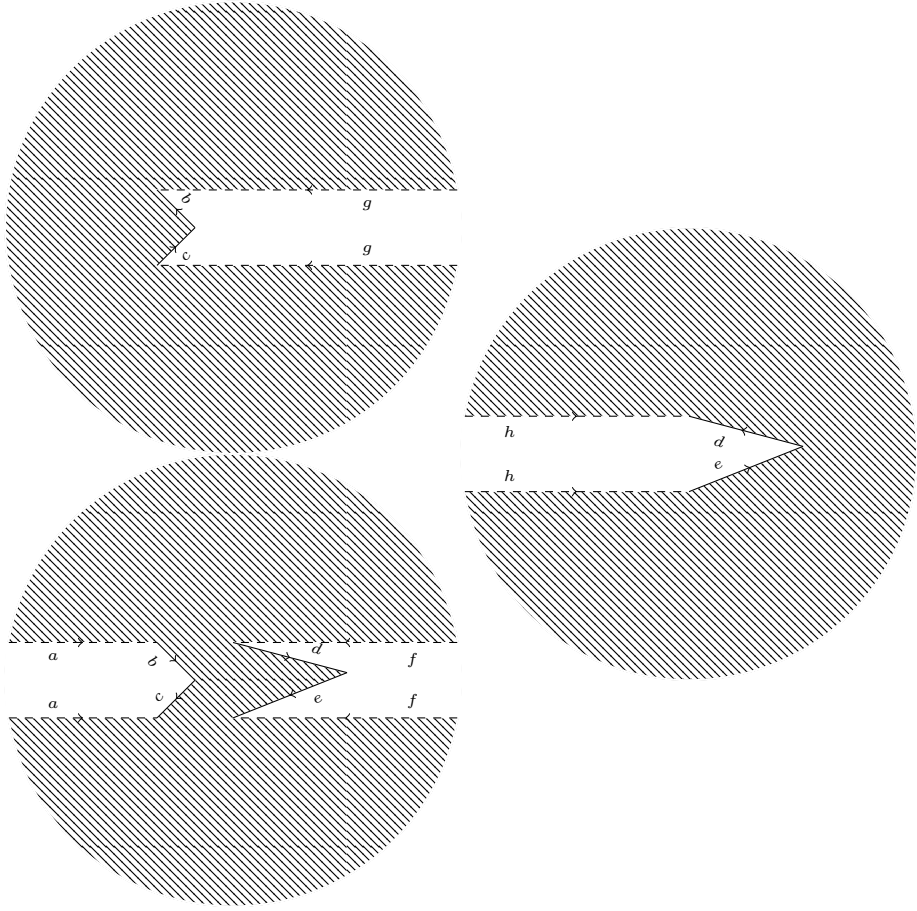
\begin{figure}	
\centering
		\begin{tikzpicture}
	\begin{scope}
\filldraw[ pattern=north west lines] (0,0) circle (3);  
\draw[color=white,thick] (0,0) circle (3);
	
	\coordinate (A) at (-2.95,0.5);
	\coordinate (B) at (-2.95,-0.5);
	\coordinate (C) at (-1,0.5);
	\coordinate (D) at (-1,-0.5);
	\coordinate (E) at (-0.5,0);
	\filldraw[color=white](-3,0.5)--(C)--(E)--(D)--(-3,-0.5)--(-3,0.5);
	
	\draw[dashed, ->-] (A) --(C)   node [pos=0.3, sloped, below] {$\scriptscriptstyle a$}; 
	\draw[->-](C)--(E)  node [pos=0.2, sloped, below] {$\scriptscriptstyle b$}; 
	\draw[->-](E)--(D)   node [pos=0.7, sloped, above] {$\scriptscriptstyle c$}; 
	\draw[dashed, ->-](B)--(D) node [pos=0.3, sloped, above] {$\scriptscriptstyle a$}; 
	\end{scope}
			\begin{scope}
		\coordinate (A2) at (2.95,0.5);
		\coordinate (B2) at (2.95,-0.5);
	\coordinate (C2) at (0,0.5);
		\coordinate (E2) at (1.5,0.1); 
		\coordinate (D2) at (0,-0.5); 
			\filldraw[color=white](3,0.5)--(C2)--(E2)--(D2)--(3,-0.5)--(3,0.5);
	
	\draw[dashed, ->-] (A2) --(C2) node [pos=0.2, sloped, below] {$\scriptscriptstyle f$};  
	\draw[->-](C2)--(E2)  node [pos=0.7, sloped, above ] {$\scriptscriptstyle d$}; 
	\draw[->-](E2)--(D2)   node [pos=0.3, sloped, below ] {$\scriptscriptstyle e$}; 
	\draw[dashed, ->-](B2)--(D2) node [pos=0.2, sloped, above] {$\scriptscriptstyle f$};  
\end{scope}

	\begin{scope}[shift={(0,6)}]

\filldraw[ pattern=north west lines] (0,0) circle (3);  
\draw[color=white,thick] (0,0) circle (3);
	
		\coordinate (A) at (2.95,0.5);
		\coordinate (B) at (2.95,-0.5); 
	\coordinate (C) at (-1,0.5);
	\coordinate (D) at (-1,-0.5);
	\coordinate (E) at (-0.5,0);
		\filldraw[color=white](3,0.5)--(C)--(E)--(D)--(3,-0.5)--(3,0.5);
	
		\draw[dashed, ->-] (A) --(C)   node [pos=0.3, sloped, below] {$\scriptscriptstyle g$};  
	\draw[->-](E)--(C)   node [pos=0.5, sloped, above] {$\scriptscriptstyle b$}; 
	\draw[->-](D)--(E)  node [pos=0.5, sloped, below] {$\scriptscriptstyle c$}; 
	\draw[dashed, ->-](B)--(D) node [pos=0.3, sloped, above ] {$\scriptscriptstyle g$}; 

	\end{scope}

	\begin{scope}[shift={(6,3)}]
	
	\filldraw[ pattern=north west lines] (0,0) circle (3);  
	\draw[color=white,thick] (0,0) circle (3); 
		\coordinate (A2) at (-2.95,0.5);
		\coordinate (B2) at (-2.95,-0.5);
	\coordinate (C2) at (0,0.5);
		\coordinate (E2) at (1.5,0.1); 
		\coordinate (D2) at (0,-0.5); 
			\filldraw[color=white](-3,0.5)--(C2)--(E2)--(D2)--(-3,-0.5)--(-3,0.5);
	
	\draw[dashed, ->-] (A2) --(C2) node [pos=0.2, sloped, below] {$\scriptscriptstyle h$};  
	\draw[-<-](C2)--(E2)  node [pos=0.3, sloped, below] {$\scriptscriptstyle d$}; 
	\draw[-<-](E2)--(D2)   node [pos=0.7, sloped, above] {$\scriptscriptstyle e$}; 
	\draw[dashed, ->-](B2)--(D2) node [pos=0.2, sloped, above] {$\scriptscriptstyle h$};  

	\end{scope}
\end{tikzpicture}
\caption{Gluing the three planes as shown, we obtain $\CP^1$ with three double poles and four zeroes. Dashed lines are semi-infinite. \label{fig:3poles} }
\end{figure}

The flat surface corresponding to this one-form has $n$ regions which are a plane with a semi-infinite strip removed, where the two boundaries of the strip are glued together. See \fig\ref{fig:3poles} for the surface in the case $n = 3$. 

The resulting theory is
\begin{equation} 
	\mbf{IFT}(\CP^1,\omega, \mc{P}(q_1^w), \dots , \mc{P}(q_{n-1}^w), \br{\mc{P}}(q_1^{\br{w}}), \dots, \br{\mc{P}}(q_{n-1}^{\br{w}}) ) \;. \label{eqn:ift_cp1} 
\end{equation}
We show that this model yields a generalization of the principal chiral model, on the manifold $G^{n} / G$ (we quotient by the diagonal right action).  The model has a $G^n$ global symmetry, given by the left action. We explicitly calculate the Lagrangian: it is given by a certain $G^n$ invariant metric and WZ three-form on $G^n / G$.  

The expression we derive for the Lagrangian matches precisely an expression derived in \cite{Delduc:2018hty,Delduc:2019bcl} in a study of the affine Gaudin model.

We have explained the results of our construction for genus $0$ curves. We have not computed the most general integrable field theory one can build using the disorder operators considered so far. One constraint we have imposed is that  we always consider an equal number of chiral and anti-chiral disorder operators.  Another restriction is that we have mostly focused on one-forms with only second order poles, and no first order poles. 

One can show, however, that a first-order pole with a chiral Dirichlet boundary condition is obtained from a collision of a second-order pole, with a full Dirichlet boundary condition, and a first-order zero with a chiral disorder operator.   Therefore the model \eqref{eqn:ift_cp1} encompasses all genus $0$ models build from this class of disorder operators, which have the same number of chiral and anti-chiral degrees of freedom.

A further generalization which we consider in section \ref{sec:general_PCM} is given by splitting the Dirichlet boundary condition at a second-order pole into two ``trigonometric'' boundary conditions $\mc{T}_+$, $\mc{T}_-$ at first order poles.   Earlier, we studied the theory on $(\CP^1,\d z / z)$ with trigonometric boundary conditions $\mc{T}_{+}(0)$, $\mc{T}_-(\infty)$  at $0,\infty$, and with order defects placed at points $z_0,z_1$. We found that this led to models such as the sausage model discussed in section \ref{subsec:sausage}.  All of the models listed above, such as those in \eqn\eqref{eqn:ift_cp1}, admit trigonometric deformations.  We do not study the trigonometric deformations of the most general models, but in section \ref{sec:general_PCM} we do find explicit formulae for the Lagrangian of the trigonometric deformation of the principal chiral model.

\subsection{Symmetric Spaces}

We have not yet encountered some of the most familiar models with a genus $0$ spectral curve: the Riemannian symmetric space models.  To engineer these models, we need to introduce a new class of disorder operators in sections \ref{sec:symmetric}, \ref{sec:gss}.  We suppose that our group $G$ has an action of $\Z_n$, generated by a transformation $\rho : G \to G$.  The subgroup fixed by $\rho$ will be denoted $G_0$.  

Because $\rho$ acts on $G$, it also acts on the Lie algebra $\g$, which decomposes as $\g = \g_0 \oplus \dots \oplus \g_{n-1}$, where $\rho$ acts on $\g_k$ by $e^{2 \pi \i k/n}$. The new class of disorder operator is obtained by asking that the component $A_w^{(k)}$ of the gauge field which is in $\g_k$ acquires a pole of fractional order at $z = z_0$:
\begin{equation} 
	A_w^{(k)} \sim (z-z_0)^{-\frac{k}{n}} \;. 
\end{equation}
We call this defect $\mc{P}^{\rho}(z_0)$. Similarly, we let $\br{\mc{P}}^{\rho}(z_0)$ be the defect where $A_{\wbar}$ has a fractional pole.  

Clearly, the disorder operator $\mc{P}^{\rho}(z_0)$ can not live by itself: it lives at the end of a topological line defect, where as we cross the line we apply the automorphism $\rho$. If we have a line connecting $z_0$ and $z_1$, the  line defect can end on $\mc{P}^{\rho}(z_0)$ and on $\br{\mc{P}}^{\rho^{-1}}(z_1)$.  The simplest theories constructed with this class of disorder operators are
\begin{equation} 
	\mbf{IFT}_{G,k}(\CP^1, \d z, \mc{P}^{\rho}(z_0), \br{\mc{P}}^{\rho^{-1}}(z_1)) \;. 
	\label{eqn:symmetricft}
\end{equation}

In section \ref{sec:symmetric}, we show that, if $\rho$ generates a $\Z_2$ action on $G$, the model  \eqref{eqn:symmetricft} is the integrable symmetric space $\sigma$-model on the coset $G/G_0$.  The Lax matrix is the standard one, in a coordinate where it has a branch cut as it crosses the line connecting $z_0$ and $z_1$.

We analyze the case when  $\rho$ generates a $\Z_n$ action on $G$ in section \ref{sec:gss}. In this case, \eqref{eqn:symmetricft} is an integrable $\sigma$-model on the quotient $G/G_0$, which is a type of homogeneous space known as an $n$-symmetric space. 

The most famous case of such a model is when $G = PSU(2,2\, | \, 4)$, equipped with a certain $\Z_4$ action.  In this case, the model $\mbf{IFT}_{PSU(2,2\mid 4),k}(\CP^1, \d z, \mc{P}^{\rho}(z_0), \br{\mc{P}}^{\rho^{-1}}(z_1))$ is the integrable $AdS_5 \times S^5$ $\sigma$-model, in the formulation that is natural for the pure-spinor formulation of string theory.  The Lax matrix is precisely the pure-spinor Lax matrix, in a coordinate where there is a branch cut on the line connecting $z_0$ and $z_1$.  

\subsection{Higher Genus Models}

Our construction allows one to build generalizations of this model where the spectral curve is an arbitrary Riemann surface. We investigate these models in section \ref{sec:higher_genus}.  We consider a Riemann surface $C$ of arbitrary genus $g$, equipped with a meromorphic one-form $\omega$. We assume that $\omega$ has $n$ second-order poles at points $p_k$, and $2g-2+2n$ simple zeroes, divided into two groups $q_i^w$, $q_i^{\wbar}$ of size $g-1+n$.  At $q_i^w$ we place chiral disorder operators, and at $q_i^{\wbar}$ we place anti-chiral disorder operators.  The resulting theory is
\begin{equation} 
	\mbf{IFT}_{G,k}(C,\omega, \mc{P}(q_i^w), \br{\mc{P}}(q_i^{\wbar}) ) \;. 
\end{equation}
(In this analysis, we exclude the special case that $C$ is of genus $1$ and $\omega$ has no poles. This model gives rise to a two-dimensional topological field theory (TFT)).

\begin{figure}
\centering
\begin{tikzpicture}
	\filldraw[ pattern=north west lines] (1,0) circle (3);  
	\draw[color=white,thick] (1,0) circle (3); 
	\coordinate (A) at (0,0);
	\coordinate (B) at (60:1);
	\coordinate (C) at ($(B) + (0:1)$);
	\coordinate (D) at ($(C) + (-60:1)$);
	\coordinate (E) at (-60:1);
	\coordinate (F) at ($(E) + (0:1)$);
	\filldraw[color=white] (A) --(B) --(C) --(D) --(F) --(E) --(A);
	\draw [->-](A) --(B) node [midway, sloped, below] {$\scriptscriptstyle a$}; 
	\draw [->-](B) --(C) node [midway, sloped, below] {$\scriptscriptstyle b$};  
	\draw [->-](C) --(D) node [midway, sloped, below] {$\scriptscriptstyle c$}; 

	\draw [->-](A) --(E) node [midway, sloped, above] {$\scriptscriptstyle c$};  
	\draw [->-](E) --(F) node [midway, sloped, above] {$\scriptscriptstyle b$};  
	\draw [->-](F) --(D) node [midway, sloped, above] {$\scriptscriptstyle a$};  
	
\end{tikzpicture}
	\caption{Identifying parallel sides of the hexagon gives a torus with a one-form with one double pole and two zeroes.  }
	\label{fig:torus}
\end{figure}
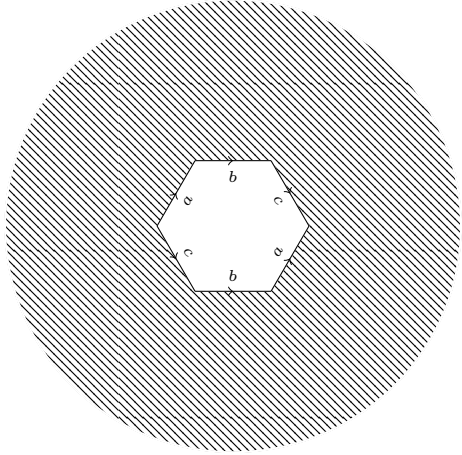

An interesting example is given by the elliptic curve with a one-form with a second order pole and two zeroes.  The corresponding flat surface is illustrated in \fig \ref{fig:torus}. 

This class of models is the most general that we can build using our disorder operators, subject to the constraint that there are the same number of chiral and anti-chiral degrees of freedom. (We do not analyize trigonometric boundary conditions, but these can of course be included).  

We find that $\mbf{IFT}_{G,k}(C,\omega, \mc{P}(q_i^w), \br{\mc{P}}(q_i^{\wbar}) )$ can be described as a $\sigma$-model whose target space is a certain moduli of bundles on $C$.  We let $\op{Bun}_{\GC}^0(C, \omega)$ be the open subset of the moduli space of holomorphic $\GC$ bundles on $C$, trivialized at the poles of $\omega$, with the stability constraint that the bundle admits no holomorphic gauge transformations which vanish at $p_k$ and have at most first-order poles at $q_i^w$.  If we choose an anti-holomorphic involution on the Riemann surface $C$, then we induce one on $\op{Bun}_{\GC}^0(C,\omega)$; the fixed points are a certain manifold $\op{Bun}_{G}^0(C,\omega)(\R)$. 

In the case that $C = \CP^1$ and $\omega$ has $n$ second-order poles, then $\op{Bun}_G^0(C,\omega)(\R) = G^{n} / G$, and we are in the situation studied above. 

In section \ref{sec:higher_genus} we show that $\mbf{IFT}_{G,k}(C,\omega, \mc{P}(q_i^w), \br{\mc{P}}(q_i^{\wbar}) )$ is  the $\sigma$-model on the manifold $\op{Bun}_G^0(C,\omega)(\R)$, where this manifold is equipped with a certain metric and three-form.  The metric and three-form are described explicitly in terms of an integral kernel on $C$, the \emph{Sz\"ego kernel}.  Unfortunately, we were unable to obtain a closed-form expression for the Sz\"ego kernel in general, although it is uniquely determined by the differential equations that it satisfies.  We also find an expression for the Lax matrix in terms of certain geometric data on the moduli space $\op{Bun}_G^0(C,\omega)(\R)$. 

In the case that $C = E$ is of genus $1$, and 
\begin{equation} 
	\omega =  a \wp(z) \d z + b  \d z \;,
\end{equation}
where $\wp$ is the Weierstrass $\wp$-function,  then we can be more explicit. The corresponding flat surface is illustrated in \fig\ref{fig:torus}.  The three parameters in this model --- the elliptic modulus $\tau$ and the parameters $a,b$ in the one-form $\omega$ --- correspond, after a non-trivial transformation, to the three vectors $a,b,c$ in \fig\ref{fig:torus}.  

In this case, the target space of the $\sigma$-model is the group manifold $G$. Because $\omega$ has a single second-order pole, the model has a $G$-global symmetry; this is given by the adjoint action.

The model is specified by a metric and a three-form on the group manifold $G$, invariant under the adjoint action. In the case $G = SU(2)$ we compute the metric and three-form reasonably explicitly, in terms of certain elliptic functions.  

\subsection{Gluing}
We have explained how to build an integrable field theory from a Riemann surface $C$ with a one-form.  In section \ref{sec:gluing} we will show that gluing of Riemann surfaces at a node corresponds to gauging the corresponding integrable field theories.

\begin{figure}
\centering
	\subfloat[Gluing two genus $1$ surfaces, each with two cylindrical boundaries, to get a genus $2$ surface.  In the center, we have a conformal WZW model. $\mc{D}$ and $\br{\mc{D}}$ indicate chiral and anti-chiral boundary conditions. ]{
		\label{fig:gluing_a}
	\begin{tikzpicture}[scale=0.85]

\begin{scope}[shift={(-1,0)}]

	\node at (-1.5,1.5) {$\cdots$}; 
	\node at (-1.5,-1.5) {$\cdots$};

	\node at (-1.2,0) {$\scriptstyle \mc{D}$};
	\coordinate (A) at (3,1.5);
	\coordinate (B) at (3,-1.5);
	\coordinate (C) at (-1,1.5);
	\coordinate (D) at (-1,-1.5);
	
	\filldraw[pattern=north west lines](A)--(C)--(D)--(B)--(A);
	\draw[color=white,thick](A)--(C)--(D)--(B)--(A);
	\draw[dashed, ->-] (A) --(C);

	\draw[dashed, ->-](B)--(D);

	\begin{scope}[shift={(0,0)}]
	\coordinate (A) at (0,0);
	\coordinate (B) at (60:1);
	\coordinate (C) at ($(B) + (0:1)$);
	\coordinate (D) at ($(C) + (-60:1)$);
	\coordinate (E) at (-60:1);
	\coordinate (F) at ($(E) + (0:1)$);
	\filldraw[color=white] (A) --(B) --(C) --(D) --(F) --(E) --(A);
	\draw [->-](A) --(B) node [midway, sloped, below] {$\scriptscriptstyle a$}; 
	\draw [->-](B) --(C) node [midway, sloped, below] {$\scriptscriptstyle b$};  
	\draw [->-](C) --(D) node [midway, sloped, below] {$\scriptscriptstyle c$}; 

	\draw [->-](A) --(E) node [midway, sloped, above] {$\scriptscriptstyle c$};  
	\draw [->-](E) --(F) node [midway, sloped, above] {$\scriptscriptstyle b$};  
	\draw [->-](F) --(D) node [midway, sloped, above] {$\scriptscriptstyle a$};  
	\end{scope}

\end{scope}
	
	\node (n1) at (2.2,0) {$\scriptstyle \br{\mc{D}}$};
	\node (n2) at (3.8,0){$\scriptstyle \br{\mc{D}}$};
	\draw (n1) -- (n2) node[midway,below]{\tiny BRST};
	  
\begin{scope} 
	
	\coordinate (A2) at (4,1.5);
	\coordinate (B2) at (4,-1.5);
	\coordinate (C2) at (7,1.5);
	\coordinate (D2) at (7,-1.5);	
	\filldraw[pattern=north west lines](A2)--(C2)--(D2)--(B2)--(A2);
	\draw[color=white,thick](A2)--(C2)--(D2)--(B2)--(A2);
	\draw[dashed, -<-] (A2) --(C2);
	\draw[dashed, -<-](B2)--(D2);
	
\end{scope}
	
	\node(n3) at (7.2,0) {$\scriptstyle {\mc{D}}$};

	
	\node(n4) at (8.8,0){$\scriptstyle {\mc{D}}$};
	\draw (n3) -- (n4) node [midway, below]{\tiny BRST}; 

\begin{scope}[shift={(1,0)}] 
	
	\begin{scope}[shift={(9,0)}]

	\coordinate (A) at (3,1.5);
	\coordinate (B) at (3,-1.5);
	\coordinate (C) at (-1,1.5);
	\coordinate (D) at (-1,-1.5);
	
	\filldraw[pattern=north west lines](A)--(C)--(D)--(B)--(A);
	\draw[color=white,thick](A)--(C)--(D)--(B)--(A);
	\draw[dashed, ->-] (A) --(C);

	\draw[dashed, ->-](B)--(D);

	\end{scope}

	\begin{scope}[shift={(8.5,0)}]
	\coordinate (A) at (0,0);
	\coordinate (B) at (30:1);
	\coordinate (C) at ($(B) + (-10:1)$);
	\coordinate (D) at ($(C) + (-40:1)$);
		\coordinate (E) at ($(D) + (-150:1)$);
	\coordinate (F) at ($(E) + (170:1)$);
	\filldraw[color=white] (A) --(B) --(C) --(D) --(E) --(F) --(A);
	\draw [->-](A) --(B) node [midway, sloped, below] {$\scriptscriptstyle x$}; 
	\draw [->-](B) --(C) node [midway, sloped, below] {$\scriptscriptstyle y$};  
	\draw [->-](C) --(D) node [midway, sloped, below] {$\scriptscriptstyle z$}; 

	\draw [-<-](D) --(E) node [midway, sloped, above] {$\scriptscriptstyle x$};  
	\draw [-<-](E) --(F) node [midway, sloped, above] {$\scriptscriptstyle y$};  
	\draw [-<-](F) --(A) node [midway, sloped, above] {$\scriptscriptstyle z$};  
	\end{scope}

	\node at (12.5,1.5) {$\cdots$}; 
	\node at (12.5,-1.5) {$\cdots$};

	\node at (12.2,0) {$\scriptstyle \br{\mc{D}}$};
	
\end{scope}

\end{tikzpicture}}

\subfloat[As $L \to \infty$ we approach configuration (a). ]{
	\label{fig:gluing_b}
\begin{tikzpicture}[scale=0.85]

	\node at (-1.5,1.5) {$\cdots$}; 
	\node at (-1.5,-1.5) {$\cdots$};	

	\node at (-1.2,0) {$\scriptstyle \mc{D}$};
	\coordinate (A) at (12,1.5);
	\coordinate (B) at (12,-1.5);
	\coordinate (C) at (-1,1.5);
	\coordinate (D) at (-1,-1.5);
	
	\filldraw[pattern=north west lines](A)--(C)--(D)--(B)--(A);
	\draw[color=white,thick](A)--(C)--(D)--(B)--(A);
	\draw[dashed, ->-] (A) --(C);

	\draw[dashed, ->-](B)--(D);

	\begin{scope}[shift={(0,0)}]
	\coordinate (A) at (0,0);
	\coordinate (B) at (60:1);
	\coordinate (C) at ($(B) + (0:1)$);
	\coordinate (D) at ($(C) + (-60:1)$);
	\coordinate (E) at (-60:1);
	\coordinate (F) at ($(E) + (0:1)$);
	\filldraw[color=white] (A) --(B) --(C) --(D) --(F) --(E) --(A);
	\draw [->-](A) --(B) node [midway, sloped, below] {$\scriptscriptstyle a$}; 
	\draw [->-](B) --(C) node [midway, sloped, below] {$\scriptscriptstyle b$};  
	\draw [->-](C) --(D) node [midway, sloped, below] {$\scriptscriptstyle c$}; 

	\draw [->-](A) --(E) node [midway, sloped, above] {$\scriptscriptstyle c$};  
	\draw [->-](E) --(F) node [midway, sloped, above] {$\scriptscriptstyle b$};  
	\draw [->-](F) --(D) node [midway, sloped, above] {$\scriptscriptstyle a$};  
	\end{scope}

	\begin{scope}[shift={(8.5,0)}]
	\coordinate (A) at (0,0);
	\coordinate (B) at (30:1);
	\coordinate (C) at ($(B) + (-10:1)$);
	\coordinate (D) at ($(C) + (-40:1)$);
		\coordinate (E) at ($(D) + (-150:1)$);
	\coordinate (F) at ($(E) + (170:1)$);
	\filldraw[color=white] (A) --(B) --(C) --(D) --(E) --(F) --(A);
	\draw [->-](A) --(B) node [midway, sloped, below] {$\scriptscriptstyle x$}; 
	\draw [->-](B) --(C) node [midway, sloped, below] {$\scriptscriptstyle y$};  
	\draw [->-](C) --(D) node [midway, sloped, below] {$\scriptscriptstyle z$}; 

	\draw [-<-](D) --(E) node [midway, sloped, above] {$\scriptscriptstyle x$};  
	\draw [-<-](E) --(F) node [midway, sloped, above] {$\scriptscriptstyle y$};  
	\draw [-<-](F) --(A) node [midway, sloped, above] {$\scriptscriptstyle z$};  
	\end{scope}

	\node at (12.5,1.5) {$\cdots$}; 
	\node at (12.5,-1.5) {$\cdots$};

	\node at (12.2,0) {$\scriptstyle \br{\mc{D}}$};
	\draw[<->] (2,-1.7) -- (8.5,-1.7)  node [midway,below] {$\scriptstyle L$};
\end{tikzpicture}}

\caption{\label{fig:gluing} }
\end{figure}

Suppose we have a family $(C_L, \omega_L)$ of flat surfaces which has a cylinder of length $L \to \infty$ and of fixed radius (see \fig\ref{fig:gluing}).  When $L \to \infty$, we can cut the surface along the cylinder to find a new surface $\til{C}$ with a one-form $\til{\omega}$.  This has two special punctures $z_1,z_2$ where the one-form $\til{\omega}$ has a first-order pole, with opposite residue.  In the natural flat metric, $\til{C}$ has two infinite cylinders which are glued together to give $\lim_{L \to \infty} C_L$. 

The surface $\til{C}$ may have other special points, where the one-form $\til{\omega}$ has other poles or zeroes, or where we have other defects.  We consider the integrable field theory
\begin{equation} 
	\mbf{IFT}(\til{C}, \til{\omega}, \dots, \mc{D}(z_0),\br{\mc{D}}(z_1) ) \label{eqn:ift_normalized} 
\end{equation}
with chiral and anti-chiral Dirichlet boundary conditions at $z_0,z_1$, where the other boundary conditions and defects are indicated by $\ldots$.

The Dirichlet boundary conditions give the theory in \eqn \eqref{eqn:ift_normalized} an action of the chiral and anti-chiral Kac-Moody at the same level. 

The theory on the cylinder is the conformal WZW model. This also has chiral and anti-chiral Kac-Moody algebra actions. By choosing the residue of the one-form on the cylinder correctly, we can ensure that the Kac-Moody actions have levels opposite to those of $\mbf{IFT}(\til{C}, \til{\omega})$.  We can then glue the conformal WZW model to $\mbf{IFT}(\til{C}, \til{\omega})$ by doing both chiral and anti-chiral BRST reduction.

The main result of section \ref{sec:gluing} is that this is the same as the $L \to \infty$ limit of the theory $\mbf{IFT}(C_L, \omega_L)$. That is, 
\begin{align} 
	 &\lim_{L \to \infty} \mbf{IFT}(C_L,\omega_L,\dots ) \nonumber\\
	 & \qquad \qquad =\Bigg\{ \mbf{IFT}(\til{C}, \til{\omega}, \dots, \mc{D}(z_1), \br{\mc{D}}(z_2)) \times \mbf{IFT}\left(\CP^1,\lambda \tfrac{\d z}{z} \right) \Bigg\} 
	 \Bigg/\!\!\!\Bigg/ (G_L \times G_R)  \;.
\end{align}
See \fig\ref{fig:gluing} for a geometric representation of this equality.  In this way, the models we construct on a higher genus curve are built from a small number of simple genus $0$ models, by gluing and then performing certain marginal deformations. 

Note that, if on $\til{C}$, we have the same number of chiral and anti-chiral degrees of freedom, then the glued theory also has the same number of chiral and anti-chiral degrees of freedom.  This gluing procedure therefore preserves this class of models. 

\section{Conformal WZW Model}
\label{sec:WZW}

The examples we have introduced so far do not include the principal chiral model, and its cousin with the 
Wess-Zumino term \cite{Witten:1983ar}, namely the Wess-Zumino-Witten (WZW) model. Since these are one of the 
most commonly studied integrable field theories, one can ask if they can be realized by the four-dimensional Chern-Simons theory.
 
Because the (compact) group manifold $G$ (as opposed to its complexification $\GC$) is not K\"ahler, this model can not be realized by coupling to a $\beta-\gamma$ system. 

In \cite{Costello:2017dso}, four-dimensional Chern-Simons theory was studied on $\R^2\times \CP^1$,
where the  spectral curve $\CP^1$ is equipped with the one-form $\d z /z$.  In these works, we introduced boundary conditions at $0$ and $\infty$ on $\CP^1$ which are topological along $\R^2$. Here we will introduce different boundary conditions, which are either chiral or anti-chiral along $\R^2$.

The boundary condition we take at $z = 0$ is given by asking $A_{\wbar}$ and $A_{\zbar}$ to be divisible by $z$. We also require gauge transformations to be divisible by $z$.  The Lagrangian $\d z / z \, \textrm{CS}(A)$ now has no poles at $z = 0$, because every term in the Lagrangian contains $A_{\wbar}$ or $A_{\zbar}$. This boundary condition is the \emph{chiral Dirichlet boundary condition}.  

One can also check that the kinetic term in the Lagrangian remains elliptic modulo gauge transformation. We will explain this later in section \ref{subsec.omega_zero} in a more general setting.

\begin{figure}[htbp]
\centering\includegraphics[scale=0.50]{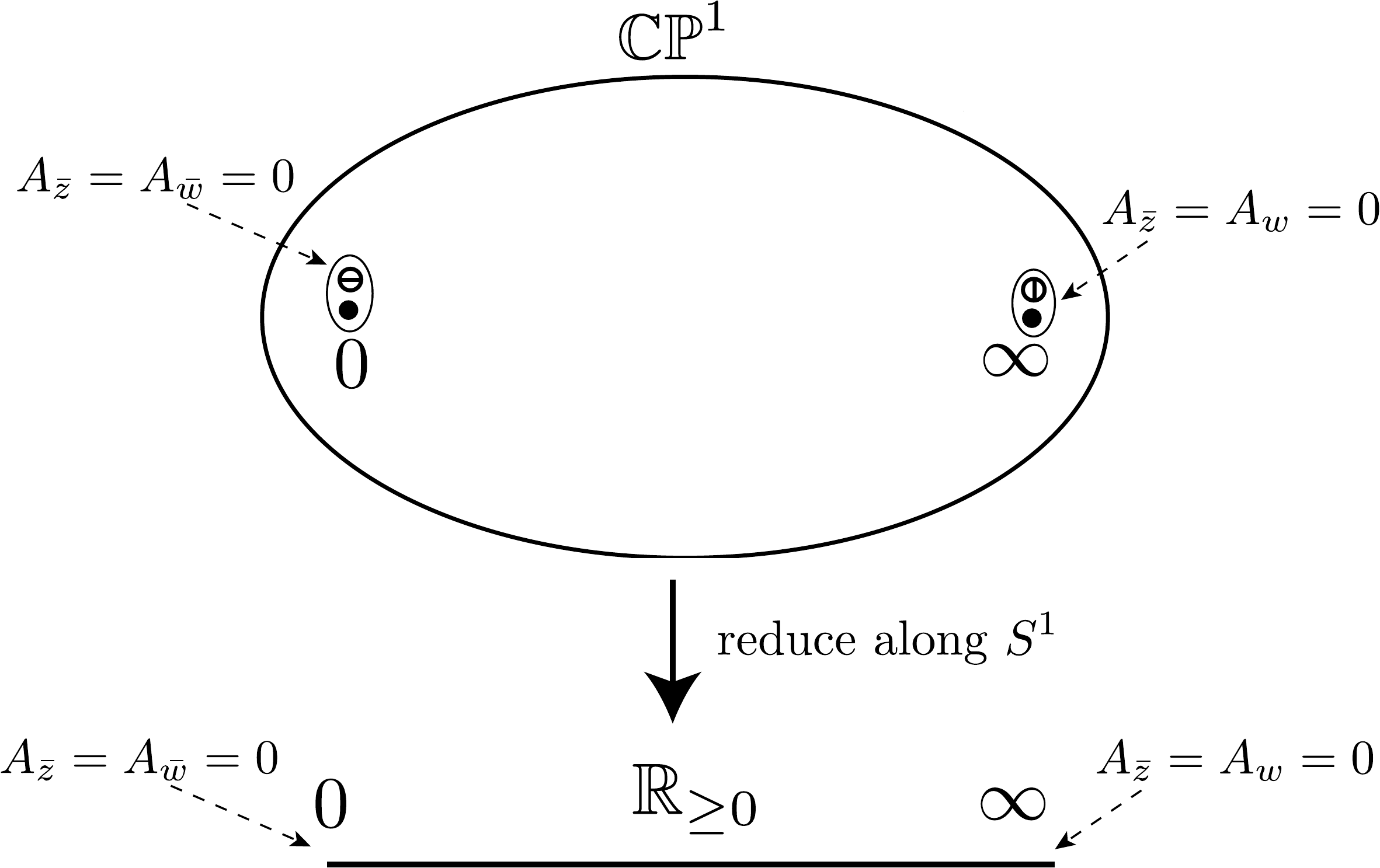}
\caption{The boundary condition used in this section. At the two poles $z=0$ ($\infty$) of the one-form $\omega =\d z/z$, we require that $A_{\zbar}=A_{w}=0$ ($A_{\zbar}=A_{\wbar}=0$). If we regard
$\mathbb{CP}^1$ as a cylinder and reduce the theory along $S^1$, we obtain
three-dimensional CS theory with holomorphic/anti-holomorphic boundary conditions in the 
``axial gauge'' $A_{\zbar}=0$. As we shall see in the next section, the setup can be regarded as a degeneration of a more general setup, which is why $0$ and $\infty$ are represented each as a collision of two special points.}
\label{Fig.III_14}
\end{figure}

Similarly, at $z=\infty$, we ask that $A_w$, $A_{\zbar}$ and all gauge transformations are divisible by $1/z$ (see \fig\ref{Fig.III_14}). This is the \emph{anti-chiral Dirichlet boundary condition}. 

We claim that four-dimensional Chern-Simons theory, compactified on $\CP^1$ with these boundary conditions, yields the conformal WZW model---i.e.\ the principal chiral model with Wess-Zumino term chosen so that it is conformal.

This result is very closely related to a well-known statement about ordinary three-dimensional Chern-Simons theory \cite{Elitzur:1989nr}.  Suppose we take Chern-Simons theory on $\Sigma \times [0,1]$, where $\Sigma$ is a Riemann surface.  At one of the endpoints $0$ we impose the boundary condition that $A_{\wbar} = 0$, and also all gauge transformations vanish (where $w$ is a local holomorphic coordinate on $\Sigma$).  This is the boundary condition corresponding to the chiral WZW model: boundary operators built from the spin one field $A_w$ give the Kac-Moody currents. At the other endpoint $1$, we impose the boundary condition that $A_w = 0$, corresponding to the anti-chiral WZW model  

We can compactify Chern-Simons theory on $\Sigma \times [0,1]$ to get a theory on $\Sigma$.  It is well-known that the resulting theory is the full WZW model, where the bulk Chern-Simons theory glues the chiral and anti-chiral parts together. 

Our four-dimensional situation can be turned into this familiar three-dimensional situation by dimension reduction.  We view $\CP^1 \setminus \{0,\infty\}$ as a cylinder with coordinates $t,\theta$.  Dimensionally reducing four-dimensional Chern-Simons theory along the circle $S^1_{\theta}$ with coordinate $\theta$ gives an analytic continuation of the three-dimensional Chern-Simons theory, with level $k$ is the period of the one-form $\frac{1}{\hbar} z^{-1} \d z$ along the circle $S^1$ we are reducing on (see section 7.8 of \cite{Costello:2018gyb} and \cite{Yamazaki:2019prm} for related discussion): 
\begin{equation}
\frac{k}{4\pi} \, \d t=\frac{1}{2\pi \hbar} \int_{S^1_{\theta}}  \frac{\d z}{z} =\frac{2\pi \i }{2\pi \hbar} \d t \;,
\label{eq.level}
\end{equation}
where the factor of $1/(2 \pi \hbar)$ comes from the normalization of the four-dimensional action \eqref{eq.action}. Note that this parameter $k$ is not necessarily an integer (and in fact not even real for real $\hbar$),
since the three-dimensional Chern-Simons theory, and hence the associated WZW model, are analytically continued.

The boundary conditions at $z = 0$ and $z = \infty$ give boundary conditions for three-dimensional Chern-Simons theory. At $z = 0$, this is the boundary condition giving the chiral WZW model, and at $z = \infty$, we find the boundary condition giving the anti-chiral WZW model. 

It will be useful, however, to give a direct proof that the four-dimensional construction yields the WZW model. 
This is what we discuss next.

\subsection{Direct Derivation of the WZW Model}

\subsubsection{Boundary Conditions}

We first note that the moduli space of stable holomorphic $G$-bundles on $\CP^1$, trivialized at $0$ and $\infty$, is $G$.  This is because all stable bundles are trivial, so that we can compare the trivializations at $0$ and $\infty$ to yield an element of $G$. 

A field of four-dimensional Chern-Simons theory yields, in particular, such a stable bundle on $\CP^1$ for every point in $\R^2$. We thus find a map $\sigma : \R^2 \to G$.  

Conversely, every such map $\sigma: \R^2 \to G$ will yield a gauge equivalence class of a field of four-dimensional Chern-Simons theory, where only the $\zbar$ component is non-zero.  The field $A_{\zbar}^{\sigma}$ associated to $\sigma$ will define, for each $(x,y) \in \R^2$, a holomorphic structure $A_{\zbar}^{\sigma}(x,y)$ on the  trivial principal $G$-bundle on $(x,y) \times \CP^1$.  This holomorphic structure has the property that if $\phi$ is a holomorphic section which takes value $g$ at $z = 0$, then it takes value $\sigma(x,y) g$ at $z = \infty$.  Notice that we have here broken the symmetry between points $z=0$ and $z=\infty$.

This constraint characterizes $A_{\zbar}^{\sigma}$ up to a gauge transformation vanishing at $z = 0$ and $z = \infty$. To find an explicit formula for some $A_{\zbar}^{\sigma}$, we need to extend $\sigma: \R^2 \to G$ to a map $\what{\sigma} : \R^2 \times \CP^1 \to G$, where $\what{\sigma} = \op{Id}_G$ in a neighborhood of $z = 0$, and $\what{\sigma} = \sigma$ in a neighborhood of $z = \infty$. We can assume without losing generality that $\what{\sigma}$ is invariant under the $U(1)$ action rotating $z$.  

The map $\sigma: \R^2 \to G$ gives us the $\zbar$ component of the four-dimensional gauge field up to gauge equivalence. One explicit choice of $A_{\zbar}^{\sigma}$ is given by 
\begin{equation} 
	A_{\zbar}^{\sigma}= \what{\sigma}^{-1} \partial_{\zbar} \what{\sigma} \;.  
\end{equation}

The remaining fields are $A_{w}$, $A_{\wbar}$. These are adjoint-valued $(1,0)$ and $(0,1)$ forms on $\R^2$ with values in $\Omega^{0,0}(\CP^1,\Oo(-D))$, where $D$ is the divisor $\infty$ for $A_w$ or $0$ for $A_{\wbar}$.
We will impose the equations of motion $F_{\zbar w} = 0$, $F_{\zbar \wbar} = 0$ to solve for $A_{w}$, $A_{\wbar}$ in terms of $\what{\sigma}$. Reinserting these fields into the action will yield the action for the WZW model. We will carry out this computation in the rest of this subsection.

We note that if $A_{\wbar}$, $A'_{\wbar}$ both satisfy $F_{\zbar w} = 0$ and both vanish at $z = 0$, then the difference between them satisfies
\begin{equation} 
	(\partial_{\zbar} + A_{\zbar}^{\sigma}) (A_{\wbar} - A'_{\wbar} ) = 0 \;.   
\end{equation}
That is, the difference between them is a holomorphic section of the adjoint bundle associated to the holomorphic bundle on $\CP^1$ given by $A_{\zbar}^{\sigma}$, which vanishes at $z = 0$. There are no such holomorphic sections, so we conclude there is a unique $A^{\sigma}_{\wbar}$ such that $F_{\zbar \wbar}= 0$ and which vanishes at $z=0$. In a similar way, there is a unique $A_{w}^\sigma$.


Evidently, if we set 
\begin{equation} 
A_{\wbar}^{\sigma} = \what{\sigma}^{-1} \partial_{\wbar} \what{\sigma}  \;, 
 \end{equation}
then both $A_{\zbar}$ and $A_{\wbar}$ are in the pure gauge and we have $F_{\zbar \wbar} = 0$.  Because $\what{\sigma}$ is constant near $z = 0$, $A_{\wbar}^{\sigma}$ also vanishes at $z = 0$, satisfying the boundary condition.

We can not simply set $A^{\sigma}_w = \what{\sigma}^{-1} \partial_w \what{\sigma}$, however, because although this satisfies $F_{\zbar w} = 0$, it does not vanish at $z = \infty$. Of course, the setup is manifestly symmetric under the 
simultaneous exchange of $z=0, z=\infty$ and $w, \wbar$, which suggests that 
there is democracy between $A_w$ and $A_{\wbar}$.
This is apparently broken now 
only because the parametrization by $\what{\sigma}$ breaks this symmetry:
$\what{\sigma}$ is an identity around $z=0$,  but not around $z=\infty$.

To recover this symmetry, let us set $\til{\sigma}= \sigma^{-1} \what{\sigma} $.  This $\til{\sigma}$ is an identity at $z = \infty$ and is $\sigma^{-1}$ near $z = 0$. Furthermore,  
\begin{equation} 
	\til{\sigma}^{-1} \partial_{\zbar} \til{\sigma} = \what{\sigma}^{-1} \partial_{\zbar} \what{\sigma} 
	\label{eq.sigmatilde}
\end{equation}
because $\sigma$ is independent of $\zbar$.  This means that the $\zbar$ component of the gauge field takes the same form whether we use $\til{\sigma}$ or $\what{\sigma}$.  

The equation $F_{\zbar w} = 0$ only depends on $A^{\sigma}_{\zbar} = \til{\sigma}^{-1} \partial_{\zbar} \til{\sigma}$, so we can solve this by setting
\begin{equation} 
	A^{\sigma}_w = \til{\sigma}^{-1} \partial_{w} \til{\sigma} = \what{\sigma}^{-1} \partial_w \what{\sigma} -  \what{\sigma}^{-1} (\partial_{w} \sigma) \sigma^{-1} \what{\sigma}\;. 
\end{equation}
This vanishes at $z = \infty$, and is the unique solution to the equation $F_{{\zbar} w} = 0$, given $A_{\zbar}$. 

Summarizing, letting $\what{A}_i = \what{\sigma}^{-1} \partial_i \what{\sigma}$, for $i = w,\wbar,\zbar$, we have
\begin{equation} 
A^{\sigma}= \what{A}+A' \;, \quad A'= -  \what{\sigma}^{-1} (\partial_{w} \sigma) \sigma^{-1} \what{\sigma}\d w
\label{eq.A_sum}  \;.
 \end{equation}

\subsubsection{Lagrangian}

Having solved for $A^{\sigma}_{\zbar}, A^{\sigma}_{\wbar}$ and $A^{\sigma}_w$, we can now reinsert them into the action. 

The Chern-Simons three-form for the sum of two connections $\what{A}$ and $A'$ in general takes the form 
\begin{equation} 
\textrm{CS}(\what{A} + A') = \, \textrm{CS}(\what{A}) + 2\op{Tr} (F(\what{A})A') - \d \op{Tr} (\what{A} A') + 2\op{Tr} (\what{A} A' A' ) +\textrm{CS}(A') \;.
\label{CS_sum_3}
 \end{equation}
In our case $F(\what{A}) = 0$, and moreover we have $\textrm{CS}(A')=\op{Tr} (\what{A} A' A' )=0$ since $A'$ has only the $w$ component, so that
\begin{align} 
\textrm{CS}(A^{\sigma}) &=\textrm{CS}(\what{A} + A') = \, \textrm{CS}(\what{A})  - \d \op{Tr} (\what{A} A') \nonumber\\
&= \textrm{CS}(\what{A}) + \d    \op{Tr} \left( \what{A}   \what{\sigma}^{-1} (\partial_{w} \sigma) \sigma^{-1} \what{\sigma}\d w \right)
   \;.
\label{eq.CS_sum_conformal}
 \end{align}

We need to integrate this over $\R^2 \times \CP^1$ against the one-form $\omega=\d z / z$.  

The integrand in the first term in \eqn \eqref{eq.CS_sum_conformal}  
is the Chern-Simons functional in pure gauge:
\begin{equation} 
\textrm{CS}(\what{A}) = -\frac{1}{3}\op{Tr} \left( \what{\sigma}^{-1} \d \what{\sigma} \right)^3  \;.
\label{eq.CS_trivial}
 \end{equation}
Since $\what{\sigma}$ is chosen to be invariant under the $U(1)$ rotation of $z$, the integrand 
is also invariant under the rotation, and the integral over the angular parameter simply yields $2 \pi \i$, leading to
an integral over $\mathbb{R}^2$ as well as the radial direction $\R_{\ge 0}$ of the $\CP^1$:
\begin{equation} 
  \int_{\R^2 \times \CP^1}\textrm{CS}(\what{A})
= 
   -\frac{2 \pi \i}{3} \int_{\R^2 \times \R_{\ge 0} }  \op{Tr} (\what{\sigma}^{-1} \d \sigma)^3  \;. 
\end{equation}
Here the three-manifold $\R^2 \times \R_{\ge 0}$ bounds the $\R^2$,
and we are viewing 
$\what{\sigma}$ as an extension of $\sigma$ to a map $\R \times \R_{\ge 0} \to G$ which is the identity near $r = 0$.
 The final expression is manifestly the Wess-Zumino term.

The second term in \eqn \eqref{eq.CS_sum_conformal} is a total derivative in $\zbar$, however we need to careful since the $\zbar$ derivative acts non-trivially on the one-form $\d z / z$ when integrated by parts. 
We can use the fact $\dbar (\d z / z) = 2 \pi \i (\delta_{z = 0} - \delta_{z = \infty} )$. Since 
$\what{\sigma}= \textrm{Id}$ and hence $\what{A}=0$ near $z = 0$, and $\what{\sigma} = \sigma$ near $z=\infty$, we find
\begin{equation}
\begin{split} 
& \int_{\R^2 \times \CP^1} \frac{\d z}{z} \wedge
\d   \op{Tr} \left( \what{A}   \what{\sigma}^{-1} (\partial_{w} \sigma) \sigma^{-1} \what{\sigma}\d w \right) \\
&=2 \pi \i \int_{\R^2 } 
  \op{Tr} \left( \what{A}   \what{\sigma}^{-1} (\partial_{w} \sigma) \sigma^{-1} \what{\sigma}\d w \right)\Big|_{z=\infty}  = 2 \pi \i \int_{\R^2} \op{Tr} (\sigma^{-1} \partial_w \sigma) (\sigma^{-1} \partial_{\wbar} \sigma)  \;.\\
\end{split}
\end{equation}  

In sum, we have found that our construction yields the principal chiral model with the Wess-Zumino term:
\begin{align} 
S_{\rm PCM+WZW}&= \frac{k}{4\pi} \int_{\R^2} \op{Tr} (\sigma^{-1} \partial_w \sigma) (\sigma^{-1} \partial_{\wbar} \sigma) - 
   \frac{k}{12\pi}  \int_{\R^2 \times \R_{\ge 0} }  \op{Tr} (\what{\sigma}^{-1} \d \sigma)^3 
 \\
& = \frac{k}{8\pi} \int_{\R^2} \op{Tr} (j \wedge \star j) -
   \frac{k}{12\pi}  \int_{\R^2 \times \R_{\ge 0} }  \op{Tr} (\what{j} \wedge \what{j}\wedge \what{j}) 
   \;,
   \label{eq.WZW_conformal}
\end{align}
where we defined $j=\sigma^{-1} \d  \sigma$ and $\what{j}=\what{\sigma}^{-1} \d \what{\sigma}$,
and $k/(4\pi)=\i/\hbar$ is the level introduced previously in \eqn \eqref{eq.level}.
It is known in the literature that this combination gives the conformal WZW model.

\subsection{The Lax Operator} 

To construct the Lax operator of this model, we follow the prescription we gave earlier: the Lax operator is simply the expectation value of the four-dimensional gauge field, at some point $z \in \C^\times$.  

We have written the gauge field explicitly as a function of $\sigma$, so the Lax operator will be simply the value of the $A_{w}^{\sigma}$, $A_{\wbar}^{\sigma}$ at some point $z$. As we discussed before in \eqn \eqref{eq.Lax_holomorphic}, we should evaluate the Lax operator in a gauge in which $A_{\zbar} = 0$.
This means to apply the inverse gauge transformation by $\what{\sigma}$,
so that the gauge field \eqref{eq.A_sum} becomes $A_w^{\sigma} = 0$ and $A_{\wbar}^{\sigma} = \sigma^{-1} \partial_{\wbar} \sigma = J_{\wbar}$.

We conclude that in this gauge
\begin{equation} 
\begin{split}
	\Lax_{\wbar}(z) &= J_{\wbar} \;, \\
	\Lax_w(z) &= 0 \;.
\end{split}	
\end{equation}
The Lax operator is quite trivial. However, the Lax equation tells us something interesting: $J_{\wbar}$ is conserved by itself (and therefore $J_w$ is also conserved by itself).  

We have found that the Lax operator is compatible with this construction yielding the conformal WZW model. 

\section{\texorpdfstring{Disorder Operators at Zeroes of $\omega$}{Disorder Operators at Zeroes of Omega}}\label{subsec.omega_zero}

Almost all of the models we will study from now on involve one-forms with zeroes. Equivalently, this means that the flat surface with metric $\omega \br{\omega}$ has a conical singularity with angle $4 \pi$.

This gives us a new ingredient in the construction. We need to carefully prescribe the behavior of the fields of our theory at the zeroes. Otherwise, the equations of motion -- which, near a zero of the one-form, state that $z \d z F(A) = 0$ -- are not elliptic modulo gauge, and we can not write down a sensible propagator needed for a perturbation theory.

What we can do is to ask that at a zero, we allow $A_{\wbar}$ to have a first-order pole (alternatively, we could allow $A_w$ to have a first-order pole).   Because we are allowing the gauge field to have a pole, we refer to this kind of defect as a \emph{disorder operator}.  

Let us first analyze the problem that forces us to introduce these disorder operators. If we choose a local coordinate $z$ around a zero so that $\omega = z \d z$, then the quadratic term of the Lagrangian takes the form
\begin{equation} 
	\int z \d z A \d A = \int z \d z \d w \d \wbar \d \zbar  \left(- A_w \partial_{\zbar} A_{\wbar} + A_{w} \partial_{\wbar} A_{\zbar} - A_{\wbar} \partial_w A_{\zbar} \right)  \;.
\end{equation}
This kinetic term is degenerate: it does not define an elliptic equation modulo gauge.  Therefore we can not write down a propagator and the theory is ill-defined.
In fact this was the reason why we excluded higher-genus curves in the perturbative analysis of integrable lattice models in \cite{Costello:2017dso}.

To rectify this problem, we must allow some of the field components to have poles at $z = 0$.  We will choose to allow $A_w$ to have a first-order pole at $z = 0$.  If we do this, and write $A_w = z^{-1} \til{A}_w$ (so that $\tilde{A}_w$ is non-singular at the zero), then we find that the kinetic term becomes 
\begin{equation} 
	\int  \d z \d w \d \wbar \d \zbar \left(- \til{A}_w \partial_{\zbar} A_{\wbar} + \til{A}_{w} \partial_{\wbar} A_{\zbar}\right) 
	-\int z \d z \d w \d \wbar \d \zbar  A_{\wbar} \partial_w A_{\zbar} \;.
\end{equation}
This kinetic term is elliptic modulo gauge. To see this, note that the terms in the first line are elliptic by themselves, since they are given by the $\dbar$ operator on $\C^2$.  The second line has a triangular form which does not spoil ellipticity.   

Note that $A_w$ only enters once into the cubic term in the Lagrangian, so allowing $A_w$ to have a first order pole does not introduce any singularities into the cubic term. 

We can check more carefully that this modification of our field theory is elliptic if we introduce ghosts and anti-fields. (The reader uncomfortable with this language will not lose much by skipping this paragraph). In this language, the ellipticity of the kinetic term becomes the statement that the linearized BRST operator defines an elliptic complex. The full complex of fields consists of the quotient $\Omega^{\ast,\ast}(\C^2)\otimes \g[1]$ of the de Rham complex of $\C^2$, by the subspace of those forms divisible by $\d z$.  The symbol $[1]$ is a cohomological shift, placing one-forms in degree $0$.  The degree $0$ fields are the three components of the gauge field.  The linearized BRST operator is given by the expression
\begin{equation} 
	Q_{\rm BRST} = \dbar + z \partial  \;,
\end{equation}
which is the sum of the $\dbar$ operator and $z$ times the operator $\partial$. Since we remove all forms divisible by $\d z$, $\partial$ is the same as $\d w \partial_w$.   
The operator $\dbar$ defines an elliptic complex, and $z \partial$ is upper-triangular with respect to the grading given by the number of $\d w$'s.  From this it follows that $Q_{\rm BRST}$ defines an elliptic complex.

\section{Principal Chiral Model with Wess-Zumino Term} 
\label{sec:PCM}

Now that we now how to consider a one-form with a pole, we are ready to explain how to
deform the conformal WZW model away from the conformal point. 
This will create the principal chiral model together with the Wess-Zumino term, where the latter has a general coefficient. One can call this theory the non-conformal WZW model.

Since the theory is non-conformal, our construction involves the introduction of a scale. In our constructions, a scale is typically set by having a one-form whose integral between the two points corresponding to the defects is non-zero and finite.

The principal chiral model is constructed from a one-form has two poles of order two and two zeroes of order one. We will take the poles to be at $z = 0,\infty$ and the zeroes at $z_0,z_1$. We can take our one-form to be 
\begin{equation} 
	\omega = z^{-2}(z-z_0)(z-z_1)\d z  \;,
	\label{eq.omega_8}
\end{equation}
where we have chosen the normalization of the one-form such that the coefficient of $z^{-2}$ is $1$. 

\begin{figure}
\centering
		\begin{tikzpicture}
\begin{scope}
	\filldraw[ pattern=north west lines] (0,0) circle (3);  
	\draw[color=white,thick] (0,0) circle (3);
	
	\coordinate (A) at (-2.95,0.5);
	\coordinate (B) at (-2.95,-0.5);
	\coordinate (C) at (-1,0.5);
	\coordinate (D) at (-1,-0.5);
	\coordinate (E) at (-0.5,0);
	\filldraw[color=white](-3,0.5)--(C)--(E)--(D)--(-3,-0.5)--(-3,0.5);
	
	\draw[dashed, ->-] (A) --(C)   node [pos=0.3, sloped, below] {$\scriptscriptstyle a$}; 
	\draw[->-](C)--(E)  node [pos=0.2, sloped, below] {$\scriptscriptstyle b$}; 
	\draw[->-](E)--(D)   node [pos=0.7, sloped, above] {$\scriptscriptstyle c$}; 
	\draw[dashed, ->-](B)--(D) node [pos=0.3, sloped, above] {$\scriptscriptstyle a$}; 
\end{scope}
	
\begin{scope}[shift={(6,0)}]

	\filldraw[ pattern=north west lines] (0,0) circle (3);  
	\draw[color=white,thick] (0,0) circle (3);
	
	\coordinate (A) at (2.95,0.5);
	\coordinate (B) at (2.95,-0.5); 
	\coordinate (C) at (-1,0.5);
	\coordinate (D) at (-1,-0.5);
	\coordinate (E) at (-0.5,0);
		\filldraw[color=white](3,0.5)--(C)--(E)--(D)--(3,-0.5)--(3,0.5);
	
		\draw[dashed, ->-] (A) --(C)   node [pos=0.3, sloped, below] {$\scriptscriptstyle g$};  
	\draw[->-](E)--(C)   node [pos=0.5, sloped, above] {$\scriptscriptstyle b$}; 
	\draw[->-](D)--(E)  node [pos=0.5, sloped, below] {$\scriptscriptstyle c$}; 
	\draw[dashed, ->-](B)--(D) node [pos=0.3, sloped, above ] {$\scriptscriptstyle g$}; 

	\end{scope}
\end{tikzpicture}
	\caption{The flat surface corresponding to a one-form on $\CP^1$ with two second order poles. The residue of the pole is the sum of the vectors $b+c$, and the integral of the one-form between the two zeroes is $b$ or $c$, depending on the cycle chosen.\label{fig:twopoles}} 
\end{figure}
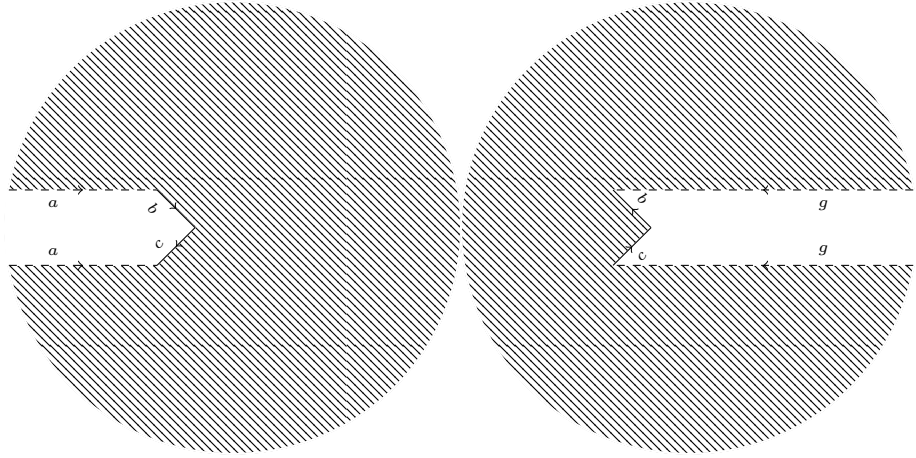

The flat surface corresponding to this one-form has two planes glued along two intervals; the ends of these intervals are conical singularities, corresponding to the zeroes of the one-form. See \fig\ref{fig:twopoles}. 

\subsection{Boundary Conditions}

\begin{figure}[htbp]
\centering\includegraphics[scale=0.50]{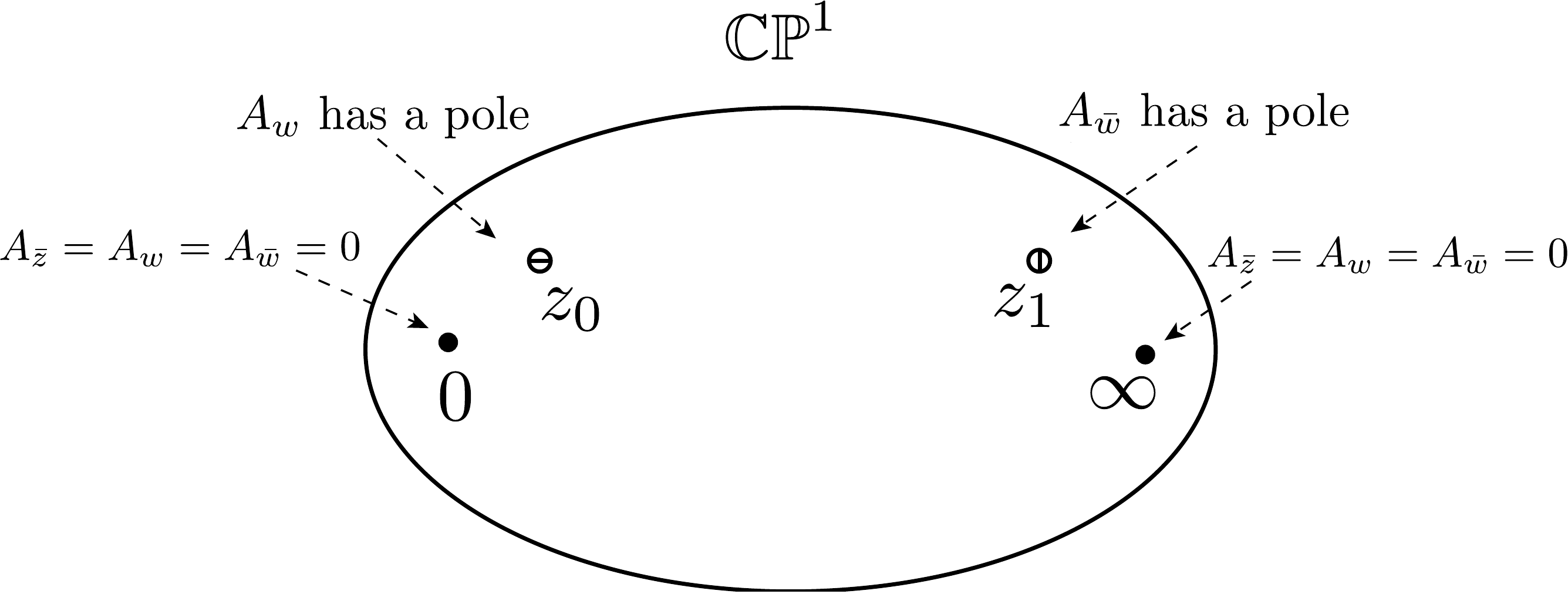}
\caption{The boundary condition used in this section. At the two zeros $z=z_0$ ($z_1$) of the one-form $\omega$, we require that $A_{w}$ ($A_{\wbar}$) has a pole. At the two poles $z=0$ ($\infty$) of the one-form
we impose the topological boundary condition, such that all fields and gauge transformations are 
divisible by $z$ ($1/z$).}
\label{Fig.III_10}
\end{figure}

To realize the principal chiral model, we will impose the following constraints on our fields at the zeroes and poles of $\omega$ (see Figure \ref{Fig.III_10}).
\begin{enumerate}
	\item We will use a topological boundary condition at $z = 0$, given, as in \cite{Costello:2017dso}, by asking all fields and gauge transformations are divisible by $z$ near $z = 0$. Similarly, near $z = \infty$, all fields and gauge transformations are divisible by $1/z$.  
	\item At the zero  $z=z_0$ of the one form near $z = 0$, we allow $A_w$ to have a first order pole. 	At the zero $z=z_1$ of the one form near $ z= \infty$, we allow $A_{\wbar}$ to have a first order pole. 
\end{enumerate}
In the limit as $z_0 \to 0$ or $z_1 \to \infty$, so that each zero collides with a pole, we end up with boundary condition where at $z = 0$, $A_{\wbar} = A_{\zbar} =  0$, and at $z = \infty$, $A_w = A_{\zbar} = 0$.  (Gauge transformations also vanish at $z = 0,\infty$). 
These are the boundary conditions which gave us the conformal WZW model in the previous section.  We find that, as expected, the parameter $a$ controls deformations away from the conformal model.

\subsection{Recovering the Lagrangian for the Principal Chiral Model}

To recover the Lagrangian of the principal chiral model, we take our four-dimensional gauge theory with the boundary conditions listed above.  We will show that the value of $A_{\zbar}$, up to gauge equivalence, defines a map $\sigma : \R^2 \to G$. The equations $F_{\zbar \wbar} = 0$ and $F_{\zbar w} = 0$ fix $A_w, A_{\wbar}$ uniquely in terms of $\sigma$.  We will find that the Lagrangian of four-dimensional Chern-Simons is the same as that of the principal chiral model with the Wess-Zumino term added.

\subsubsection{Solving for Gauge Fields}

First, we note that the $A_{\zbar}$ component of the four-dimensional gauge field gives a family of holomorphic bundles on $\CP^1$, parameterized by $\R^2$.  These bundles are trivialized at $0$ and $\infty$.  The moduli of such bundles is $G$, so the $A_{\zbar}$ component, modulo gauge, defines a map $\sigma : \R^2 \to G$.

To explicitly write $A_{\zbar}$ in terms of $\sigma$, we proceed as we did for the conformal WZW model.  We chose some map $\what{\sigma} : \R^2 \times \CP^1 \to G$ so that $\what{\sigma} = \sigma$ in a neighborhood of $z = \infty$ and $\what{\sigma} = \op{Id}$ in a neighborhood of $z = 0$.
Then, we can set
\begin{equation} 
	A_{\zbar}^{\sigma} = \what{\sigma}^{-1} \partial_{\zbar} \what{\sigma} \;.
\end{equation}
This vanishes at $0$ and $\infty$, as required.

We let $z_0, z_1$ be the two zeroes of our one-form, with $z_0 z_1 = 1$. We assume $z_0$ is near $0$, $z_1$ near $\infty$. We need to find $A_w^{\sigma}$, with a pole near $z_0$ and a zero at $0,\infty$, and $A_{\wbar}^{\sigma}$, with a pole at $z_1$ and a zero at $0,\infty$, so that the equations of motion $F_{\zbar w} = 0$, $F_{z w} = 0$ are satisfied.

We should note that there is a unique $A_{w}^{\sigma}$ such that $F_{\zbar w} = 0$, and which has the required poles and zeroes (and the same holds for $A_{\wbar}$).  The point is that if $A_{w}^{\sigma} + \gamma$ solves $F_{\zbar w}=0$ , then $\gamma$ satisfies $(\partial_{\zbar} + A_{\zbar}^{\sigma}) \gamma = 0$. Thus, $\gamma$ is a holomorphic section of the adjoint bundle associated to the holomorphic bundle given by $A_{\zbar}$. Since this bundle is trivial, we view $\gamma$ as a holomorphic section of the trivial bundle, with two zeroes and one pole.  All such sections are zero. 

Because the curvature is gauge covariant, and $A_{\zbar}^{\sigma}$ can be made trivial with the gauge transformation given by $\what{\sigma}$,
we can solve the flat connection equation $F_{\zbar \wbar}=0$ by the pure gauge expression
\begin{equation} 
	A_{\wbar} \overset{?}{=} \what{\sigma}^{-1} \partial_{\wbar} \what{\sigma} \;. 
\end{equation}
This solution, however, does not have the correct structure of poles and zeroes.   Although it vanishes near $0$, it does takes value $\sigma^{-1} \partial_{\wbar} \sigma$ near $z = \infty$. This is one of the differences from the previous section.

We can add to this solution any quantity which when conjugated by $\what{\sigma}$ is holomorphic.  The solution with the correct poles and zeroes is
\begin{equation} 
	A_{\wbar}^{\sigma} = \what{\sigma}^{-1} \partial_{\wbar} \what{\sigma} - \frac{z}{z - z_1} \what{\sigma}^{-1}  (\partial_{\wbar} \sigma) \sigma^{-1} \what{\sigma} \;.   
	\label{eq.A_wbar_eg}
\end{equation}
This vanishes at $z = 0,\infty$ and has a pole at $z = z_1$, and is a solution to the equation $F_{\zbar \wbar} = 0$.

Similarly, we can solve for the $w$-component to be 
\begin{equation} 
	A_{w}^{\sigma} =  \what{\sigma}^{-1} \partial_{w}\what{\sigma} - \frac{z}{z - z_0} \what{\sigma}^{-1} (\partial_{w} \sigma) \sigma^{-1} \what{\sigma} \;.
	\label{eq.A_w_eg}   
\end{equation}

Putting this together, we have
\begin{align} 
A^{\sigma}&= \what{A}+A' \;, \label{eq.Asigma_8} \\
\what{A}_i &= \what{\sigma}^{-1} \partial_i \what{\sigma}\;, \\
A'&= - \frac{z}{z - z_0} \what{\sigma}^{-1} (\partial_{w} \sigma) \sigma^{-1} \what{\sigma}\d w - \frac{z}{z - z_1} \what{\sigma}^{-1}  (\partial_{\wbar} \sigma) \sigma^{-1} \what{\sigma} \d \wbar \;.
 \end{align}

\subsubsection{Lax Operator}

The expressions for the gauge field derived above immediately determines the Lax operator. 
Let us choose the gauge $A_{\zbar}=0$, as explained around \eqn \eqref{eq.Lax_holomorphic}.
This means to apply the inverse gauge transformation by $\what{\sigma}$,
so that the gauge field \eqref{eq.Asigma_8} becomes
\begin{equation} 
A= - \frac{z}{z - z_0} \sigma^{-1}(\partial_{w} \sigma) \d w - \frac{z}{z - z_1}  \sigma^{-1} (\partial_{\wbar} \sigma) \d \wbar \;, 
 \end{equation}
 and hence
\begin{equation} 
\begin{split}
	\Lax_{w}& = - \frac{z_0}{z - z_0} J_w  \;, \\
	\Lax_{\wbar} & = -\frac{z_1}{z - z_1} J_{\wbar} \;.
\end{split}	
\end{equation}
Setting $z_0 = 1$, $z_1 = -1$ gives us 
\begin{equation} 
	\Lax = \frac{(J_w+J_{\wbar}) + z (J_w-J_{\wbar}) } {1- z^2  }  \;, 
\end{equation}
which is equation (3.11) of \cite{Zarembo:2017muf}.  

\subsubsection{Lagrangian}

Now let us turn to computing the action functional.

We can again use the same formula \eqref{CS_sum_3}, where 
we can now set $F(\what{A}) = \textrm{CS}(A')=0$ since $\what{A}$ is flat and $A'$ has no $\zbar$ component.
This gives
\begin{equation} 
\textrm{CS}(\what{A} + A') = \, \textrm{CS}(\what{A})  - \d \op{Tr} (\what{A} A') + 2\op{Tr} (\what{A} A' A' ) \;. 
\label{eq.CS_sum}
 \end{equation}
and hence
\begin{equation}
\begin{split} \label{eq.Asigma_tmp}
\textrm{CS}(A^{\sigma}) &= \textrm{CS}(\what{A})  \\
&+\d  \frac{z}{z - z_0}  \op{Tr} \left( \what{A}   \what{\sigma}^{-1} (\partial_{w} \sigma) \sigma^{-1} \what{\sigma}\d w \right)
+ \d  \frac{z}{z - z_1}  \op{Tr} \left( \what{A}   \what{\sigma}^{-1} (\partial_{\wbar} \sigma) \sigma^{-1} \what{\sigma}\d \wbar \right)  \\ 
&+ \frac{2 z^2}{(z-z_0)(z-z_1)}\d \zbar \d w \d \wbar  \op{Tr} \left( (\partial_{\zbar} \what{\sigma})  \what{\sigma}^{-1} [ (\partial_{w}\what{\sigma}) \sigma^{-1}, (\partial_{\wbar}\what{\sigma}) \sigma^{-1}  \right).
 \end{split}
 \end{equation}

We are interested in the integral of $\textrm{CS}(A^{\sigma})$ against the one-form $\omega$ on $\CP^1$.  

The first term in \eqn \eqref{eq.Asigma_tmp} is
\begin{equation} 
\int \frac{ (z-z_0)(z - z_1)}{z^2} \d z \, \textrm{CS}(\what{A}) \;. 
 \end{equation}
As in the previous section, since $\what{A}$ is a gauge trivial connection trivialized by $\what{\sigma} : \R^2 \times \CP^1 \to G$, we have \eqn \eqref{eq.CS_trivial}.
This integral can be computed by first performing an integral over the argument of $z$ ($S^1$ of the cylinder), yielding 
\begin{equation} 
-2\pi \i \frac{z_0 + z_1}{3} \int_{\R^2 \times \R_{\ge 0}} \op{Tr} \left(\what{\sigma}^{-1} \d \what{\sigma}\right)^3. 
 \end{equation}
This is a Wess-Zumino term, with a coefficient proportional to $z_0 + z_1$. 

The second and third terms in $\textrm{CS}(A^{\sigma})$ are total derivatives. Integrating by parts, and using the fact that $\omega = z^{-2}(z-z_0)(z-z_1) \d z$, we find that the second and third terms can be given by the contour integral  around $z=0$ and $z=\infty$:
\begin{equation} 
\oint    \frac{z-z_1}{z} \d z \op{Tr} \left( \what{A}   \what{\sigma}^{-1} (\partial_{w} \sigma) \sigma^{-1} \what{\sigma}\d w \right) 
+ \oint   \frac{z-z_0}{z} \d z  \op{Tr} \left( \what{A}   \what{\sigma}^{-1} (\partial_{\wbar} \sigma) \sigma^{-1} \what{\sigma}\d w \right) \;.
 \end{equation}
Since $\what{\sigma} = \sigma$ near $\infty$, and is the identity (hence $\what{A}=0$) near $0$, only the pole at $z = \infty$ contributes to the residue evaluation of the contour integral and we find 
\begin{equation} 
 2\pi \i (z_1 -z_0)\d w \d \wbar \op{Tr} ( \sigma^{-1} (\partial_{\wbar} \sigma) \sigma^{-1} (\partial_w \sigma) )   
 \;.
 \end{equation}
This is the standard kinetic term for the $\sigma$-model, with a prefactor of $\pi \i (z_1 - z_0)$. We find that the distance between $z_1$, $z_0$ controls the size of the metric.

 The last term in \eqn \eqref{eq.Asigma_tmp} gives
\begin{equation} 
 \int \d z \d \zbar \d w \d \wbar  \op{Tr} \left( (\partial_{\zbar} \what{\sigma})  \what{\sigma}^{-1} [ (\partial_{w}\what{\sigma}) \sigma^{-1}, (\partial_{\wbar}\what{\sigma}) \sigma^{-1}  \right).
 \end{equation}
The integrand of this expression is of charge $1$ under the $U(1)$ action rotating $z$, and hence integrating over the argument $\theta$ of $z$ gives us zero.  

In sum, we have found the principal chiral model with WZW term proportional to $z_0 + z_1$ and kinetic term proportional to $z_0 - z_1$:
\begin{equation} 
S_{\rm PCM+WZ}=
\frac{ k (z_1-z_0)}{8 \pi} \int_{\R^2}  \op{Tr} (j \wedge \star j ) 
 -
\frac{k(z_1 + z_0)}{12\pi} \int_{\R^2 \times \R_{\ge 0}} \op{Tr} \left(\what{j} \wedge \what{j} \wedge \what{j} \right) \;, 
\end{equation}
where as before we defined $j= \sigma^{-1} d \sigma$ and $\what{j}=\what{\sigma}^{-1} \d \what{\sigma}$
and we have introduced the overall constant, the level $k$, following \eqn \eqref{eq.level}.

This recovers the conformal WZW model  \eqref{eq.WZW_conformal} in the previous section in the limit
$z_0 \to 0$ or $z_1 \to \infty$, as expected. 

Note that in the case $z_1=z_0$ we have only the topological WZ term. This is consistent with the 
fact that there is no finite length scale when $z_1=z_0$.

\section{Riemannian Symmetric Spaces}
\label{sec:symmetric}

The $\sigma$-models on Riemannian symmetric spaces are known to be integrable \cite{Eichenherr:1979ci}. In this section we will explain how to construct these models in our setting.  The construction is along the same line as our construction of the principal chiral model and its deformations, but is slightly more subtle. A pay-off of our analysis is a construction of a very  large number of variants of these models, most of which are new.

For these models, we work on $\CP^1$ with the one-form $\d z$.  At $z = \infty$, the one-form has a double pole and we impose the boundary condition that our gauge field is trivial (and the corresponding principal $G$-bundle has a section).  We introduce defects at $z=z_0,z_1$, and also along a line segment $[z_0,z_1]$ connecting them.

To describe these defects, we recall that a Riemannian symmetric space is associated to a $\Z_2$ grading $\mf{g} = \mf{g}_0 \oplus \mf{g}_1$ of the Lie algebra $\mf{g}$.  If we let $G_0 = \op{exp} \mf{g}_0$, then the (complexified) symmetric space is $G / G_0$.  We $\rho : G \to G$ be the $\Z_2$-involution on the group $G$ which at the level of Lie algebras has $+1, -1$ eigenspaces $\mf{g}_0$, $\mf{g}_1$.  

The involution $\rho$ acts on all fields of four-dimensional Chern-Simons theory.  On the line segment $[z_0,z_1]$, we introduce a defect by asking that when we cross this line, we apply the automorphism $\rho$. This defect is codimension $1$, and is a topological domain wall.

We need to explain how to make this domain wall end at $z_0$ and $z_1$.  We will do this by passing to a double cover of the $z$-plane, and describing the boundary conditions there. We will can assume without loss of generality that $z_0 = -1-\lambda^2$ and $z_1 = 1+\lambda^2$ for some parameter $\lambda$.  We introduce a coordinate $u$ implementing the Joukowsky transform:
\begin{equation} 
\lambda^{-1} u + \lambda u^{-1}  = z \;.  
 \end{equation}
Then the $u$-plane is a double cover of the $z$-plane, branched over $z_0, z_1$.  The deck transformation on the $u$-plane sends $u \mapsto \lambda^2 u^{-1}$, and the fixed points are $u = \pm \lambda$.

In this coordinate on the double cover, the one-form $\d z$ pulls back to
\begin{equation} 
\omega = ( \lambda^{-1} - \lambda u^{-2} ) \d u = \lambda^{-1} u^{-2} (u - \lambda )(u + \lambda ) \d u \;. 
 \end{equation}
This has second-order poles at $u=0$ and $u=\infty$ (preimages of $z=\infty$), and first order zeros at $u=\pm \lambda$.  This brings us to the situation we studied in the principal chiral model (recall for example the one-form in \eqn \eqref{eq.omega_8}).

The fields of the gauge theory on the $u$-plane have an involution sending $u \mapsto \lambda^2 u^{-1}$, and also acting by the involution $\rho$ on the Lie algebra $\g$.  
Note that the two second-order poles of the one-form, $u=0$ and $u=\infty$, are exchanged under the transformation $u \mapsto \lambda^2 u^{-1}$.
This means that when we parametrize $G^2/G$ we need to exchange two copies of $G$ in $G^2$,
and instead of parametrizing $A_{\zbar}$ in terms of $\what{\sigma}$ trivialized around $u=0$
we need a different $\what{\sigma}$ trivialized around $u=\infty$ (we encountered a similar argument around
\eqn \eqref{eq.sigmatilde}). This amounts to a replacement $g$ by $g^{-1}$ for an element $g\in G\simeq G^2/G$.
The path integral is performed over those fields which are invariant under the combined involution.  

At the points $u = \pm \lambda$, i.e. $z = z_0,z_1$, we need to allow certain fields to have pole so that the kinetic part of the Lagrangian is elliptic.  As in our analysis of the principal chiral model, we let $A_{w}$ have a pole at $u =-\lambda$ and $A_{\br{w}}$ have a pole at $u = \lambda$. 

Decomposing our gauge field $A$ into $A^0 + A^1$ according to the decomposition $\mf{g} = \mf{g}_0 + \mf{g}_1$, we find that only the $A^1$ component can have a pole at $u = \pm \lambda$. The reason is that the residue of the pole must be odd under the involution $u \mapsto \lambda^2 u^{-1}$.  

From our analysis of the principal chiral model, we know that before we ask that our fields be $\Z_2$ invariant this model produces the principal chiral model with target $G$, with Wess-Zumino term set to zero 
and kinetic term proportional to $2 \lambda$.   

Once we impose the condition that our fields are $\Z_2$ invariant, we find a $\sigma$-model not with target $G$, but with target the submanifold 
\begin{equation} 
X = \left\{g \in G \,\, | \, \, \rho(g)  = g^{-1} \right\} \subset G \;,
 \end{equation}
where as explained above we used the combination of the two actions $g\to \rho(g)$ and $g\to g^{-1}$
as our $\Z_2$-transformation.
The metric on this submanifold is the restriction of the metric on $G$. 

We will now show that this manifold $X$ is isomorphic to $G / G_0$,\footnote{Assuming it is connected: if not we take the connected component of the identity in $X$.} where $G_0$ is the fixed point set of $\rho$ in $G$.  There is a $G$-action on $X$ which sends $x\in X$ to 
 \begin{equation} 
 x \mapsto \rho(g) x g^{-1} \;. 
  \end{equation}
  This preserves the defining equation $x \rho(x) = \op{Id}$ of $X$.  This action is transitive: assuming $X$ is connected we write $x = \exp(y)$ for $y \in \mf{g}_1$, so that $y$ is odd under the action of $\rho$.  Then $x$ is obtained by the identity by applying the element $\exp(-y/2) \in G$. The stabilizer of the identity is $G_0$, so that $X = G / G_0$. 

We have shown that this construction engineers the $\sigma$-model on the Riemannian symmetric space $X$. 

\section{Generalizations of the Principal Chiral Model}\label{sec:general_PCM}

\subsection{Boundary Conditions}

\begin{figure}[htbp]
\centering\includegraphics[scale=0.55]{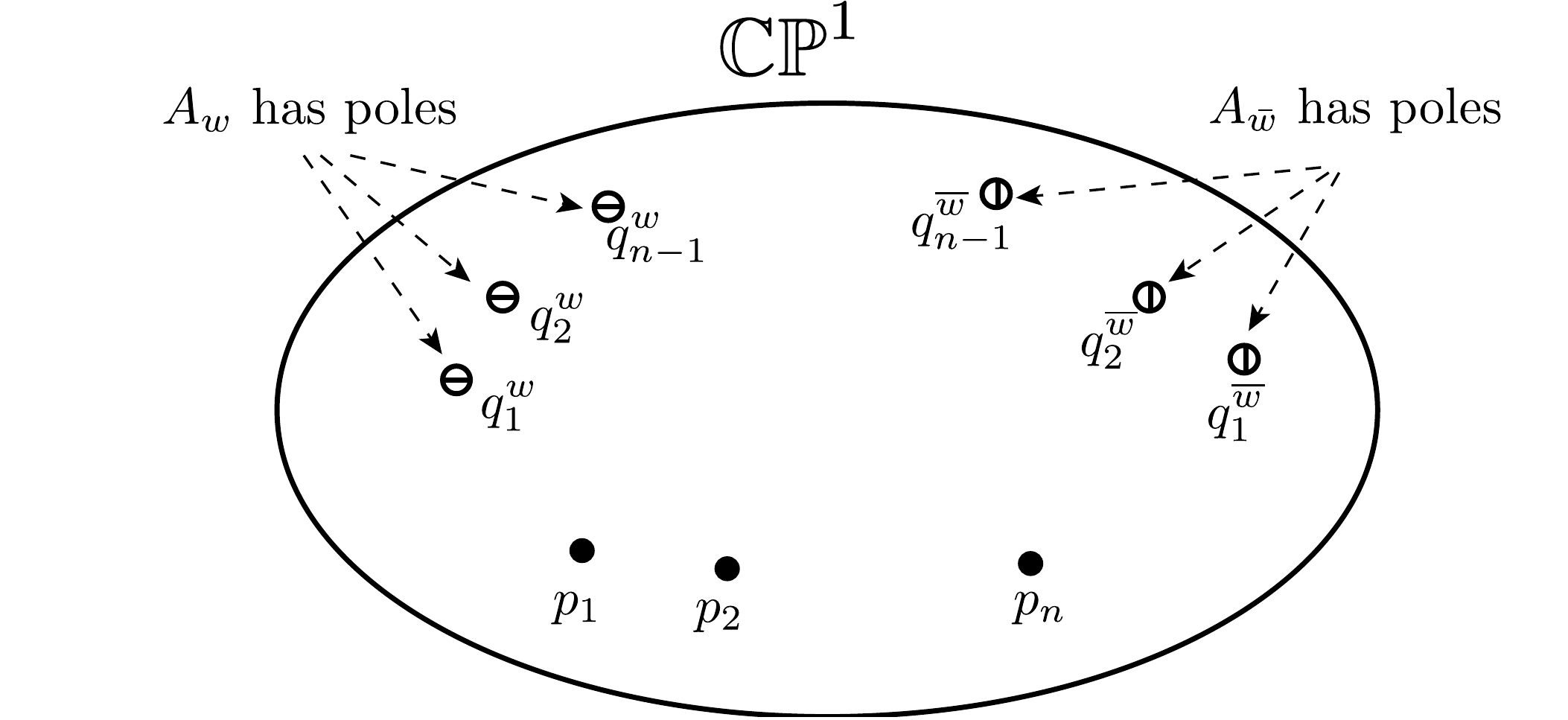}
	\caption{The boundary condition used in this section. At the $(n-1)+(n-1)$ zeros  $z=q_1^w, \dots, q_{n-1}^w$ ($q_1^{\wbar}, \dots, q_{n-1}^{\wbar}$ ) of the one-form $\omega$, we require that $A_{w}$ ($A_{\wbar}$) has a pole. At the  second-order poles $z=p_1, \dots, p_n$ of the one-form
we impose the topological boundary condition, such that all fields and gauge transformations are 
divisible by $z-p_i$).}
\label{Fig_III_9}
\end{figure}

Our construction of the principal chiral model can be generalized in a straight-forward way to cases with more general choices of the one-form. We introduce a one-form on $\CP^1$ which has $n$ second-order poles $p_{i=1,\dots, n}$, and $2n-2$ corresponding first-order zeroes.  We divide the zeroes into two groups of $n-1$, which we call $q_{i=1, \dots, n-1}^w$ and $q_{i=1, \dots, n-1}^{\wbar}$ (see Figure \ref{Fig_III_9}). Explicitly,
\begin{equation} 	\label{eq.omega_multiWZW}
	\omega = \frac{\prod_{i=1}^{n-1} (z - q_i^w) \prod_{j=1}^{n-1}(z - q_j^{\wbar} ) } {\prod_{k=1}^{n}(z - p_k)^2} \d z \;.  
\end{equation}
This is the unique such form, up to scale, with the prescribed poles and zeroes.  There are a total of $3n-4$ parameters in the choice of $\omega$: the choice of $3n-2$ points on $\CP^1$ up to the $SL_2(\C)$ action, together with the overall scale of $\omega$.  

At the poles $p_i$, all fields have a zero. At the zeroes $q_i^w$, we allow $A_w$ to have a pole, and at $q_i^{\wbar}$, we allow $A_{\wbar}$ to have a pole.  We allow the zeroes $q_i^{\wbar}$ to collide with each other, introducing higher-order zeroes at which $A_{\wbar}$ has a pole of the corresponding order. Similarly, the zeroes $q_i^w$ can collide with each other. We assume that $q_i^w \neq q_i^{\wbar}$.  

The $\zbar$ component of the gauge field, up to gauge equivalence, defines a map $\sigma : \R^2 \to G^{n} / G$, where the $n$ copies of $G$ correspond to the trivializations at $p_i$ and we quotient by the right diagonal action.  We use the constant gauge transformation to set the trivialization at $p_n$ to correspond to the identity in $G$, so that $\sigma = (\sigma_1,\dots,\sigma_{n-1})$ where $\sigma_i : \R^2 \to G$.  Then, we can write
\begin{equation} 
A_{\zbar}^{\sigma} = \what{\sigma}^{-1} \dbar \what{\sigma}  \;,
 \end{equation}
where $\what{\sigma} = \sigma_i$ near $p_i$ and $\what{\sigma} = \op{Id}$ near $p_n$.  We will end up with some $\sigma$-model with target $G^n / G$, which retains the $G^n$ symmetry coming from the left action. 

Then, as before, we set $\what{A}= \what{\sigma}^{-1} \d \what{\sigma}$ and 
\begin{equation} 
\begin{split}
A^{\sigma}_w &= \what{A}_w + \what{\sigma}^{-1} F(z) \what{\sigma} \;,\\ 
A^{\sigma}_{\wbar} &= \what{A}_{\wbar} + \what{\sigma}^{-1} G(z) \what{\sigma}\;,
\end{split}
 \end{equation}
where $F(z), G(z)$ are meromorphic functions on $\CP^1$ valued in $\mf{g}$ with certain prescribed poles and zeroes. We require that $F(z) = - \partial_w (\sigma_i) \sigma_i^{-1}$ near $p_i$, and we allow $F(z)$ to have poles at $q_i^w$. Similarly, $G(z) = - \partial_{\wbar} (\sigma_i) \sigma_i^{-1}$ and $G(z)$ is allowed to have poles at $q_i^{\wbar}$.  

The solutions to these constraints are given by 
\begin{equation} 
\begin{split}
	F(z) &= -\frac{(z - p_1) \dots (z - p_n)}{(z-q_1^w) \dots (z - q_{n-1}^w)}\left(\sum_{j  =1}^n \frac{1}{z - p_j} \frac{(p_j - q_1^w) \dots (p_j - q_{n-1}^w)   }{\prod_{k \neq j}(p_j - p_k) }   (\partial_w \sigma_j) \sigma_j^{-1}   \right) \;,\\
	G(z) &= -\frac{(z - p_1) \dots (z - p_n)}{(z-q_1^{\wbar}) \dots (z - q_{n-1}^{\wbar})}\left(\sum_{j  =1}^n \frac{1}{z - p_j} \frac{(p_j - q_1^{\wbar}) \dots (p_j - q_{n-1}^{\wbar})   }{\prod_{k \neq j}(p_j - p_k) }   (\partial_{\wbar} \sigma_j) \sigma_j^{-1}   \right)\;. 
\end{split}
\end{equation}
In these expressions, we assume that all $p_i$, $q_i^{w}$, $q_i^{\wbar}$ are not $\infty$.  Then we note that $F(z)$, $G(z)$ take a finite value at $\infty$, and are uniquely determined by the fact that they have at most first order poles at $q_i$, $q_i^{\wbar}$, have the required value at $p_i$ and are regular at $\infty$.

\subsection{The Lagrangian}

The Lagrangian can be computed using the same technique we used for the principal chiral model. 

We write the gauge field $A^{\sigma}$ as $\what{A} + A'$, where $\what{A} = \what{\sigma}^{-1} \d \sigma$. Then, we have \eqref{eq.CS_sum}:
\begin{equation} 
	\int \omega \, \textrm{CS}(A^{\sigma}) = \int \omega \, \textrm{CS}(\what{A}) - \int \omega\, \d \!\op{Tr} (\what{A} A') +2  \int \omega \op{Tr} (\what{A}A' A') \;. 
	\label{eq.CS_sum_4}
\end{equation}
The first term will yield a Wess-Zumino term.  The second will give the kinetic term, and the third term vanishes as in previous sections. 

\subsubsection{The first two terms in the Lagrangian} 

Let us first analyze the term
\begin{equation} 
	- \int \omega\, \d\op{Tr} (\what{A} A')  =- \int \d (\omega \op{Tr} (\what{A} A') ) \;.
\end{equation}
This is an integral of the exact form $ \d (\omega \op{Tr}(\what{A} A'))$ over the complement of the $p_i$, $q_i^w$, $q_i^{\wbar}$ in $\CP^1$.  By Stokes' theorem the integral reduces to a sum of contour integrals around these points. Since $\what{A} = 0 $ near $q_i^w$, $q_i^{\wbar}$, only the integrals around $p_i$ contribute, so that the second term can be evaluated by Stokes' theorem to yield a sum of contour integrals around the points $p_i$:  
\begin{equation} 
	\sum_i \oint_{\abs{z - p_i} = \eps} \omega \op{Tr} (\what{A} A') \;. 
\end{equation}
Since $\what{A} = \sigma_i^{-1} \d \sigma_i$ near $p_i$, and 
\begin{equation} 
\begin{split}
	A'_w &= \sigma_i^{-1} F(z) \sigma_i \;, \\
	A'_{\wbar}&= \sigma_i^{-1} G(z) \sigma_i \;, 
\end{split}
\end{equation}
near $p_i$, we find 
\begin{equation} 
2\pi \i \sum_i \op{Res}_{p_i}\left( \omega \op{Tr} (\partial_{\wbar} \sigma_i) \sigma_i^{-1}  F(z) \right)   - 	
2\pi \i \sum_i \op{Res}_{p_i} \left( \omega \op{Tr} ( (\partial_{w} \sigma_i) \sigma_i^{-1} G(z) ) \right). 
\end{equation}
Using the expression for $F(z)$, $G(z)$ and $\omega$, and writing $J_{i,w} = (\partial_w \sigma_i)\sigma_i^{-1}$, $J_{i,\wbar} = (\partial_{\wbar} \sigma_i) \sigma_i^{-1}$,  we can evaluate this to be
\begin{multline} 
	-2\pi \i \sum_{i,j}  \op{Tr} \left( J_{i,\wbar} J_{j,w}  \right)    \frac{\prod_k (p_j - q_k^w)    }{\prod_{k \neq j}(p_j - p_k) }  \op{Res}_{p_i}\left( \frac{\prod_k (z - q_k^{\wbar})}{\prod_k(z-p_k) }  \frac{1}{z - p_j}\right) \\ 
	+2\pi \i \sum_{i,j}  \op{Tr} \left( J_{i,w} J_{j,\wbar}  \right)    \frac{\prod_k (p_j - q_k^{\wbar}) }{\prod_{k \neq j}(p_j - p_k) }  \op{Res}_{p_i}\left( \frac{\prod_k (z - q_k^{w})}{\prod_k(z-p_k) }  \frac{1}{z - p_j}\right)  \;.
\end{multline}
We can calculate the residues appearing above explicitly. It is convenient to express the result in terms of the functions
\begin{align} 
	\varphi_{i,w} (z) &= \frac{\prod_k (z- q_k^w)}{\prod_{j \neq i}(z - p_j)} \;,\\
	\varphi_{i,\wbar} (z) &= \frac{\prod_k (z- q_k^{\wbar})}{\prod_{j \neq i}(z - p_j)}  \;,
\end{align}
%
%
we find the Lagrangian 
\begin{multline}
\frac{k}{4\pi}
\int_{\mathbb{R}^2}	 
	\left[ \sum_i J_{i,\wbar} J_{i,w} \left( - \varphi_{i,w}(p_i) \varphi_{i,\wbar}'(p_i)  + \varphi'_{i,w}(p_i) \varphi_{i,\wbar}(p_i)  \right) 	\right.	
\\
	\left.	- 2 \sum_{j \neq i}\frac{ \varphi_{i,\wbar}(p_i) \varphi_{j,w}(p_j)    }{p_i - p_j} J_{i,\wbar} J_{j,w}  \right] \;.
\end{multline}
This expression is a mixture of terms which are symmetric under the interchange of $J_{i,w}$ with $J_{i,\wbar}$ and terms which are anti-symmetric. The symmetric terms contribute to the metric on $G^{n}/G$. The anti-symmetric terms are the integral of the pull-back of a two-form on $G^{n}/G$, and so can be thought of as part of the Wess-Zumino term.

The symmetric part gives the kinetic term 
\begin{multline} \label{eq.WZWmulti_Lkin}
	S_{\rm{kin}} =  
	\frac{k}{4\pi}
\int_{\mathbb{R}^2}	 
	\left[ \sum_i J_{i,\wbar} J_{i,w} \left( - \varphi_{i,w}(p_i) \varphi_{i,\wbar}'(p_i)  + \varphi'_{i,w}(p_i) \varphi_{i,\wbar}(p_i)  \right) 	\right.	
\\
	\left.	-  \sum_{j \neq i}\frac{ \varphi_{i,\wbar}(p_i) \varphi_{j,w}(p_j)  -\varphi_{j,\wbar}(p_j) \varphi_{i,w}(p_i)     }{p_i - p_j} J_{i,\wbar} J_{j,w}  \right] \;.	
\end{multline}
The anti-symmetric part gives
\begin{equation}
	S_{2-\rm{form}} 	=
		-  \sum_{j \neq i}\frac{ \varphi_{i,\wbar}(p_i) \varphi_{j,w}(p_j)  +\varphi_{j,\wbar}(p_j) \varphi_{i,w}(p_i)     }{p_i - p_j} J_{i,\wbar} J_{j,w}  \;.
		\label{eq.WZWmulti_L2form}
\end{equation}
The anti-symmetric term is the integral of a two-form over the world-sheet.  If we let $X_{i,a}$ denote a basis of left-invariant one-forms on the $i$-th copy of $G$ in $G^n$, where $a$ denotes an orthonormal basis of $\mf{g}$, then the two-form is
\begin{equation}
B_2=	-\frac{1}{2} 
\sum_{j \neq i}\frac{ \varphi_{i,\wbar}(p_i) \varphi_{j,w}(p_j)  +\varphi_{j,\wbar}(p_j) \varphi_{i,w}(p_i)     }{p_i - p_j} X_{i,a} X_{j,a}. 
\end{equation}

\subsubsection{The WZ term}

Next let us compute the Wess-Zumino term, coming 
 from the first term in  \eqn \eqref{eq.CS_sum_4}. We use again \eqref{eq.CS_trivial} since $\til{A}$ is pure gauge. When integrating against $\omega$, this becomes
\begin{equation} 
	-\frac{2 \pi \i}{3}\sum_i \op{Res}_{p_i}(\omega)\int_{\R^2 \times \R_{\ge 0} } \op{Tr} ( (\sigma_i^{-1} \d \sigma_i)^3) \;.  
\end{equation}
That is, we find Wess-Zumino terms for the fields $\sigma_i$ with a factor of $-\frac{2\pi \i}{3}\sum_i \op{Res}_{p_i}(\omega)$, where the residue can be computed to be
\begin{equation} 
	\op{Res}_{p_i}(\omega) = \frac{\prod_k (p_i - q_k^{w}) (p_i - q_{k}^{\wbar})  }{\prod_{k \neq i} (p_i - p_k)^2  }\left( \sum_k \frac{1}{p_i - q_k^w} + \frac{1}{p_i - q_k^{\wbar}} - 2\sum_{k \neq i} \frac{1}{p_i - p_k} \right)   \;.
\end{equation}
We define $J_{r,i}$ to be $(\partial_r \what{\sigma}) \what{\sigma}^{-1}$ (where $r$ is the extra direction
along $\mathbb{R}_{\ge 0}$, namely the cylindrical direction of the $\mathbb{CP}^1$).   Then, the Wess-Zumino term is
\begin{equation} 
\label{eq.WZWmulti_LWZ}
S_{\rm WZ}=
	- \frac{k}{2 \pi} \sum_i \int_{\R^2 \times \R_{\ge 0} }  \op{Tr} (J_{\wbar,i} J_{w,i} J_{r,i} ) \left(\varphi_{i,w}(p_i) \varphi'_{i,\wbar}(p_i) +\varphi'_{i,w}(p_i) \varphi_{i,\wbar}(p_i) \right)   \;.
\end{equation}
This is associated to the three-form
\begin{equation}
	H_3=- \frac{k}{12 \pi} \sum_i \int_{\R^2 \times \R_{\ge 0} } f^{abc} X_{i,a} X_{i,b} X_{i,c} \left(\varphi_{i,w}\varphi'_{i,\wbar}+\varphi'_{i,w}\varphi_{i,\wbar}\right)   \;.
\end{equation}
on $G^{n-1} / G$, with $X_{i,a}$ as before. 

\subsubsection{The Total Lagrangian}

Summarizing, the total action is 
\begin{equation}
	S_{\rm total}=S_{\rm kin}+S_{2-\rm{form}} + S_{\rm WZ}  \label{eq.WZWmulti_Ltotal}
\end{equation}
with $S_{\rm kin}$, $S_{2-\rm{form}}$ and $S_{\rm WZ}$ given in \eqns \eqref{eq.WZWmulti_Lkin}, \eqref{eq.WZWmulti_L2form}
and \eqref{eq.WZWmulti_LWZ}, respectively.

It turns out that the model obtained this way coincides precisely with the coupled sigma model
introduced recently in connection with the affine Gaudin model \cite{Delduc:2018hty,Delduc:2019bcl}.

To see this explicitly, let us write the one-form $\omega$ in 
\eqref{eq.omega_multiWZW} as
\begin{equation}
	\omega = \varphi_w(z)  \varphi_{\wbar}(z)\d z \;, 
	\label{eq.omegaphiphi}
\end{equation}
where we defined functions $\varphi(z)$ and  $\bar{\varphi}(z)$ for chiral and anti-chiral zeros:
\begin{equation}
\varphi_w(z):=\frac{\prod (z - q_k^w)  } {\prod(z - p_i)} \;, \quad
	\varphi_{\wbar}(z):=\frac{\prod(z - q_k^{\wbar} ) } {\prod(z - p_i)} \;.
\end{equation}
In the language of \cite{Delduc:2018hty,Delduc:2019bcl} these are the twist functions for an 
affine Gaudin model; in our context these functions specify the one-form $\omega$.

We defined above
\begin{equation}
\begin{split}
	\varphi_{w,i}(z)=(z-p_i) \varphi(z) \;, \\
	\varphi_{\wbar,i}(z)=(z-p_i) \bar{\varphi}(z) \;.
\end{split}
\end{equation}
The expression for the Lagrangian given above in terms of the functions $\varphi_{w,i}$ and $\varphi_{\wbar,i}$ coincides with the formulas in \cite{Delduc:2018hty}.\footnote{Overall factor of the action is identified as
 $-k/(4\pi)=\ell_{\infty}/4$, where $\ell_{\infty}$ is the notation used in \cite{Delduc:2018hty}.}

There are many ingredients which seem common between this paper and 
\cite{Delduc:2018hty,Delduc:2019bcl}, and it would be interesting to better understand the relation
at a deeper level.

\subsection{The Lax Operator}
To calculate the Lax operator, we choose a holomorphic gauge $A_{\zbar}=0$ by applying the inverse gauge transformation by $\what{\sigma}$. Then, for $z \neq p_i$, the Lax operator $\Lax(z)$ is given by the value of $A_w^{\sigma}$ and $A_{\wbar}^{\sigma}$ at a point $z$, i.e. by the functions $F,G$ respectively. We have 
\begin{equation} 
\begin{split}
	\Lax_w(z) &= \sum \frac{ \varphi_{w,i}(p_i)}{\varphi_{w,i}(z)} J_{i,w} \;, \\
	\Lax_{\wbar}(z) &= \sum \frac{ \varphi_{\wbar,i}(p_i)}{\varphi_{\wbar,i}(z)} J_{i,\wbar} \;.
	\label{eq:Lax_phi}
\end{split}
\end{equation}

We can again verify that these formulas coincides with those from 
\cite{Delduc:2018hty,Delduc:2019bcl}.

\section{Generalized Riemannian Symmetric Spaces}\label{sec:gss}
Riemannian symmetric spaces are associated to a group which has an action of $\Z_2$. This concept has been generalized \cite{Kac, LedgerObata, Kowalski} to $n$-symmetric spaces, which come from groups which have an action of $\Z_n$.  This class of examples includes the super-homogeneous space $PSU(2,2 \, | \, 4)/(SO(4,1) \times SO(5))$ relevant for type IIB on $AdS^5 \times S^5$.  Even more generally, some authors \cite{Lutz} have introduced the notion of $\Gamma$-symmetric space, associated to an action of a finite group $\Gamma$ on some group $G$.

In this section we will study the $\sigma$-model with target an $n$-symmetric space. 
Thus, fix a simple group $G$ with an action of the finite group $\Z_n$ by isometric group automorphisms.    We let $G_0 \subset G$ be the fixed points of the $\Z_n$ action. We assume $G$ has a metric invariant under the left action of $G$ and the right action of $G_0$, and also invariant under the $\Z_n$ action on $G$. The quotient $G / G_0$ inherits a metric from that on $G$: it is an example of an $n$-symmetric space.  In this section,  we will construct the integrable $\sigma$-model with target the $n$-symmetric space, for a particular metric on $G$ (see \cite{Young:2005jv,Beisert:2012ue} for earlier study of this theory from a different angle).

\begin{figure}[htbp]
\centering
\includegraphics[scale=0.52]{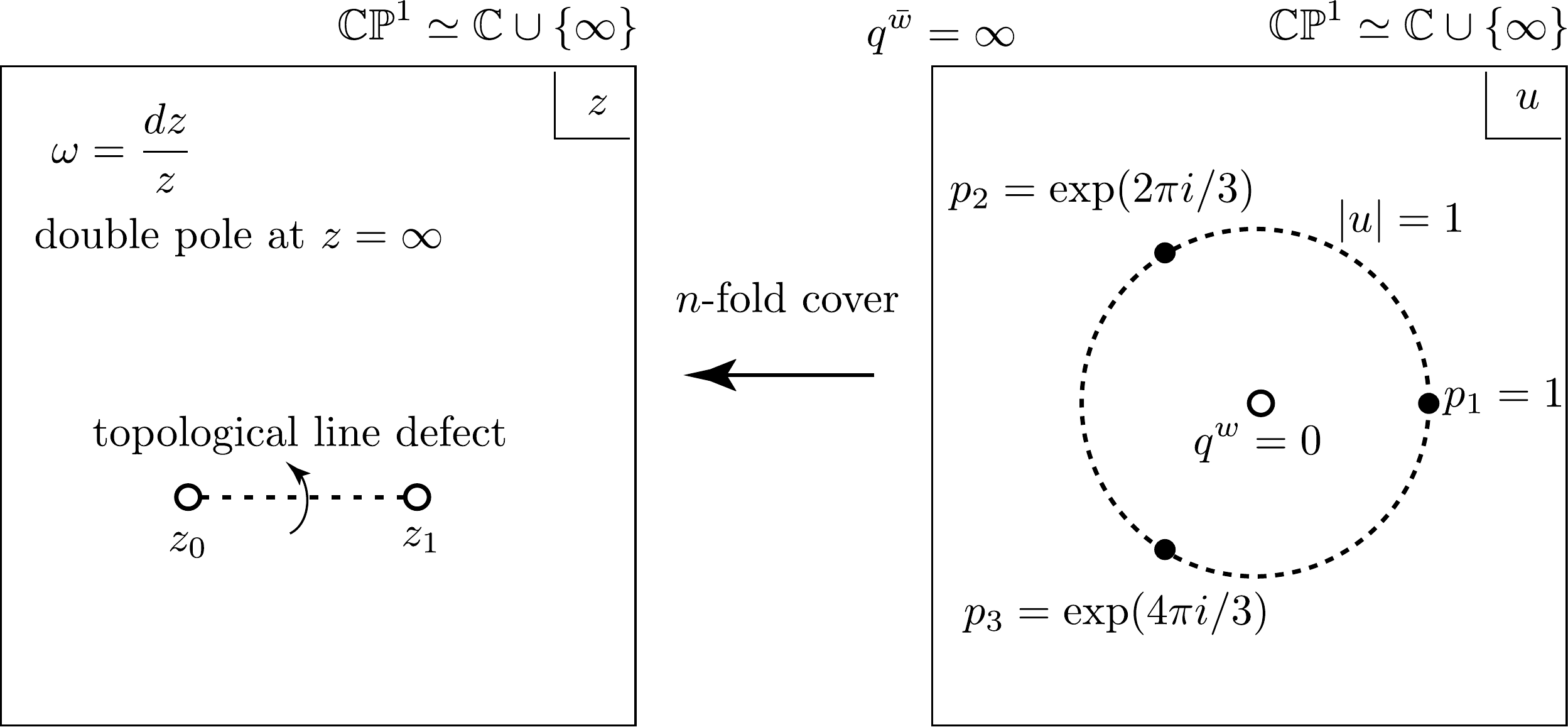}
\caption{We include a topological line defect associated with the $\Z_n$-symmetry of the gauge group $G$.
The line defect, represented by a dashed line on the left, connects two points $z_0$ and $z_1$ in the 
$\mathbb{CP}^1$ parametrized by $z$ with the one-form $\omega=dz/z$. In the $n$-fold cover (parametrized by $u$)
the inverse image of the endpoints $z_0, z_1$ of the line defects are degree $n-1$ zeros $q^w, q^{\wbar}$ of the one-form,
which we have taken to be at $0$ and $\infty$, respectively. The inverse images of the double zero $z=\infty$ are $n$ points $p_1, p_2, \dots, p_n$, which we have taken to be located at the $n$-th roots of unity. The $\Z_n$ symmetry then permutes the points.
The figure shows the case of $n=3$. See \cite{Beisert:2005bm} for a similar figure for $n=4$, in a closely related context.}
\label{Fig_III_13}
\end{figure}

Let us consider the four-dimensional gauge theory on $\CP^1$ with the one-form $\d z$. We fix points $z_0$, $z_1$ and consider a defect on the interval $[z_0,z_1]$ where as we cross this line, we apply the order $n$ automorphism $\rho$ of $G$.  We let $u$ be a coordinate on the corresponding $n$-fold cover, and let $q^w$, $q^{\wbar}$ be the inverse image of $z_0,z_1$.   We let $p_i$ be the $n$ inverse images of $z = \infty$.  The one-form $\omega$ on the $u$-plane has second order poles at $p_i$, and zeroes of order $n-1$ at $q^{w}$, $q^{\wbar}$ (see \fig\ref{Fig_III_13}).

This brings us to the limit of the situation considered above (see section \ref{sec:general_PCM} and in particular \fig\ref{Fig_III_9}) where the points $q_i^{w}$ coincide, and the points $q_i^{\wbar}$ also coincide.  In this limit, the boundary condition at $q^w$ (respectively, $q^{\wbar}$) give the field $A_w$ (respectively, $A_{\wbar}$) a pole of order $n-1$.   The two-dimensional field theory we find will be the $\sigma$-model on the $\Z_n$ fixed points of the generalized principal chiral model discussed in section \ref{sec:general_PCM}.

We will set $q^w = 0$.  We will rescale $k \mapsto k / (q^{\wbar})^{n-1}$ and then send  $q^{\wbar} \to \infty$. Finally, we  set the $p_i$ to be the $n$th roots of unity.  Then the one-form $\omega$ on the $u$-plane takes the form
\begin{equation} 
	\omega = \frac{u^{n-1}  } {\prod_{i=1}^n (u - p_i)^2} \d u =  \frac{u^{n-1}}{(u^n-1)^2} \d u\;.  
\end{equation}
We take the functions $\varphi_{i,w}$, $\varphi_{i,\wbar}$ in \eqn \eqref{eq.omegaphiphi} to be
\begin{equation} 
	\begin{split}
\varphi_{i,\wbar}(u) &= \frac{1}{\prod_{j \neq i}(u-p_j)} =\frac{u-p_i}{u^n-1} \;, \\
		\varphi_{i,w}(u)  &= \frac{u^{n-1}}{\prod_{j \neq i}(u-p_j)} = u^{n-1}\frac{u-p_i } {u^n-1 }   \;.
		\label{eq:phiphi}
	\end{split}
\end{equation}
These are the same functions we used in the generalized principal chiral model, except that we have rescaled $\varphi_{i,\wbar}$ by $1/(q^{\wbar})^{n-1}$ before sending $q^{\wbar}$ to infinity. 

The kinetic term in the Lagrangian is
\begin{multline}	
	S_{\rm{kin}}=\frac{k}{4\pi}
\int_{\mathbb{R}^2}	 
\left[ \sum_i J_{i,\wbar} J_{i,w} \left( - \varphi_{i,w}(p_i) \varphi_{i,\wbar}'(p_i)  + \varphi'_{i,w}(p_i) \varphi_{i,\wbar}(p_i)  \right) 	\right.	
\\
	\left.	-  \sum_{j \neq i}\frac{ \varphi_{i,\wbar}(p_i) \varphi_{j,w}(p_j)  -\varphi_{j,\wbar}(p_j) \varphi_{i,w}(p_i)     }{p_i - p_j} J_{i,\wbar} J_{j,w}  \right] \;.	
\end{multline}
Calculating this explicitly we find
\begin{equation}	
	S_{\rm{kin}} = \frac{k}{4\pi}
\int_{\mathbb{R}^2}	 
	\left[ \sum_i J_{i,\wbar} J_{i,w}\frac{n-1}{n^2}  	
		-  \sum_{j \neq i} \frac{1}{n^2} J_{i,\wbar} J_{j,w}  \right] \;.
		\label{eq:Skin_cover}
\end{equation}
As explained before, this is the $\sigma$-model on $G^{n-1}\simeq G^n/G$. 

Similarly, the two-form component is given by pulling back the two-form 
\begin{equation} 
	 S_{2-\rm{form}}= - \frac{k}{8 \pi n^2 } \int_{\mathbb{R}^2} \sum_{j \neq i}\frac{ p_i + p_j  }{p_i - p_j} X_{i,a} X_{j,a} \;.\label{eq:2form_cover}
\end{equation}
The Wess-Zumino term is associated to the three-form
\begin{equation} 
	S_{\rm{WZ}}=- \frac{k}{12 \pi n^2 } \sum_i \int_{\R^2 \times \R_{\ge 0} } f^{abc} X_{i,a} X_{i,b} X_{i,c} \;. 
	\label{eq:WZ_cover}
\end{equation}

\subsection{\texorpdfstring{Passing to $\Z_n$ Fixed Points}{Passing to Z(n) Fixed Points}}
Now let us turn to calculating the $\Z_n$ fixed points of the $\sigma$-model with this action.  We would like to identify the $\Z_n$ fixed points of this $\sigma$-model with the generalized symmetric space with target $G/G_0$.  The group $\Z_n$ acts on the plane with coordinate $u$ by multiplication by roots of unity, and simultaneously acts on $G$ by the given $\Z_n$ action.  

The action on the $u$-plane means that we permute the $n$ copies of $G$ in $G^{n-1}$, where we identify $G^{n-1}$ with the quotient of $G^n$ by the diagonal right $G$-action. We can then think of the $n$ copies of $G$ as being given by parallel transport from $u = 0$ to the roots of unity, which are permuted by the $\Z_n$-action on the $u$-plane.

We can represents the points of $G^{n-1}$
in terms of the ``inhomogeneous coordinates'' $(g_1, \dots, g_{n-1})$. In the ``homogeneous coordinate''
in $G^n/G$, this element can also be written as $[(g_1, \dots, g_{n-1}, 1)]$, where in the bracket notation we mean the equivalence relation $[(g_1, \dots, g_{n-1}, g_n)] \simeq [(g_1 g, \dots, g_{n-1}g, g_n g)]$
for all $g\in G$. Then the $u$-plane action permuting $n$ copies of $G$ can be identified as
$[(g_1, \dots, g_{n-1}, 1)] \to [(1, g_1, \dots, g_{n-1})] \simeq [(g_{n-1}^{-1}, g_1 g_{n-1}^{-1}, \dots, g_{n-2} g_{n-1}^{-1}, 1)]$,
or equivalently $(g_1, \dots, g_{n-1}) \to (g_{n-1}^{-1}, g_1 g_{n-1}^{-1}, \dots, g_{n-2} g_{n-1}^{-1})$
in the inhomogeneous coordinate.

If $\rho : G \to G$ is the order $n$ symmetry generating our chosen $\Z_n$ action, the total $\Z_n$ action 
on
$G^{n-1}$ is given by first applying $\rho$ to $G^{n-1}$ and then apply the $u$-plane rotation given above.
This gives, in inhomogeneous coordinate,
\begin{equation} 
	(g_1,\dots,g_{n-1}) \mapsto (\rho(g_{n-1})^{-1},\rho(g_1)\rho(g_{n-1})^{-1},\dots,\rho(g_{n-2})\rho(g_{n-1})^{-1}) \;.  
	\label{eq.Zn}
\end{equation}

We let $Y \subset G^{n-1}$ be the fixed points of this $\Z_n$ action, with the induced metric.
We claim that this manifold realizes the $\Z_n$ coset:
\begin{align}
Y:=(G^{n-1})_{\Z_n}  \simeq G/G_0 \;.
\label{eq.YG}
\end{align}

Let us first verify that $Y$ allows for a $G$-action
as implied by \eqn\eqref{eq.YG}. This $G$-action originates from a
left  $G$-action in $G^n$, where an element $g\in G$ acts on the point
$[(g_1, g_2, \dots, g_n)] \in G^n$ as
\begin{align}
[(g_1, g_2, \dots, g_n)] \mapsto [(g g_1, \rho(g) g_2, \dots, \rho^{n-1}(g) g_n)] \;.
\end{align} 
This commutes with the right $G$-action on $G^n$, and hence
descends to the $G$-action on $G^n/G\simeq G^{n-1}$, given by
\begin{equation} 
	(g_1,\dots,g_{n-1}) \mapsto (g g_1 \rho^{n-1}(g)^{-1}, \rho(g) g_2 \rho^{n-1}(g)^{-1}, \dots, \rho^{n-2}(g) g_{n-1} \rho^{n-1}(g)^{-1} ) \;.  
\end{equation}
We can further verify that this $G$-action commutes with the $\Z_n$ action given in \eqref{eq.Zn},
and hence further descends to the action on $Y=(G^{n-1})_{\Z_n}$.

In order to show \eqn\eqref{eq.YG}, let us note that 
any
 element $(y_1,\dots,y_{n-1}) \in Y$ has $y_1 = \rho(y_{n-1})^{-1}$, $y_2 = \rho(y_1)\rho(y_{n-1})^{-1}$, $y_3 = \rho(y_2)\rho(y_{n-1})^{-1}$, etc. Thus, $y_1, y_2,\dots,y_{n-1}$ are determined from $y = y_{n-1}^{-1}$.  To define an element of $Y$, the element $y \in G$ must satisfy
\begin{equation} 
	\rho^{n-1}(y) \rho^{n-2}(y) \dots y = 1\;.  \label{eqn_generalized_symmetric} 
\end{equation}
We can thus identify the fixed point manifold $Y$ with the set of solutions to this equation.

The $G$-action on $Y$ (after applying the automorphism $\rho^{n-1}$ of $G$) sends 
\begin{equation} 
	y \mapsto  \rho(g)y g^{-1} \;.  
\end{equation}
The stabilizer of the identity consists of $G_0$. We thus have a map $G_0 \to G$.  To check that this map is an isomorphism, we need to show that the $G$-action on $Y$ is transitive. To see that the action is transitive on the connected component of the identity on $Y$, it suffices to check that the map from $\mf{g}$ to the tangent space of the identity in $Y$ is surjective. 

Let us decompose $\mf{g} = \mf{g}_0 \oplus \mf{g}_1$, where $\mf{g}_0$ is the Lie algebra of $G_0$ and $\mf{g}_1$ is the direct sum of all non-trivial representations of $\Z_n$ occurring in $\mf{g}$. If $y = (y_0,y_1)$  then at the infinitesimal level the equation \eqref{eqn_generalized_symmetric} becomes the additive equation $\sum_{i = 0}^{n-1} \rho^{i}(y) = 0$, for $y \in \mf{g}$.  This equation is satisfied if and only if $y \in \mf{g}_1$, so that the tangent space to $Y$ is $\mf{g}_1$.  The infinitesimal action of $t \in \mf{g'}$  on the identity in $Y$ yields
\begin{equation} 
	\exp( \eps \rho(t)) \exp(-\eps t) = 1 + \eps \rho(t) - \eps t + O(\eps^2) \;. 
\end{equation}
The linear operator $\rho - 1 : \mf{g}_1 \to \mf{g}_1$ is invertible. Therefore the action of $G$ on $Y$ is infinitesimally transitive, and hence transitive.  This establishes \eqn\eqref{eq.YG}.

\subsection{Identification of the Metric}

We have shown that this construction yields a $\sigma$-model on the space $Y = G/G_0$ with a transitive action of $G$ by isometries.  It remains to determine the metric and WZ term of this model.    Because the $G$-action is transitive, and the metric, two-form and three-form on $Y$ are $G$-invariant, it suffices to describe these forms on the tangent space at the origin in $Y$. The cotangent space is the set of elements
\begin{align} 
\begin{split}
	&[(y_1,\dots,y_{n})] \in \mf{g}_1^{\oplus n-1}\simeq \mf{g}_1^{\oplus n}/\mf{g}  \\
	&\textrm{satisfying} \quad
	[(y_1,\dots,y_{n})] = [(\rho(y_{n}),\rho(y_1), \dots, \rho(y_{n-1})  ) ]\;.
\end{split}
\end{align}

The kinetic term in \eqn\eqref{eq:Skin_cover} now reads 
\begin{equation}	
	S_{\rm kin}= \frac{k}{8 n^2  \pi} \int_{\mathbb{R}^2} 
	\left[ (n-1) \sum_i  J_{i,a} (\star J)_{i,a} 
		-  \sum_{j \neq i} J_{i,a} (\star J)_{j,a}  \right] \;.
\end{equation}
Letting $J = J_{1}$, and letting $\ip{-,-}$ denote the Killing form on $\g$, which is invariant under $\rho$, the expression inside the bracket is 
\begin{equation}	
	(n-1) n \ip{ J,  \star J } 
		- n \sum_{j = 1}^{n-1} \ip{ J, \star \rho^{j}  J}  
		\;.
		\label{eq:ytmp}
\end{equation}
Because $J_{1} + \rho J_{1} + \dots + \rho^{n-1} J_{1} = 0$ (this is the diagonal part of $\mf{g}^{\oplus n}$,
which we quotient out when we discuss $\mf{g}^{\oplus n-1}$), the second term in \eqn\eqref{eq:ytmp} is $n \ip{J,\star J}$ so we find that the kinetic term simplifies to
\begin{equation} 
	S_{\rm kin}=  \frac{k}{8 \pi}  \int_{\mathbb{R}^2}  \ip{J, \star J}.  
\end{equation}
That is, the metric on $G$ is the bi-invariant metric.

If we decompose $J = \sum_{r=0}^{n-1} J^{(r)}$, where $\rho$ acts on $J^{(r)}$ by $e^{2 \pi \i r / n}$,
then this can be written as
\begin{equation} 
	S_{\rm kin}=  \frac{k}{8 \pi}  \int_{\mathbb{R}^2}  \sum_{r=0}^{n-1}\ip{J^{(r)},\star J^{(n-r)}}.  
\end{equation}
Here we used the condition that the metric $\langle -,-\rangle$ is compatible with $\Z_n$-grading,
i.e.\
\beq
\ip{J^{(r)},\star J^{(s)}}=0 \quad \textrm{unless} \quad r+s\equiv 0  \quad (\mathrm{mod }\,\, n) \;.
\eeq

Next, let us calculate the two-form and three-form contributions. The two-form part of the action in \eqn\eqref{eq:2form_cover} is
\begin{equation} 
	S_{\rm 2-form}=- \frac{k}{8 \pi n } \sum_{k = 1}^{n-1}\frac{ 1 + e^{\frac{2 \pi i k }{ n}}  }{1 - e^{\frac{2 \pi i k }{ n}} } \int_{\mathbb{R}^2} \ip{J, \rho^{k} J} . 
\end{equation}
When decomposed into $J^{(r)}$, this is (cf.\ \cite{Young:2005jv})
\beq
	S_{\rm 2-form}&=- \frac{k}{8 \pi n } \sum_{r = 0}^{n-1} \sum_{k = 1}^{n-1}\frac{ 1 + e^{\frac{2 \pi i k }{ n}} }{1 - e^{\frac{2 \pi i k }{ n}}}\, e^{\frac{2 \pi i k r }{ n}}  \int_{\mathbb{R}^2} \ip{J^{(n-r)},  J^{(r)}} \\
	&=\frac{k}{8 \pi } \sum_{r = 1}^{n-1} \left(1-\frac{2r}{n}\right)  \int_{\mathbb{R}^2} \ip{J^{(n-r)},  J^{(r)}} 
	\;,
\eeq
where we used 
\beq
\sum_{k = 1}^{n-1}\frac{ 1 + e^{\frac{2 \pi i k }{ n}} }{1 - e^{\frac{2 \pi i k }{ n}}}\, e^{\frac{2 \pi i k r }{ n}}
=
\begin{cases}
r-2n &  (r=1, \dots,  n-1) 
\\
0 & (r =0)  
\end{cases}
\;.
\eeq
The three-form contribution in \eqn\eqref{eq:WZ_cover} is similarly
\beq
	S_{\rm WZ}&=- \frac{k}{12 \pi n } \int_{\R^2 \times \R_{\ge 0} } \ip{ J, [J, J]} \\
	&=  - \frac{k}{12 \pi n } \sum_{\genfrac{}{}{0pt}{2}{r,s,t=0, \dots, n-1}{r+s+t\equiv 0\,  (\textrm{mod}\,\, n)}} \int_{\R^2 \times \R_{\ge 0} } \ip{ J^{(r)}, [J^{(s)}, J^{(t)}]}  \;. 
\eeq

\subsection{The Lax Operator}

Finally, it is easy to compute the Lax operator from \eqns \eqref{eq:Lax_phi} and \eqref{eq:phiphi}:
\begin{equation} 
\begin{split}
	\Lax_w(u) &= \sum_{k=0}^{n-1} \frac{u^{n}-1 }{n u^{n-1}\left(u-e^{\frac{2 \pi i k }{ n}}\right)} \rho^{k}  J_{w} \;, \\
	\Lax_{\wbar}(u) &= \sum_{k = 0}^{n-1} \frac{(u^n -1)e^{\frac{2 \pi i k }{ n}} }{n \left(u - e^{\frac{2 \pi i k }{ n}}\right)} \rho^k J_{\wbar} \;,
\end{split}\label{eq:gsm_lax}
\end{equation}
where we used $J_{k, w}=\rho^k J_{n,w}, J_{k, \wbar}=\rho^k J_{n,\wbar}$ and 
denoted $J_w=J_{n,w}$ and $J_{\wbar}=J_{n, \wbar}$.

Let us decompose $J_{w}, J_{\wbar}$ as $J_w=\sum_{r=0}^{n-1} J_{(r),w}$ and $J_{\wbar}=\sum_{r=0}^{n-1} J_{(r),\wbar}$, where the $\Z_n$-involution $\rho$ acts on $J_{(r),w}$ and $J_{(r),\wbar}$ by $e^{2 \pi i r/  n}$.  Using the identity
\begin{equation} 
	\sum_{k = 0}^{n-1} e^{\frac{2 \pi \i k (l+1) }{ n}} \frac{u^n - 1}{u -e^{\frac{2 \pi i k }{ n}}} 
	=\sum_{k = 0}^{n-1} e^{\frac{2 \pi \i k l) }{ n}} \frac{u^n - 1}{u\, e^{-\frac{2 \pi i k }{ n}}-1} 
	=  \sum_{k,s = 0}^{n-1} e^{\frac{2 \pi \i k (l-s) }{ n}} u^s = n u^{l} \;,
\end{equation}
we find
\begin{equation} 
\begin{split}
	\Lax_w(u) &=  J_{0,w} +  \sum_{l = 1}^{n-1} u^{-l} J_{n-l,w}\;,  \\
	\Lax_{\wbar}(u) &=  \sum_{l = 0}^{n-1} u^{l} J_{l,\wbar} \;.
\end{split}
\end{equation}

\subsection{\texorpdfstring{Comparison with the $AdS^5 \times S^5$ Integrable Coset Model}{Comparison with the AdS(5) times S(5) Integrable Coset Model}}
In this section we will take $\mf{g} = \mf{psu}(2,2\, |\,  4)$, with the $\Z_4$ action as discussed in the review articles \cite{Magro:2010jx, Zarembo:2017muf}.  We decompose $\mf{g}$
into eigenspaces of the $\Z_4$ action
\begin{equation} 
	\mf{g} = \mf{g}_0 \oplus \mf{g}_1 \oplus \mf{g}_2 \oplus \mf{g}_3  \;,
\end{equation}
where the generator of $\Z_4$ acts by $\i^k$ on $\mf{g}_k$.  The group $G_0$ is given by $SO(4,1) \times SO(5)$. 

There are a number of points which are considerably more subtle when discussing the super-string (as opposed to field theory) coset model. In the Green-Schwarz formulation \cite{Metsaev:1998it}
, the kinetic term of the coset Lagrangian is degenerate on the fermionic fields.  This degeneracy is resolved by introducing $\kappa$-symmetry, which is a fermionic gauge symmetry which removes some of the fermionic degrees of freedom.  

In the pure spinor formulation \cite{Berkovits:2000fe}, the kinetic term of the coset Lagrangian is non-degenerate, but additional degrees of freedom are introduced, together with an additional BRST operator.  

We make no attempt here to construct either the pure spinor or Green-Schwarz models as a string theory. We simply construct the $PSU(2,2 \, | \, 4)/ (SO(4,1) \times SO(5))$ coset model as a $\sigma$-model, with a non-degenerate kinetic term.  Because the kinetic term is non-degenerate, the model we construct is related to the pure spinor formulation of the super-string, and not to the Green-Schwarz formulation. This is reflected in the fact that the Lax matrix we find matches the pure-spinor Lax matrix, which is different from the Green-Schwarz Lax matrix. 

To write down the action and the Lax matrix of the coset model, we simply specialize the discussion above to the case $n=4$ and $G = PSU(2,2 \, | \, 4)$.  
For example, the kinetic term is given by
\begin{equation} 
	S_{\rm kin}=\frac{k}{8 \pi}\int_{\mathbb{R}^2} \left(  \ip{J_{0,w}, \star J_{0,\wbar} } +  \ip{J_{2,w}, \star J_{2,\wbar} } + \ip{J_{1,w},\star J_{3,\wbar}} \right) \;, 
\end{equation}
which is non-degenerate even for fermionic components ($J_1$ and $J_3$).
For the Lax operator from \eqref{eq:gsm_lax}, 
writing the currents as $J_{i,w}$, $J_{i,\wbar}$, $i = 0,\dots,3$, we find 
\begin{equation} 
\begin{split}
	\Lax_w(u) &= J_{0,w} + u^{-1} J_{3,w} + u^{-2} J_{2,w} + u^{-3} J_{1,2} \;,\\
	\Lax_{\wbar}(u) &= J_{0,\wbar} + u J_{1,\wbar} + u^2 J_{2,\wbar} + u^3 J_{3,\wbar} \;. 
\end{split}\label{eq:gsm_lax_2}
\end{equation}
This is the same as the pure-spinor Lax matrix \cite{Vallilo:2003nx,Magro:2010jx}.\footnote{The relation
with the Bena-Polchinski-Roiban Lax matrix \cite{Bena:2003wd} was discussed in \cite{Vallilo:2003nx}.}


\section{Trigonometric Deformation of the Principal Chiral Model}
\label{sec:PCM_deformed}

In this section we discuss yet another generalization of the principal chiral model---trigonometric deformation
(the previous cases correspond in this language to the rational cases). Our discussion can be 
thought of as a field-theory analog for lattice model counterparts in section 9.1 of \cite{Costello:2017dso}, whose primary examples 
is the so-called magnetic field deformation of the six-vertex model.

\subsection{Boundary Conditions}

In section 9.1 of \cite{Costello:2017dso} we explained how to introduce topological boundary conditions when the one-form $\omega$ has a first-order pole.  The construction is as follows. We introduce an extra copy $\til{H}$ of the Cartan of our group, which is isomorphic to the original Cartan $H$, but where the inner product on the Lie algebra of $\til{H}$ has the opposite sign to that on the Lie algebra of $H$. 

We define two complementary Lagrangian subalgebras $\mf{l}_{\pm} \subset \mf{g} \oplus \mf{h}$ as follows. First, we decompose $\mf{g} = \mf{n}_- \oplus \mf{h} \oplus \mf{n}_+$ into lower-triangular, diagonal, and upper-triangular pieces. Then we define $\mf{l}_+$ to be $\mf{n}_+$, together with the Lagrangian subspace 
\begin{equation} 
(X, X) \subset \mf{h} \oplus \til{\mf{h}} 
 \end{equation}
 (for $X \in \mf{h}$).  We define $\mf{l}_-$ similarly to consist of $\mf{n}_-$ together with the subspace $(X, X)$ in $\mf{h} \oplus \til{\mf{h}}$. 

When the one-form $\omega$ has a first-order pole, we can define a boundary condition by asking that the gauge field $A$ lives in $\mf{l}_+$ or $\mf{l}_-$ at the pole. Gauge transformations also must take values in $\mf{l}_+$ or $\mf{l}_-$ at the pole.

To define the principal chiral model, we took a one-form which has a second order pole at $0$ and at $\infty$ on $\CP^1$, and also two zeroes.  We can introduce a two-parameter deformation of this model by separating each second order pole into a pair of first order poles.

The one-form we took to define the principal chiral model was $z^{-2} (z - z_0)(z-z_1) \d z$. The one-form associated to this deformation will be
\begin{equation} 
\omega = \frac{(z - z_0)(z - z_1)}{z(z-z_+)(1 - z/z_-) } \d z    \;.
\label{eq.omega_trig_PCM}
 \end{equation}
This has first-order poles at $0,\infty,z_+,z_-$, and if we send $z_+ \to 0$, $z_- \to \infty$ we recover the one-form defining the principal chiral model (up to overall normalization).

We impose the following boundary conditions at the poles of the one-form.  At $z = 0$ and $z = z_-$, we ask that the gauge field is in $\mf{l}_-$, and at $z = \infty$, $z = z_+$ the gauge field is in $\mf{l}_+$. 

At the zeroes $z_0,z_1$ of the one-form, we do what we did in the case of the rational principal chiral model: at $z_0$, we allow $A_w$ to have a first-order pole, and at $z_1$ we allow $A_{\wbar}$ to have a first-order pole.

When we send $z_+ \to 0$ to reintroduce a second-order pole, these boundary conditions imply that that $A$ is in both $\mf{l}_+$ and $\mf{l}_-$ at $z = 0$.  Since $\mf{l}_+\cap \mf{l}_- = 0$, this means that $A = 0$ at $z = 0$, so we reintroduce the Dirichlet boundary conditions that led to the principal chiral model.  Similarly, sending $z_- \to \infty$ gives us a Dirichlet boundary condition at $\infty$.

In this way, we see that the model constructed from a one-form on $\CP^1$ with four first-order poles and two zeroes does indeed limit to the principal chiral model when pairs of poles collide.  However, it is \emph{not} the principal chiral model for $G$, rather it is the model for the extended gauge group
\begin{equation} 
\til{G} =  G \times \til{H} \;. 
 \end{equation}

Now let us compute the Lagrangian of the deformed model following the method employed previously. 

The field $A_{\zbar}$ defines a $\til{G}$ bundle on $\CP^1$, with a reduction to $L_+$ at $\infty,z_+$ and a reduction to $L_-$ at $0,z_-$.  We will assume, as usual, that $A_{\zbar}$ defines a trivial bundle on $\CP^1$. In that case, the data of the reductions at the four points is $(\til{G}/L_+)^2 \times (\til{G}/L_-)^2$.  This must be divided by the action of $\til{G}$, acting as gauge transformations of the trivial bundle.  So we find that the gauge equivalence class of $A_{\zbar}$ defines a map
\begin{equation} 
\sigma: \R^2 \to  \til{G} \setminus\left(  \til{G}/L_+    \times \til{G}/L_+  \times  \til{G}/L_- \times \til{G}/L_- \right) \;.
 \end{equation}

To get good coordinates on the target space in which to describe the Lagrangian, we will use the Bruhat decomposition of $G$. 

Recall that there is an open subset -- the big Bruhat cell -- $U \subset G$ such that there is a triangular decomposition
\begin{equation} 
U = N_- \cdot H \cdot N_+ \;. 
 \end{equation}
That is, every element of $U$ can be written uniquely in the form $a_- X b_+$ where $a_- \in N_-$, $X \in H$ and $b_+ \in N_+$.  We let $\til{U} = U \times \til{H}$.

 Clearly, $L_+ = H_+ \cdot N_+$ and $L_- = N_- \cdot H_-$ where $H_{+}$ is the diagonal $(X,X)$ in $H \times \til{H}$ and $H_-$ is the anti-diagonal $(X,X^{-1})$.

 From this we see that 
\begin{equation} 
L_- \cdot L_+ = N_- \cdot H_- \cdot H_+ \cdot N_+ \;. 
 \end{equation}
The multiplication map
\begin{equation} 
H_- \times H_+ \to H \times \til{H}  
 \end{equation}
 is a $2^r:1$ cover, sending $(X,X,Y,Y^{-1})$ to $(XY, XY^{-1})$.  Indeed, given $A = XY$, $B = XY^{-1}$, we find $X$ as a square root of $AB$ and $Y$ as a square root of $AB^{-1}$. In a rank $r$ torus there are $2^r$ possible square roots. 

This tells us that $L_- \times L_+$ is a $2^r$ fold cover of the open subset $\til{U} = U \times \til{H} \subset \til{G}$.

The target of the $\sigma$-model is
\begin{equation} 
 \til{G} \setminus\left(  \til{G}/L_-    \times \til{G}/L_+  \times  \til{G}/L_- \times \til{G}/L_+ \right) \;.
 \end{equation}
 The four copies of $\til{G}/L_{\pm}$ correspond to the reductions of the gauge group to $L_{\pm}$ at $z = 0,z_+,z_-,\infty$.   We can use the overall $\til{G}$-symmetry to move the point at $\til{G}/L_-$ corresponding to $z_-$ to the base point. We are left with an overall $L_-$ gauge symmetry. By assuming that the reduction to $L_+$ at $z_+$ comes from the open Schubert cell which is the image of $L_- \cdot L_+ \subset \til{G}$ in $\til{G}/L_+$, we can use the remaining $L_-$ gauge symmetry to move this reduction to the base point.  We find that an open subset of the target of the $\sigma$-model is given by 
 \begin{equation} 
 \til{G}/L_- \times \til{G}/L_+  \;,
  \end{equation}
corresponding to a reduction of gauge group of the trivial bundle to $L_-$ at $z = 0$ and to $L_+$ at $z = \infty$. We will further assume that each reduction lies in the top Schubert cell, which is the open orbit of $L_-$ or $L_+$. The target is then $L_+ \times L_-$, which is of course an open subset of $\til{G}$. 

We write a map
\begin{equation} 
 \sigma : \R^2 \to L_+ \times L_- 
 \end{equation}
 as $\sigma = (\sigma_+, \sigma_-)$ where $\sigma_{\pm}$ are maps to $L_{\pm}$.

In this gauge, the field $A_{\zbar}$ corresponding to a map $\sigma = (\sigma_+, \sigma_-)$ has the following characterization: it can be brought to the trivial gauge field by a gauge transformation which vanishes at $z_{\pm}$, takes value $\sigma_+$ at $0$, and value $\sigma_-$ at $\infty$. We can thus choose some $\what{\sigma}$ which is $0$ near $z_{\pm}$, and is $\sigma_+$ near $0$ and $\sigma_-$ near $\infty$. We set
\begin{equation} 
A_{\zbar}^{\sigma} = \what{\sigma}^{-1} \partial_{\zbar} \what{\sigma} \;.
 \end{equation}

We need to find $A_{\wbar}^{\sigma}$ so that $F_{\zbar \wbar} = 0$.  Using the decomposition $\til{\mf{g}} = \mf{l}_+ \oplus \mf{l}_-$, we write
\begin{equation} 
A_{\wbar}^{\sigma} = A_{\wbar}^+ + A_{\wbar}^- \;. 
 \end{equation}
 Then we can solve the equation $F_{\zbar \wbar} = 0$ by setting
\begin{equation} 
A_{\wbar}^{\pm} = \what{\sigma}^{-1} \partial_{\wbar} \what{\sigma} + \what{\sigma}^{-1} F(z)^{\pm} \what{\sigma} \;,
 \end{equation}
where $F(z)^{\pm}$ are meromorphic functions of $z$ with values in $\mf{l}_{\pm}$, and with a first order pole at $z = z_1$.  To satisfy the boundary conditions that $A_{\wbar}^+ = 0$ at $0,z_-$ and $A_{\wbar}^- = 0$ at $\infty,z_+$ we require that 
\begin{equation} 
\begin{split}
F(z_-)^+ &= 0\;, \\
F(z_+)^- &= 0 \;,\\
F(\infty)^- &= - (\partial_{\wbar} \sigma_- ) \sigma_-^{-1}\;,\\
F(0)^+ &= - (\partial_{\wbar} \sigma_+) \sigma_+^{-1}\;.
\end{split}
 \end{equation}
(These equations are derived using the fact that $\what{\sigma} = \sigma_+$ near $0$ and $\sigma_-$ near $\infty$.)

The unique solution is given by
\begin{equation} 
\begin{split}
F(z)^- &=  -  \frac{z - z_+}{z - z_1}   (\partial_{\wbar} \sigma_- ) \sigma_-^{-1} \;,\\
F(z)^+ &= -  \frac{1- z/z_-}{1 - z/z_1} (\partial_{\wbar} \sigma_+) \sigma_+^{-1}\;.
 \end{split}
 \end{equation}
This tells us that
\begin{equation} 
\begin{split}
A_{\wbar}^{-} &=  \what{\sigma}^{-1} \partial_{\wbar} \what{\sigma} - \frac{z-z_+}{z-z_1}  \what{\sigma}^{-1} (\partial_{\wbar} \sigma_-) \sigma_-^{-1}    \what{\sigma} \;,\\
 A_{\wbar}^{+} &=  \what{\sigma}^{-1} \partial_{\wbar} \what{\sigma}-  \frac{1- z/z_-}{1 - z/z_1} \what{\sigma}^{-1} (\partial_{\wbar} \sigma_+) \sigma_+^{-1} \what{\sigma}\;.
 \end{split}
 \end{equation}
Similarly, 
\begin{equation} 
\begin{split}
A_{w}^{-} &=  \what{\sigma}^{-1} \partial_{w} \what{\sigma} - \frac{z-z_+}{z-z_0}  \what{\sigma}^{-1} (\partial_{w} \sigma_-) \sigma_-^{-1}    \what{\sigma} \;,\\
 A_{w}^{+} &=  \what{\sigma}^{-1} \partial_{w} \what{\sigma}-  \frac{1- z/z_-}{1 - z/z_0} \what{\sigma}^{-1} (\partial_w \sigma_+) \sigma_+^{-1} \what{\sigma}\;.
\end{split}
 \end{equation}
 
\subsection{The Lax Operator}

As before, we can work in the holomorphic gauge, to obtain the holomorphic Lax operator 
\begin{equation} 
\begin{split}
\Lax_w^{-} &=   -  \frac{z - z_+}{z - z_0}   (\partial_{\wbar} \sigma_- ) \sigma_-^{-1} \;,\\
\Lax_w^+ &= - \frac{1- z/z_-}{1 - z/z_0} (\partial_{\wbar} \sigma_+) \sigma_+^{-1}\;,\\
\Lax_{\wbar}^{-} &=   -  \frac{z - z_+}{z - z_1}   (\partial_{\wbar} \sigma_- ) \sigma_-^{-1}\;,\\
\Lax_{\wbar}^+ &= - \frac{1- z/z_-}{1 - z/z_1} (\partial_{\wbar} \sigma_+) \sigma_+^{-1}\;,
 \end{split}\end{equation}
where $\Lax^{\pm}$ represents the $\mf{l}_{\pm}$ part of the Lax operator.
 
 \subsection{The Lagrangian}

The Lagrangian of the model can be calculated using the same techniques as before. We  write 
 \begin{equation} 
 A^{\sigma} = \what{A} + A'  \;,
  \end{equation}
  where $\what{A} = \what{\sigma}^{-1} \d \what{\sigma}$ and $A'$ includes all the terms involving are rational functions of $z$.  
  We can then evaluate the action from \eqref{eq.CS_sum}.  
We will assume, as before, that $\what{\sigma}$ is invariant under the $U(1)$ action rotating $z$.  

Using the explicit expressions for $A'$, we have
\begin{multline} 
	\op{Tr}(\what{A} A' A' ) = \op{Tr}  \left((\partial_{\zbar} \what{\sigma}) \what{\sigma}^{-1}(\partial_w \sigma_-) \sigma_-^{-1} (\partial_{\wbar} \sigma_+) \sigma_+^{-1}    \right)   \frac{z - z_+}{z - z_0}   \frac{1- z/z_-}{1 - z/z_1} \\	
-	 \op{Tr}  \left((\partial_{\zbar} \what{\sigma}) \what{\sigma}^{-1}(\partial_w \sigma_+) \sigma_+^{-1} (\partial_{\wbar} \sigma_-) \sigma_-^{-1}    \right)     \frac{z - z_+}{z - z_1}   \frac{1- z/z_-}{1 - z/z_0} \;.
\end{multline}
It follows that $\omega \op{Tr}(\what{A} A' A')$ is of charge $-1$ under the $U(1)$ action which rotates $z$ (using the fact that $\what{\sigma}$ is invariant under this $U(1)$ action). Therefore the integral of this term over the $z$-plane is zero.

We are left with the two terms:
\begin{equation} 
	\textrm{CS}(A^{\sigma})= \textrm{CS}(\what{A}) -  \d \op{Tr} (\what{A} A') \;.
\end{equation}
These will give a Wess-Zumino-type term and a kinetic term, respectively.

The first term, since $\what{A}$ is pure gauge (recall \eqn \eqref{eq.CS_trivial}), gives
the integral
\begin{equation} 
- \frac{1}{3} \int \omega \op{Tr} \left( \what{\sigma}^{-1} \d \what{\sigma} \right)^3  \;.
 \end{equation}
We will  use again the fact that $\what{\sigma}$ is invariant under the $U(1)$ action rotating $z$, and is the identity except near $0,\infty$.  Then the integral 
gral defined near $0$ and one near $\infty$. We can perform each integral by integrating first over the argument of $z$.  The result just picks up the residue of $\omega$ at $0$ or $\infty$, times a Wess-Zumino term for $\sigma_{\pm}$: 
\begin{equation} 
-\frac{2\pi \i }{3}\op{Res}_0 \omega  \int_{M^3}\op{Tr}( \what{\sigma}_+^3)  + \frac{2\pi i }{3} \op{Res}_\infty \omega \int_{M^3} \op{Tr} ( \what{\sigma}_-^3) \;. 
 \end{equation}
Here $M^3$ is a $3$-manifold bounding the $w$-plane, and $\what{\sigma}_{\pm}$ are extensions of $\what{\sigma}$ to this manifold. (The difference in signs between the two terms comes from a difference of orientation in the integral over the radial direction $r = \abs{z}$). 

For the second term $- \d \op{Tr} (\what{A} A')$ is integrated against $\omega$, we obtain by integration by parts, a sum of contributions from the poles of $\omega$, each accompanied by the residue at these poles times the value of the rational functional appearing in $A_{w,\wbar}^{\pm}$ at these poles. Since we work in a gauge where $\what{A} = 0$ near the poles at $z_{\pm}$, only the poles at $0$ and $\infty$ contribute.  Since $\what{\sigma} = \sigma_+$ near $0$, only the terms of $A'$ involving $\sigma_-^{-1} \d \sigma_-$ contribute near $0$, and similarly only the terms involving $\sigma_+^{-1} \d \sigma_+$ contribute near $\infty$.

The contribution from $0$ is:
\begin{multline} 
	\pi \i \op{Res}_0 (\omega) \frac{z_+}{z_1}\int_{\R^2} \sigma_+^{-1} (\partial_w \sigma_+)  \sigma_+^{-1} (\partial_{\wbar} \sigma_-) \sigma_-^{-1}\sigma^+  \\
	- \pi \i \op{Res}_0(\omega) \frac{z_+}{z_0}  \int_{\R^2} \sigma_+^{-1} (\partial_{\wbar} \sigma_+)  \sigma_+^{-1} (\partial_{w} \sigma_-) \sigma_-^{-1}\sigma^+   \;.
 \end{multline}
Note that because we are taking the trace, the $\sigma_+^{-1}$ at the beginning of each expression cancels with the $\sigma_+$ at the end. 

Let
\begin{equation} 
 J_i^{\pm} = (\partial_i \sigma_{\pm} ) \sigma_{\pm}^{-1} 
 \end{equation}
 for $i = w,\wbar$.   We can write the contribution from $0$ as 
\begin{equation} 
	 \pi \i  \op{Res}_0(\omega)z_+ \left\{  \frac{1}{z_1} \int\op{Tr} (J_w^+ J_{\wbar}^-) 
-	\frac{1}{z_0}  \int_{\R^2}\op{Tr}(J_w^- J_{\wbar}^+) \right\}  \;.
 \end{equation}
The residue at $\infty$ similarly yields
\begin{equation}
 \pi \i  \op{Res}_{\infty}(\omega)\frac{1}{z_-} \left\{ z_1  \int_{\R^2}\op{Tr} (J_w^- J_{\wbar}^+)
-z_0    \int_{\R^2}\op{Tr}(J_w^+J_{\wbar}^- ) \right\} \;.
 \end{equation}
 
Summarizing, the total action is given by
\begin{align} 
S&=S_{\rm kin}+S_{\rm WZ} \;, \\
S_{\rm kin} &=	 \frac{k}{8 \pi}  \Bigg[  \op{Res}_0(\omega)z_+ \left\{  \frac{1}{z_1} \int_{\R^2} \op{Tr} (J_w^+ J_{\wbar}^-) 
-	\frac{1}{z_0}  \int_{\R^2}\op{Tr}(J_w^- J_{\wbar}^+) \right\} \nonumber\\
&\qquad +
  \op{Res}_{\infty}(\omega)\frac{1}{z_-} \left\{ z_1  \int_{\R^2}\op{Tr} (J_w^- J_{\wbar}^+)
-z_0    \int_{\R^2}\op{Tr}(J_w^+J_{\wbar}^- ) \right\}  \Bigg]\;, \\
S_{\rm WZ}&=\frac{k}{12\pi} \left[-\op{Res}_0 \omega  \int_{M^3}\op{Tr}( \what{\sigma}_+^3)  + \op{Res}_\infty \omega \int_{M^3} \op{Tr} ( \what{\sigma}_-^3)\right] \;,
 \end{align}
 where we introduced the overall normalization factor following \eqn \eqref{eq.level}.

\subsection[Trigonometric Deformations of the Generalized Symmetric Space Models]{Trigonometric Deformations of the \\ Generalized Symmetric Space Models}

It is straightforward, but slightly tedious, to repeat our analysis of generalized symmetric spaces from section \ref{sec:gss} to include the trigonometric deformation.  To do this, we take a group $G$ with a $\Z_n$ action, and consider the gauge theory on $\CP^1$ with gauge group $G \times H$ with the one-form $\d z /z$.  We take the $n$-fold branched cover of the $z$-plane, branched at $z_0,z_1$.  Along the branch cut $[z_0,z_1]$ we apply the automorphism $\rho$ generating the $\Z_n$  action on $G$.  If $u$ is a coordinate on the $n$-fold cover, the one-form $\omega$ on the $u$-plane has two zeroes $q^w$, $q^{\wbar}$ of order $n-1$, and two collections $p_i^{\pm}$ of $n$ first order poles.  We take coordinates so that $q^{\wbar} = \infty$, $q^w = 0$, $p_k^{\pm} = \lambda^{\pm} e^{2 \pi \i k / n}$, where $\lambda$ is the parameter deforming us to the trigonometric situation. 

On the $u$-plane, we are in the trigonometric situation discussed above.  We will decompose $\mf{g} \oplus \mf{h} = \mf{l}_{+} \oplus \mf{l}_-$, as before.  This decomposition may or may not be compatible with the action of $\Z_n$ given by $\rho$.  If it is not, we let $\mf{l}_{\pm}^{(k)}$ be the image of $\mf{l}_{\pm}$ under the automorphism $\rho^k$.   

We let $A_{\pm}^{(k)}$ be the component of the gauge field that lives in $\mf{l}_{\pm}^{(k)}$.   We impose the boundary condition that $A_{\pm}^{(k)}$ vanishes at $\lambda^{\pm} e^{2 \pi \i k / n}$.  This boundary condition is compatible with the $\Z_n$ action that simultaneously applies the automorphism $\rho$ to $G$ and multiplies $u$ by a root of unity.   At $\infty$, we ask that $A_{\wbar}$ has a pole of order $n-1$, and at $0$ we ask that $A_w$ has a pole of order $n-1$.  

This gives us a certain trigonometric integrable $\sigma$-model deforming that in $(G \times H)^{n} / (G \times H)$. Passing to the $\Z_n$ fixed points of the target, we get a trigonometric deformation of the generalized symmetric space $\sigma$-model, with target $(G \times H) / (G_0 \times H) = G/G_0$.  We quotient by $G_0 \times H$ because the $\Z_n$ action on $G \times H$ is the identity on $H$, so the fixed subgroup is $G_0 \times H$. 

\begin{conjecture}
	The construction sketched above yields the known \cite{Delduc:2013fga} trigonometric deformations of the symmetric space $\sigma$-models. 
\end{conjecture}

\section{Higher-Genus Spectral Curves}
\label{sec:higher_genus}

The discussion to this point was restricted to cases where the spectral curve has either genus $0$ or $1$.
However, as it hopefully clear by now, our construction can be generalized to general higher-genus spectral curves. In this section we obtain a novel class of integrable two-dimensional field theories whose spectral curve is of higher genus.\footnote{See e.g.\ \cite{Krichever:2001zg} for previous study of zero-curvature equations on higher-genus spectral curves.}

Higher-genus spectral curves were not considered in 
the discussion of integrable lattice models in \cite{Costello:2017dso,Costello:2018gyb}. This is because the zeros of the 
one-form $\omega$ requires non-topological boundary conditions
and hence breaks the topological invariance, which was crucial for the explanation of the Yang-Baxter equation.
As explained in \cite{Costello:2017dso}, this is consistent with the classification of the quasi-classical $r$-matrix \cite{Belavin-Drinfeld}. Such a restriction, however, does not apply to integrable field theories---we indeed have 
used non-topological (holomorphic and anti-holomorphic) boundary conditions in previous sections.
For this reason our framework 
generates even richer class of theories when applied to integrable field theories.

One can ask why models with higher-genus spectral curves will be useful.  It turns out that the very simplest generalizations of our construction of the Riemannian symmetric space model leads us immediately to models of this type.

\subsection{Further Generalizations of Riemannian Symmetric Space Models} 

Our constructions of symmetric space models in section \ref{sec:symmetric} (as well as generalized symmetric space
models in section \ref{sec:gss}) involved the spectral curve $\CP^1$, with the one-form $\d z$ and a branch-cut connecting points $z_0,z_1$.  We can try to generalize this in the obvious way: we can place many more points $z_i$ with branch-cuts connecting them. 

In the simplest case, we have a group $G$ with a $\Z_2$ action.  Then we can place $2n$ defects at points $q^{w}_i, q^{\wbar}_i$.  As in our construction of the symmetric space $\sigma$-model, we connect $q^{w}_i$ to $q^{\wbar}_i$ by a line $[q^w_i,q^{\wbar}_i]$.  We assume that these lines do not intersect.  As we cross the line $[q^w_i,q^{\wbar}_i]$ we apply the involution $\rho$ of the gauge group $G$.  

This data gives us a double cover $\Sigma$ of $\CP^1$, branched at $q^w_i, q^{\wbar}_i$.  We let $\til{q}^w_i$, $\til{q}^{\wbar}_i$ be the inverse image of these points on the double cover $\Sigma \to \CP^1$. The inverse image of $\infty$ gives two points $p_1,p_2 \in \Sigma$. The pull-back of the one-form $\d z$ on $\CP^1$ gives a one-form $\omega_{\Sigma}$ on $\Sigma$, with first-order zeroes at $\til{q}^{w}_i$, $\til{q}^{\wbar}_i$, and second order poles at $p_1,p_2$.

The multi-defect generalization of the symmetric space model is given by studying the $\sigma$-model engineered from the gauge theory on $\Sigma$, and passing to the $\Z_2$ fixed points under the action which applies $\rho$ to $G$ and the deck-transformation automorphism to $\Sigma$. 

The surface $\Sigma$ has genus $g = n-2$, if there are $n$ pairs of branch points $q_i^{w}$, $q^{\wbar}_i$.  In order to understand these models, we need to first understand how to build integrable field theories from four-dimensional Chern-Simons theory compactified on a higher-genus surface. 

We could also try to understand elliptic versions of Riemann symmetric space models, where we put four-dimensional Chern-Simons theory on an elliptic curve with no poles or zeroes, and introduce a branch-cut where we apply the involution on the group.  This leads to a model living on a double cover of the elliptic curve branched at two points: this is a genus $2$ curve.  

Similarly, we could consider the elliptic generalization of the $AdS_5 \times S^5$ $\sigma$-model. In this example, we would have four-dimensional Chern-Simons theory with gauge group $PSU(2,2 \, | \, 4)$ on a four-fold branched cover of the elliptic curve, branched at two points. The total space of the branched cover is of genus $4$, with a one-form with two order $3$ zeroes.  As before, we introduce chiral and anti-chiral boundary conditions at the zeroes, and consider the fields invariant under the $\Z_2$ action which simultaneously acts on $PSU(2,2 \, | \, 4)$ and on the $4$-fold cover. It would be interesting to consider the implications of the integrability of the resulting sigma model
to (an elliptic deformation of) the AdS/CFT correspondence.

\subsection{Gauge Theory Set-up for Higher Genus Models}

Consider our four-dimensional gauge theory on the product of some Riemann surface $C$ of genus $g > 1$ with $\R^2$.  We must equip $C$ with a holomorphic one-form, which for now we assume has only simple zeroes and double poles. Away from the zeroes of $\omega$, the theory is defined as before, with Lagrangian $\int \omega \wedge {\rm CS}(A)$ and gauge-field $A$ which has no $(1,0)$ component in the $C$ direction. 

At the zeroes of $\omega$ we choose the boundary condition as in section \ref{subsec.omega_zero}: one of $A_w$ or $A_{\wbar}$ has a simple pole. At the poles, we ask that all components of the gauge field vanish.  

We let $q_i^{w}$, $q_i^{\wbar}$ be the zeroes of the one-form $\omega$, where $A_w$ (respectively, $A_{\wbar})$ has a first-order pole at $q_i^w$, $q_i^{\wbar}$.  We let $p_i$ be the location of the second-order poles of the one-form $\omega$, and we ask that all components of the gauge field vanish at each $p_i$. We let 
\begin{equation} 
	\begin{split}
		D_w &= \sum q_i^w - \sum p_j \;, \\
		D_{\wbar} &= \sum q_i^{\wbar} - \sum p_j  \;,\\
		D_{\zbar} &=  \sum p_j \;.
	\end{split}
\end{equation}
Thus, the fields $A_{w}$ (respectively $A_{\wbar}$) are sections of $\Oo(D_w)$, $\Oo(D_{\wbar})$ and $A_{\zbar}$ is a $(0,1)$ form with values in $\Oo(-D_{\zbar})$.   The one-form $\omega$ is a nowhere-vanishing section of $K_C (-D_w - D_{\wbar})$ ($K_C$ is the canonical bundle of $C$), so that $K_C \iso \Oo(D_w + D_{\wbar})$. 

We were unable to find a clean description of the integrable field theory for the most general choice of boundary conditions. We leave this as a challenge for future work. Here we will focus on the case that 	both $D_w$ and $D_{\wbar}$ have degree $g-1$ (recall that $K_C$ has degree $2g-2$). In this case we find the (analytically-continued) $\sigma$-model on a moduli space of $G$-bundles on $C$, trivialized on $D_{\zbar}$.  This moduli space is equipped with a certain complex-analytic metric and closed three-form, which we will describe explicitly.

\subsection{\texorpdfstring{The Case $D_w$, $D_{\wbar}$ Have Degree $g-1$}{The Case D(w), D(wbar) Have Degree g-1}}
Let $\op{Bun}_G(C,D_{\zbar})$ be the moduli space of holomorphic $G$-bundles on $C$, trivialized on the points $p_i$ which form the divisor $D_{\zbar}$.  

\begin{definition}
Let $\op{Bun}_G^0(C,D_{\zbar})$ be the open subset of the moduli space of $G$-bundles trivialized on $D_{\zbar}$ with the property that $H^0(C, \g_P(D_{\wbar})) = 0$.  
\end{definition} 
Here $\g_P=(P\times \g)/G$ is the adjoint bundle of $G$ associated with the bundle $P$, and 
$\g_P(D_{\wbar})=\g_P \otimes \Oo(D_{\wbar})$ allows for poles along the divisor $D_{\wbar}$.

Note that by the Riemann-Roch theorem and Serre duality\footnote{The Riemann-Roch theorem states
\begin{equation} 
	\op{dim} H^0(C, \g_P(D_{\wbar}))- \op{dim} H^0(C, \g_P\otimes K_C(-D_{\wbar})) =\mathrm{deg}(D_{\wbar})-g+1\;.
\end{equation}
The right hand side of this equation is zero in our case since $\mathrm{deg}(D_{\wbar})=g-1$.}
\begin{equation} 
	\op{dim} H^0(C, \g_P(D_{\wbar}))= \op{dim} H^0(C, \g_P\otimes K_C(-D_{\wbar})) \;.
\end{equation}
Further, $K_C(-D_{\wbar}) = \Oo(D_w)$, so if $P$ is in $\op{Bun}_G^0(C,D_{\zbar})$ we have $H^0(C,\g_P(D_w)) = 0$ also.  
This is why we did not explicitly impose the condition $H^0(C,\g_P(D_w)) = 0$ in the definition above.

To write down the $\sigma$-model with target $\op{Bun}_G^0(C,D_{\zbar})$, we need to endow this complex manifold with a complex-analytic metric and with a complex-analytic closed three-form.  Everything is complex analytic because we are working in an analytically continued setting. An ordinary real $\sigma$-model target will be obtained by passing to an appropriate real slice. 

A metric and a three-form are the only possibilities because the two-dimensional theory we are constructing is classically conformal.

\subsection{The Metric and Three-Form}

To describe the metric and three-form, we will calculate their value on the tangent space to any point $P \in \op{Bun}_G^0(C,D_{\zbar})$.  We do this by analyzing the effective two-dimensional theory in perturbation theory around the constant map to $P$. Therefore we assume that the two-dimensional field 
\begin{equation} 
	\sigma : \R^2 \to \op{Bun}_G^0(C,D_{\zbar}) 
\end{equation}
is a perturbation of the constant map with value $P$. We will represent this by a gauge field 
\begin{equation} 
	A^{\sigma}_{\zbar} \in   \Omega^{0,1}\left(C, \g_P(-D_{\zbar})\right) \;.
\end{equation}
We will choose the gauge where $A^{\sigma}_{\zbar}$ is anti-holomorphic in $z$.
In this gauge, there is no remaining gauge symmetry.  

We will follow our usual procedure, and use the equations $F_{w\zbar} = 0$, $F_{\wbar \zbar} = 0$ to solve for the fields $A_w, A_{\wbar}$ in terms of $A^{\sigma}_{\zbar}$.  Reinserting these fields into the Lagrangian, we  will obtain the effective action.  It is important to note that, since $H^0(C,\g_P(D_w)) = H^0(C,\g_P(D_{\wbar})) = 0$, there is a unique $A_w,A_{\wbar}$ solving the equations of motion, given $A_{\zbar}$.

To write the expression for $A_w$, $A_{\wbar}$, we need to introduce the Sz\"ego kernel. Given any kernel  
\begin{equation} 
	\mc{G} \in H^0 \left(C \times C, \g_P \otimes \g_P \otimes \Oo (D_w \times C + C \times D_{\wbar} + \tr_C )  \right)\;,
\end{equation}
where $\tr_C$ is the diagonal, we can define the residue
\begin{equation} 
	\op{Res}_{\tr} \mc{G} \in H^0(C, \g_P \otimes \g_P) \;. 
\end{equation}
The definition of the residue uses the fact that $D_w + D_{\wbar} = K_C$, so that in local coordinates $z_1,z_2$ on the two copies of $C$, the coefficient of $\frac{1}{z_1 - z_2}$ in $G$ transforms as a section of the canonical bundle of $C$.

The quadratic Casimir $c_{\g}$ is $G$-invariant, and so defines an element
\begin{equation} 
	c_{\g} \in H^0(C, \g_P \otimes \g_P) \;.
\end{equation}
\begin{definition}
	The Sz\"ego kernel $\mc{G}$ is the unique element 
	\begin{equation} 
		\mc{G} \in H^0 (C \times C, \g_P \otimes \g_P \otimes \Oo (D_w \times C + C \times D_{\wbar} + \tr_C )  )
	\end{equation}
	such that
\begin{equation} 
	\op{Res}(\mc{G}) = c_{\g} \;. 
\end{equation}
\end{definition}
Existence and uniqueness of the Sz\"ego kernel follow from the long exact sequence of sheaf cohomology groups associated to the short exact sequence of sheaves on $C \times C$
\begin{align}
\begin{split} 
	&0 \to  \g_P \otimes \g_P \otimes \Oo (D_w \times C + C \times D_{\wbar}) \\
	&\qquad \to  \g_P \otimes \g_P \otimes \Oo (D_w \times C + C \times D_{\wbar} + \tr_C) \xrightarrow{\op{Res}} \Oo_{\tr} \otimes \g_P \otimes \g_P \to 0 \;.
\end{split}  
\end{align}
The main point is that our assumptions on $P$ guarantee that the first sheaf in this sequence has no sheaf cohomologies of degree $0$ and $1$.

Kernels similar to $\mc{G}(z,z')$ were studied in Fay \cite{fay1992nonabelian, fay2006theta}, who succeeded in expressing them explicitly in terms of $\theta$-functions. 

As an example, if $P$ is the trivial bundle, then we can view $\mc{G}$ as being a meromorphic function on $C \times C$, multiplied by the quadratic Casimir $c_{\g}$.  The meromorphic function $\mc{G}(z_1,z_2)$ has first order poles at $z_1 = z_2$, $z_1 = q_i^w$, $z_2 = q_i^{\wbar}$ and zeroes at $z_1 = p_i$ or $z_2 = p_i$.  The residue
\begin{equation} 
	\oint_{\abs{u} = \eps} \mc{G}(z+u, u) \, \omega_u  
\end{equation}
is normalized to be $2 \pi \i$. 

Now, once we have the Sz\"ego kernel, we can solve for $A_w$, $A_{\wbar}$ in terms of $A_{\zbar}^{\sigma}$, in series in $A_{\zbar}^{\sigma}$.  To linear order in $A_{\zbar}^{\sigma}$ we have
\begin{equation} 
	\begin{split}
		A_w(z) &= \int_{z'} \mc{G}(z,z') \omega(z') \partial_w A^{\sigma}_{\zbar'} \;,\\
		A_{\wbar}(z) &= -\int_{z'} \mc{G}(z',z') \omega(z') \partial_{\wbar} A^{\sigma}_{\zbar'} \;.
	\end{split} \label{eqn_hg_soln}
\end{equation}
To see that these are the correct expressions, we note that $A_w(z_1)$ has a first-order pole at $z_1 = q_i^w$ and and vanishes at $z_1 = p$.  If we apply the $\dbar$-operator to $A_w(z)$, twisted by the bundle $\g_P(D_w)$, we find
\begin{equation} 
	\dbar_{\zbar} A_w(z) =  \int_{z'} \dbar \mc{G}(z,z') \omega(z') \partial_w A^{\sigma}_{\zbar'} \;.
\end{equation}
Only the pole in $\mc{G}(z,z')$ at $z = z'$ contributes to this expression, because the $\dbar$ operator is twisted by $\g_P(D_w)$ and, as a section of $\g_P(D_w) \boxtimes \g_P(D_{\wbar})$, the only poles in $\mc{G}(z,z')$ are at $z = z'$.

The statement that the residue of $\mc{G}(z,z')$ along $z = z'$ is $1$ tells us that this recovers $\partial_w A^{\sigma}_{\zbar}$.  Indeed,  away from the points $p_i, q_i^w,q_i^{\wbar}$ we can choose a coordinate $z$ so that $\omega = \d z$, in which case 
\begin{equation} 
	\mc{G}(z,z')\omega(z') = \frac{1}{2 \pi \i } \mbfn{t}_a \mbfn{t}_a (z - z')^{-1} \d z'  + \text{ non-singular terms }  \;,
\end{equation}
where $\mbfn{t}_a$ is an orthonormal basis for the Lie algebra. This makes it clear that we have 
$\dbar_{z} \mc{G}(z,z') \omega(z') = \delta_{z = z'}$. 

This gives us expressions for $A_w,A_{\wbar}$ in terms of $A_{\zbar}^{\sigma}$ to linear order in $A_{\zbar}^{\sigma}$.  If we were to work to all orders in $A_{\zbar}^{\sigma}$, we would find a more complicated expression of the form
\begin{equation} 
	A_w(z) = \int_{z'} \mc{G}(z,z') \omega(z') \partial_w A^{\sigma}_{\zbar'} - \int_{z', z''}  \mc{G}(z,z') A^{\sigma}_{\zbar'} \omega(z') \mc{G}(z',z'') \partial_w A^{\sigma}_{\zbar''} \omega(z'') + \dots \;,
\end{equation}
where we have omitted terms cubic and higher in $A_{\zbar}^{\sigma}$. There is a similar expression for $A_{\wbar}(z)$. 

To obtain the effective two-dimensional interaction, we should insert our expressions for $A_{w}, A_{\wbar}$ into the Chern-Simons Lagrangian.  We are here only interested in the metric and the three-form on the tangent space of our chosen point $P \in \op{Bun}_G^0(C,D_{\zbar})$. These are given by the terms in the Chern-Simons Lagrangian, when expanded as a series in $A_{\zbar}^{\sigma}$, which are quadratic and symmetric, or cubic and anti-symmetric, in $A^{\sigma}_{\zbar}$.  

Let us choose a basis of $H^1(C,\g_P(-D_{\zbar}))$.   Let $A^{\alpha}_{\zbar}$ be representatives of the basis elements.  Let $A_w^{\alpha}$, $A_{\wbar}^{\alpha}$ be the fields solving $F_{w \zbar} = 0$, $F_{\wbar \zbar} = 0$, as above.  We write $A_{w}^{\alpha}$, $A_{\wbar}^{\alpha}$ to linear order in $A^{\alpha}_{\zbar}$. This analysis tells us that the kinetic term is given by 
\begin{equation} 
\int_{z \in C} A_{w}^{\alpha} \dbar_{\zbar} A_{\wbar}^{\beta} \omega(z) \;.
\end{equation}
Using the equation $\dbar_{\zbar} A_{\wbar}^{\beta} = \partial_{\wbar} A^{\beta}_{\zbar}$, we can rewrite this as
\begin{equation} 
	\int_{z \in C} A_{w}^{\alpha} \partial_{\wbar} A_{\zbar}^{\beta} \omega(z) = 	 	\int_{z_1,z_2 \in C} \omega(z_1) \omega(z_2) \mc{G}(z_2,z_1) \partial_w A^{\alpha}_{\zbar}(z_1) \partial_{\wbar} A^{\beta}_{\zbar}(z_2)   \;.
\end{equation}%
This tells us that the metric is
\begin{equation} 
	g^{\alpha \beta} =  	\int_{z_1,z_2 \in C} \omega(z_1) \omega(z_2) \mc{G}(z_1,z_2) A^{\alpha}_{\zbar}(z_1) A^{\beta}_{\zbar}(z_2) \;.  \label{eqn_metric} 
\end{equation}

To understand the three-form, let us construct solutions $A_w, A_{\wbar}$ to the equations $F_{\wbar \zbar} = F_{w \zbar} = 0$, in series in $A_{\zbar}$.  The series expansion is of the form
\begin{equation} 
	A_w = A_w^{(1)} + A_{w}^{(2)} + \dots \;,
\end{equation}
where $A_w^{(k)}$ is an order $k$ expression in $A_{\zbar}^{\sigma}$. Define $A_{\wbar}^{(k)}$ similarly.  There is a unique series expansion of this form satisfying the equations $F_{w \zbar} = F_{\wbar \zbar} = 0$.

Let us insert the fields $A_w, A_{\wbar}$ into the Chern-Simons action. Keeping only the terms cubic in $A_{\zbar}^{\sigma}$, we get
\begin{multline} 
	- \int \omega A_{w}^{(2)} \partial_{\zbar} A_{\wbar}^{(1)} +   \int \omega A_{\wbar}^{(2)} \partial_{\zbar} A_{w}^{(1)} \\
	+  \int\omega A_{w}^{(2)} \partial_{\wbar} A_{\zbar}^{(\sigma)}  - \int\omega A_{\wbar}^{(2)} \partial_{w} A_{\zbar}^{(\sigma)}\\
	+ \int \omega A_w^{(1)}2 A_{\wbar}^{(1)} A_{\zbar}^{\sigma} \;.
\end{multline}
Using the equations $F_{w \zbar} = 0$, $F_{\wbar \zbar} = 0$, the first four terms cancel, and we have simply $\int \omega A_{w}^{(1)} A_{\wbar}^{(1)} A_{\zbar}^{\sigma})$.   

Using the expression for $A_{w}^{(1)}$, $A_{\wbar}^{(1)}$ \eqref{eqn_hg_soln}, we deduce that the three-form is given by 
\begin{align}
\begin{split}
	C_{\alpha \beta \gamma} &= \text{anti-symmetrization of }\\
	& \quad \int_{z_1,z_2,z_3 \in C} \omega(z_1) \omega(z_2) \omega(z_3) \mc{G}(z_1,z_2) \mc{G}(z_3,z_1)\left(  A^{\alpha} (z_1) A^{\beta}(z_2) A^{\gamma}(z_3) \right). 
\end{split}
\end{align}

\subsection{Gauge-Invariance of the Metric}
It is not obvious that the expression \eqref{eqn_metric} is invariant under linearized gauge transformations of the fields $A^{\alpha}_{\zbar}$. Gauge-invariance of the metric means it descends to cohomology $H^1(C,\g_P(-D_{\zbar}))$. We know that this must be the case for abstract reasons we explained before, but here we will check gauge invariance explicitly as a sanity check.

If we perform a gauge transformation
\begin{equation} 
	A^{\alpha}_{\zbar} \mapsto A^{\alpha}_{\zbar} + \dbar_{\zbar} \chi  \;,
\end{equation}
then the integral varies to
\begin{equation} 
	\delta g_{\alpha \beta} = - \int_{z_1,z_2 \in C}   \chi (z_1) A^{\beta}_{\zbar}(z_2)  \dbar \left(   \omega(z_1) \omega(z_2) \mc{G}(z_1,z_2) \right) \;.
\end{equation}
This integral picks up $\delta$-function contributions from the locations of the poles of $\omega(z_1) \omega(z_2) \mc{G}(z_1,z_2)$. Only the poles in the $z_1$ variable contribute, because of the presence of $\d \zbar_2$ in $A^{\beta}(z_2)$.  

The poles can occur at the points where $z_{1} = p_i$, $z_{1} = q_i^{w}$ or $z_1 = z_2$. Let us look at the poles where $z_{1} = p_i$ first. The one-form $\omega(z_1)$ has a second-order pole at $p_i$, and $\mc{G}(z_1,z_2)$ has a first-order zero at $z_1 = p_i$, leaving us with a first-order pole at $z_1 = p_i$. This means that we pick up a $\delta$-function at $z_1 = p_i$.  However, the result vanishes, because $\chi(z_1)$ vanishes where $z_1 = p_i$. 
 
The first-order poles in $\mc{G}$ at $z_{1} = q_i^w$ cancel with the first-order zeroes in $\omega(z_1)$ there.  

The only remaining poles and zeroes are those in $\mc{G}(z_1,z_2)$ at $z_1 = z_2$, where there is a first-order pole. These will cancel for symmetry reasons.  The analysis is local, so we will work in a coordinate patch away from poles and zeroes of $\omega$. We can write locally $\mc{G} = \delta_{ab} \frac{1}{z_1 - z_2}$ (where $a,b$ are Lie algebra indices) and $\omega = \d z$.  We will change notation a little, and write the $(0,1)$ form whose inner product we are computing as $A^a \mbfn{t}_a$ where $\mbfn{t}_a$ is an orthonormal basis of the Lie algebra. Then, in these coordinates, the expression we need to check is gauge invariant is
\begin{equation} 
	\int_{z_1,z_2} \frac{1}{z_1 - z_2} A^a(z_1) A^a(z_2) \d z_1 \d z_2 \;.
\end{equation}
This is unchanged under the variation $A^a \mapsto \dbar \chi^a$, for sign reasons.   
One can show that the three-form is gauge invariant by similar computations.

\subsection{Reality Conditions}
So far we have described the two-dimensional theory associated to the Riemann surface $C$ as a $\sigma$-model with an analytically-continued target.  To write this as an ordinary $\sigma$-model, we need to choose a real slice of the target.  Let us assume that $C$ is equipped with an anti-holomorphic involution $\rho$, and that the one-form $\omega$ is fixed by $\rho$.  For simplicity, let us assume that the points $p_i$, $q_i^w$, $q_i^{\wbar}$ are all fixed by $\rho$ (this condition can be relaxed).  Let us fix the anti-holomorphic involution on the complex group $G_{\C}$ so that the fixed points are the compact group $G_c$.  

The anti-holomorphic involutions $\rho$ on $C$ and on the group $G_{\C}$ induce one on the moduli space $\op{Bun}_G(C)$.   We can then study, as in \cite{BiswasHuismanHurtubise}, the moduli space of holomorphic $G$-bundles on $C$ fixed by this involution. This will be denoted $\op{Bun}_G(C)(\R)$.  Since the points $p_i$ are fixed by $\rho$, one can also study the moduli space $\op{Bun}_G(C,D_{\zbar})(\R)$ of $G$-bundles on $C$, trivialized at $p_i$, and fixed by this involution.

There is a nice description of the manifold $\op{Bun}_G(C)(\R)$ given in \cite{BiswasHuismanHurtubise}.  The result of Narasimhan-Seshadri \cite{NarasimhanSeshadri} states that the moduli of stable $G_{\mathbb{C}}$-bundles on $C$ is given by conjugacy classes of homomorphisms $\pi_1(C,p) \to G$ from the fundamental group of $C$ to the compact group $G$.

The real locus $\op{Bun}_G(C)(\R)$ is, according to \cite{BiswasHuismanHurtubise}, the space of conjugacy classes of homomorphisms $f : \pi_1(C,p) \to G$ such that $f \circ \rho = f$, where $\rho : \pi_1(C,p) \to \pi_1(C,p)$ is the map induced by the complex conjugation on $C$.  (We are assuming here that $\rho(p) = p$). A similar description holds when the bundle is trivialized at the points $p_i \in D_{\zbar}$.

\subsection{Properties of the Lax Matrix}
As usual, the expectation value of the four-dimensional gauge field defines the Lax matrix:
\begin{equation}
	\begin{split}
		\Lax_w(z) &= \ip{A^{\sigma}_w(z)} \;, \\
		\Lax_{\wbar}(z) &= \ip{A^{\sigma}_{\wbar}(z)} \;, \\
	\end{split}
\end{equation}
where $\sigma: \R^2 \to \op{Bun}_G^0(C,D_{\zbar})$ is the fundamental field.  The equation $F_{w \wbar}$ of the four-dimensional gauge theory implies immediately that the zero-curvature equation for the Lax matrix holds once we impose the equations of motion for $\sigma$.

In these models, we can not choose a gauge in which $A_{\zbar}^{\sigma} = 0$. For a general point of $\op{Bun}_G^0(C,D_{\zbar})$, the corresponding $G$-bundle on $C$ is not trivial.   

This means that we should not try to write the Lax matrix as being a one-form defining a connection on a trivial bundle.   In these more elaborate examples, it is more convenient to formulate the Lax operator in a slightly more abstract form. 

The Lax operator is a quantity that associates to every field configuration
\begin{equation} 
	\sigma : \R^2 \to \op{Bun}_G^0(C,D_{\zbar})
\end{equation}
a principal $G$-bundle $P_{\sigma}$ on $\R^2 \times C$.  This principal $G$-bundle is holomorphic on $C$, meaning that it has holomorphic transition functions. 

This bundle is equipped with a connection $\nabla_{\sigma}$ along the $\R^2$ direction, compatible with the holomorphic structure in the $C$ direction. The connection has poles along the zeroes of $\omega$. 

The Lax equation is then the statement that $\nabla_{\sigma}$ is flat whenever $\sigma$ satisfies the field equations.   

It is obvious, abstractly, how we associate such a principal bundle with connection to a field configuration $\sigma : \R^2 \to \op{Bun}_G^0(C,D_{\zbar})$.  The bundle is the pull-back of the universal bundle on $\op{Bun}_G^0(C,D_{\zbar}) \times C$.  We obtain the connection $\nabla_{\sigma}$ using the fact that every field configuration of the two-dimensional model gives rise to a gauge equivalence class of a field configuration of four-dimensional Chern-Simons theory satisfying $F_{\zbar w} = F_{\zbar \wbar} = 0$.  The Lax equation holds because $F_{w \wbar} = 0$ is equivalent to the two-dimensional  field equations. 

\subsection{The Lax Connection as Geometric Structure on the Target Space}

An important feature of the Lax connection $\nabla_{\sigma}$ associated to a field configuration $\sigma$  is that it only depends on the first derivative $\d \sigma$ of $\sigma$ (this follows from the fact that classically, four-dimensional Chern-Simons theory is scale invariant on the $w$-plane). 

Let us trivialize the universal bundle $P$ in some patch $U \times V$ of $\op{Bun}_G^0(C,D_{\zbar}) \times C$, and consider a field configuration $\sigma$ lying in $U$.  Then the connection $\nabla_{\sigma}$ is described by a $\g$-valued one-form on $\R^2$, depending holomorphically on $z \in V$.   At each point in $\R^2$, this one-form only depends on the first derivative of $\sigma$, i.e.\ the pair of holomorphic tangent vectors $\partial \sigma, \dbar \sigma \in T U$. 

This tells us that, in terms of the target space geometry, the Lax connection is described by a pair of adjoint-valued one-forms on $U$, depending holomorphically on $z \in V$.  If we change trivializations, both of these one-forms transform as connections.  

We conclude that in terms of the target space geometry,  the $w$ and $\wbar$ components of the Lax connection are each pulled back from an independent connection on the target space $\op{Bun}_G^0(C,D_{\zbar})$.

This means that the full connection $\nabla_{\sigma}$ can be written in terms of the following data:
\begin{enumerate} 
	\item A principal $G$-bundle $P$ on $\op{Bun}_G^0(C,D_{\zbar}) \times C$.  This is the universal bundle.  Along $C$, this bundle is holomorphic, i.e.\ has holomorphic transition functions.  Since we are working in an analytically continued context, in which we treat $\op{Bun}_G^0(C,D_{\zbar})$ as a complex manifold, the bundle $P$ is also holomorphic on $\op{Bun}_G^0(C,D_{\zbar})$. 
	\item A pair of holomorphic connections $\nabla^+$, $\nabla^-$ in the $\op{Bun}_G^0(C,D_{\zbar})$ directions in this bundle. We expect that $\nabla^+$ is  defined in the complement of the points $q_i^w$, where $\Lax_w(z)$ has poles, and $\nabla^-$ is defined in the complement of $q_i^{\wbar}$.
\end{enumerate}
Everything here, and throughout this section, is holomorphic on $\op{Bun}_G^0(C,D_{\zbar})$ because we are working in an analytically continued setting.  

This target-space data gives rise to the Lax connection $\nabla_{\sigma}$ as follows.
If we have a map $\sigma : \R^2 \to \op{Bun}_G^0(C,D_{\zbar})$, we get a principal $G$-bundle $\sigma^\ast P$ on $\R^2 \times C^0$, by pulling back the universal bundle on $\op{Bun}_G^0(C,D_{\zbar}) \times C^0$. Here $C^0 \subset C$ is the complement of the points $q_i^w$, $q_i^{\wbar}$.   This bundle is holomorphic along $C$. We define a connection $\nabla_{\sigma}$ on $\sigma^\ast P$ by saying that 
\begin{equation} 
	\begin{split}
		\nabla_{\sigma,w} &= \sigma^\ast \nabla^+_w \;,\\
		\nabla_{\sigma,\wbar} &= \sigma^\ast \nabla^-_{\wbar} \;.
	\end{split}
\end{equation}
The Lax equation states that the connections $\nabla^+$, $\nabla^-$ are such that $\nabla_{\sigma}$ is flat whenever $\sigma$ satisfies the equations of motion of the $\sigma$-model. 

\subsection{\texorpdfstring{Constructing the Connections $\nabla^+$, $\nabla^-$}{Constructing the connections nabla(+), nabla(-)}}
As the final step in our definition of the Lax connection, let us explain how to construct the connections $\nabla^+$, $\nabla^-$.  We will do so in a special basis for the tangent space of $\op{Bun}_G^0(C,D_{\zbar})$.

Fix a point $x$ of $\op{Bun}_G^0(C,D_{\zbar})$ corresponding to a principal $G$-bundle $P= P_x$ on $C$.  There is a short exact sequence of coherent sheaves on $C$
\begin{equation} 
	0 \to \g_P(-D_{\zbar}) \to \g_P(D_w) \to \bigoplus_j  \left(\g_P \otimes \Oo_{q_j^w}(q_j^w)  \right) \to 0 \;.\label{eqn:ses_lax} 
\end{equation}
Recall the divisor $D_w$ is $\sum q_j^w - \sum p_i$, and the points $q_i^w$ are where the field $A_w$ has a first order pole. The last term in this exact sequence is the sum of the stalks of $\g_P$ at $q_j^w$, twisted by the stalk of the line bundle $\Oo(q_j^w)$. 

We defined our moduli space $\op{Bun}_G^0(C,D_{\zbar})$ to consist of bundles with the property that $H^0(C, \g_P(D_w) ) = 0$. Since $D_w$ is of degree $g-1$, this implies (by Riemann-Roch theorem) that $H^1(C,\g_P(D_w)) = 0$ as well.  Now, taking the long exact sequence associated to the short exact sequence above, we find an isomorphism 
\begin{equation} 
	\bigoplus_j  \left(\g_P \otimes \Oo_{q_j^w}(q_j^w) \right) \iso H^1(C, \g_P(-D_{\zbar})) = T_x \op{Bun}_G^0(C,D_{\zbar}) \;.	 
\end{equation}
This isomorphism allows us to find an explicit collection of $(0,1)$ forms representing a basis of $T_x \op{Bun}_G^0(C,D_{\zbar})$.  

One way to find such representatives is as follows. Around each point $q_j^w$ we can choose a coordinate $z_j$, with the feature that the one-form $\omega$ is
\begin{equation} 
	\omega = z_j \d z_j 
\end{equation}
in this coordinate.  This coordinate is uniquely determined up to a sign change $z_j \to - z_j$.

Let us trivialize the bundle $\g_P$ near each $q_i^w$.  Then we can define $(0,1)$ forms on $C$, valued in $\g_P$, by the expressions
\begin{equation} 
	A_{\zbar}^{j,a} = \frac{1}{z_j} \delta_{\abs{z_j} = \eps} \mbfn{t}_{a} \;,
\end{equation}
where $\mbfn{t}_a$ is a basis of $\g$.   
The exact sequence given above implies that these elements form a basis for $H^1(C, \g(-D_{\zbar}))$.   

We will use this gauge to write down the connections $D_+$, $D_-$, and so the Lax operator.  If we vary the point $x \in \op{Bun}_G^0(C,D_{\zbar})$ to first order by
\begin{equation} 
	x \mapsto x + \sum \lambda_{j,a} [A_{\zbar}^{j,a}]  \;,
\end{equation}
then the principal $G$-bundle $P_x$ on $C$ is unchanged away from the points $q_j^w$. That is, for $z \in C \setminus \{q_j^w\}$, we have an isomorphism
\begin{equation} 
	P_{x,z} \iso P_{x + \sum \lambda_{j,a} [A_{\zbar}^{j,a}], z} \;.
\end{equation}

This isomorphism is precisely what is needed to give a connection on the holomorphic principal $G$-bundle $P_z$ on $\op{Bun}_G^0(C,D_{\zbar})$. After all, one abstract definition of a connection is that it gives an isomorphism between the fibre of a bundle at any point and the fibre at a first-order variation of that point.

This connection is manifestly holomorphic, and in fact algebro-geometric.  All that we have really used to defined it is the short exact sequence \eqref{eqn:ses_lax}.  Let us call this connection $\nabla^+$.

To show that this is indeed the connection giving rise to the $w$-component of the Lax operator, let us calculate the $A_w$  in this gauge, using the kernel given above. 

The linearized equation of motion for $A_w$ tells us that
\begin{equation} 
	\partial_{\zbar} A_w = \partial_w A_{\zbar} \;.
\end{equation}
If $A_{\zbar} = \sum \lambda_{j,a} A_{\zbar}^{j,a}$, where $\lambda_{j,a}$ are functions of $w,\wbar$, this equation is solved by
\begin{equation} 
	A_w = \sum \delta_{\abs{z_j} \le \eps} \frac{1}{z_j} \partial_w \lambda_{j,a} \mbfn{t}_a \;.
\end{equation}
Therefore $A_w = 0$ away from the points $q_j^w$.  We conclude that in this gauge the $w$-component of the Lax operator vanishes away form $q_j^w$.  Since this is the gauge in which $\nabla^+$ is trivial (to leading order), we find that $\nabla^+$ is indeed the connection which gives rise to the Lax operator $\Lax_w$. 

The connection $\nabla^-$ is defined in a similar way, except using the points $q_i^{\wbar}$ instead of $q_i^w$.  

\subsection{ An Explicit Example}
So far, we have explained how to construct an integrable $\sigma$-model with target the real locus of the moduli of $G$-bundles on a Riemann surface $C$.  Our discussion was a little abstract, and we did not present a closed global expression for the metric and three-form.  Instead, we gave their value on the tangent space at each point of the target space.  

In this section, we will discuss an example where things can be made somewhat more explicit.   Let us take our curve $C$ to be an elliptic curve $E$, equipped with an anti-holomorphic involution which fixes the circle where the coordinate $z$ is real.  This means the modulus $\tau$ is purely imaginary. We take the one-form to be $\omega = \wp(z) \d z$, where $\wp(z)$ is the  Weierstrass $\wp$-function. This has a second order pole at the origin, and two zeroes at $q,-q$ where $\br{q} = -q$.  It is known \cite{eichler1982zeros} that the real part of $q$ is $1/2$. 

We will take the gauge group to be $SU(2)$.  We are interested in the real locus of the moduli of holomorphic $SL_2(\C)$ bundles on $E$, trivialized at the origin. According to \cite{BiswasHuismanHurtubise}, such a bundle is described by a homomorphism $\rho$ from $\pi_1(E)$ to $SU(2)$, so that $\rho(A) = \rho(A)^{-1}$. This condition arises because the $A$-cycle -- which we take to be represented by the imaginary axis-- is reversed under the anti-holomorphic involution on $E$.   The solution $\rho(A) = 1$ is an isolated solution to the equation $\rho(A)^2 = 1$, so that a connected component of the moduli space is provided by $\rho(A) = 1$, $\rho(B)$ arbitrary.   

We have deduced that a connected component of the real locus of the moduli of $SL_2(\C)$-bundles on $E$, trivialized at $0$, is given by the group manifold $SU(2)$. Explicitly, these bundles can be parameterized by 
\begin{equation} 
	A_{\zbar}^{\sigma} = \tfrac{1}{2} (\log \sigma) \d \zbar \;.
\end{equation}
This is the $(0,1)$ component of the flat connection whose $z$-component is 
\begin{equation} 
	A_z^{\sigma} = \tfrac{1}{2} (\log \sigma) \d z \;.
\end{equation}
Note that $A_x$ component of this connection is $\log \sigma \d x$, and the $A_y$ component vanishes, so that the connection has no monodromy on the $A$-cycle as desired.

To do our computations, we will very slightly modify the boundary conditions of the gauge field at $z = 0$. This point is where the one-form $\wp(z) \d z$ has a double pole. Normally we ask that all components of the gauge field and of the gauge transformation are divisible by $z$ near this point. Here we will modify this to allow $A_{\zbar}$ to be arbitrary at $z = 0$, but allow the gauge transformation to be divisible by $z$ or $\zbar$ near $z = 0$.  This modified boundary condition is equivalent to the original one, because we can use the extra gauge transformations we have introduced to modify $A_{\zbar}$ to that it is divisible by $z$.  It is these modified boundary conditions that allow us to work in the gauge where $A_{\zbar} = \tfrac{1}{2} \log \sigma \d \zbar$.  

Recall that our metric and three-form are defined on an open subset $\op{Bun}_{SL_2(\C)}^0(E,0)$ of the moduli of bundles on $E$, trivialized at $0$. This is the open subset where the adjoint bundle has no sections with a first order pole at $0$ and a zero at $q$. Since $SU(2)$ is a connected component of the real locus of $\op{Bun}_{SL_2(\C)}(E,0)$, one can ask which elements of $SU(2)$ lie in the open subset $\op{Bun}_G^0(\sigma)$.    For $\lambda \in \mf{su}(2)$, the bundle defined by $\lambda \d \zbar$ is in this locus if the there are no sections of the adjoint bundle on $E$, holomorphic with respect to the twisted $\dbar$-operator $\dbar + \lambda \d \zbar$, which have a first order pole at $q$ and a first order zero at $0$.   This condition is equivalent to asking that there is no section with a first-order pole at $0$ and a zero at $q$. 

We will see that this condition is satisfied for all $\lambda$. To see this, let us diagonalize $\lambda$ by a gauge transformation, write $\lambda = \i \lambda_0 h$ for $h \in \mf{sl}_2(\C)$ being the standard basis element $\op{Diag}(1,-1)$, and $\lambda_0$ some real constant. The holomorphic bundle on $E$ associated to the adjoint representation breaks up as $\mc{L}_{\lambda} \oplus \Oo \oplus \mc{L}_{\lambda}^{-1}$,  according to the decomposition of $\mf{sl}_2(\C)$ into eigenspaces for $\lambda$.  Since $\lambda$ is proportional to $h \in \mf{sl}_2(\C)$, the line bundles $\mc{L}_{\lambda}$, $\mc{L}_{\lambda}^{-1}$ correspond to $e,f$.  Because the complex anti-linear involution on $\mf{sl}_2(\C)$ switches $\i e$ and $\i f$, the line bundle $\mc{L}$ is such that $\sigma^\ast \mc{L}_{\lambda} = \br{\mc{L}}_{\lambda}^{-1}$, where $\sigma$ is the anti-holomorphic involution on $E$.

The degree $0$ line bundle $\mc{L}_{\lambda}$ is the line bundle associated to some point $\lambda_0 p  \in E$, identifying in the standard way $E$ with its Jacobian. The reality condition on $\mc{L}_{\lambda}$ tells us that $\lambda_0 \br{p} = - \lambda_0  p$, so that $p$ is represented by a purely imaginary element in $\C$. 

We conclude that $\lambda_0 p \neq q$ for any value of $\lambda_0$ (where $q$ as before is one of the zeroes of $\wp(z)$) because the real part of $q$ is $1/2$.  This implies the condition that the adjoint bundle associated to $\lambda$ has no meromorphic sections with a pole at $q$ and a zero at the origin in $E$. 

From this, we deduce that our construction will give rise to some metric and three-form defined globally on the group manifold $SU(2)$. This will be invariant under the adjoint action of $SU(2)$.

We will describe the metric in coordinates provided by the exponential map from $\mf{su}_2$ to $SU_2$.
To write the metric, we will use, as before, the Sz\"ego kernel.   For each $\lambda \in \mf{su}_2$, the Sz\"ego kernel 
\begin{equation} 
	\mc{G}_{ab} (z_1,z_2, \lambda)
\end{equation}
(where $a,b$ are Lie algebra indices) is uniquely characterized by the following features. 
\begin{enumerate} 
	\item It has first order poles at $z_1 = q^w$, $z_2 = q^{\wbar}$, $z_1 = z_2$ 		and first order zeroes at $z_1 = 0$, $z_2 = 0$.  The residue at $z_1 = z_2$, defined using the one-form $\wp(z) \d z$, is $1$.
	\item It is doubly periodic in $z_1$ and $z_2$ separately.
	\item In each variable, is holomorphic for the $\dbar$ operator twisted by $\lambda$:
		\begin{equation} 
			\begin{split}
				\partial_{\zbar_1}  \mc{G}_{ab} + f_{acd} \lambda_{c}\mc{G}_{db} &= 0 \;,\\
						\partial_{\zbar_2}  \mc{G}_{ab} + f_{bcd} \lambda_{c}\mc{G}_{ad} &= 0  \;.
			\end{split}
		\end{equation}
\end{enumerate}
We were unfortunately unable to find a closed-form expression for $\mc{G}_{ab}(z_1,z_2,\lambda)$, even if abstractly we know it exists and is unique. 

To write down the $\sigma$-model Lagrangian, we view the fundamental field as $\lambda: \R^2 \to \mf{su}_2$ instead of $\sigma = \exp( 2 \lambda ) : \R^2 \to SU(2)$.  Then we have simple closed-form expressions for $A_w^\lambda$, $A_{\wbar}^{\lambda}$:
\begin{equation} 
	\begin{split}
		A_{w,a}(z_1)  &= \int_{z_2 \in E} \mc{G}_{ab}(z_1,z_2,\lambda) (\partial_w \lambda_b) \, \wp(z_2) \d z_2  \d \zbar_2 \;,\\
		A_{\wbar,a}(z_1)  &= -\int_{z_2 \in E} \mc{G}_{ab}(z_2,z_1,\lambda) (\partial_{\wbar} \lambda_b) \, \wp(z_2) \d z_2  \d \zbar_2 \;.
	\end{split}
\end{equation}
The zero-curvature equation
\begin{equation} 
	\partial_{\zbar} A_{w,a} + f_{abc} \lambda_b A_{w,c} = \partial_{w} \lambda_a 
\end{equation}
follows from the fact that $\mc{G}(z_1,z_2,\lambda)$ is holomorphic for the $\dbar$ operator twisted by $\lambda$, and has a first-order pole at $z_1 = z_2$ with residue $1$.

Using the equations $F_{\zbar w} = 0$, $F_{\zbar \wbar} = 0$, the Chern-Simons action takes the form
\begin{equation} 
	A_{\wbar} \partial_w A_{\zbar} + \tfrac{1}{2} A_w [A_{\wbar}, A_{\zbar} ] \;.
\end{equation}
Inserting our expressions for $A_{w}, A_{\wbar}, A_{\zbar}$ in terms of $\lambda : \R^2 \to \mf{su}_2$, we find that the effective Lagrangian is
\begin{align}
\begin{split} 
	S&=-    \partial_{\wbar} \lambda_a \partial_{w} \lambda_a  \int_{z_1,z_2 \in E \times E} \wp(z_1)  \wp(z_2)   \mc{G}_{ab}(z_1,z_2,\lambda) \d z_1 \d z_2 \d \zbar_1 \d \zbar_2  
	\\ 
	&+ \tfrac{1}{2}  \lambda_c \partial_w \lambda_{d}   \partial_{\wbar} \lambda_e      \int_{z_1,z_2,z_3 \in E^3} \wp(z_1) \wp(z_2) \wp(z_3) f_{abc} \mc{G}_{ad} (z_1,z_2,\lambda) \mc{G}_{be} (z_3,z_1,\lambda) \textstyle \prod \d z_i \textstyle \prod \d \zbar_i 
	\;.
\end{split}
\end{align}

\section{Gluing}
\label{sec:gluing}

So far, we have constructed new integrable field theories with a higher-genus spectral curve.   In our construction the holomorphic curve $C$ with one-form $\omega$ allows for complex structure
deformations. In particular we can consider degenerations of the curve
towards the boundary of the moduli space, where the curve degenerates into 
two components $C_1$ and $C_2$ connected by the cylinder (recall the discussion around \eqn\eqref{eq.level}
in section \ref{sec:WZW}). In this section we will show that, in the limit as the curve $C$ becomes degenerate, we can built the integrable field theory associated to $(C,\omega)$ from that associated to $(C_1,\omega_1)$ and $(C_2,\omega_2)$ by a certain BRST reduction. 

This procedure is reminiscent of the famous construction of four-dimensional $N=2$ theories of class $\mathcal{S}$ theories \cite{Gaiotto:2009we} by cutting and gluing.

\begin{figure}[htbp]
\centering
\includegraphics[scale=0.55,trim=30 0 0 0]{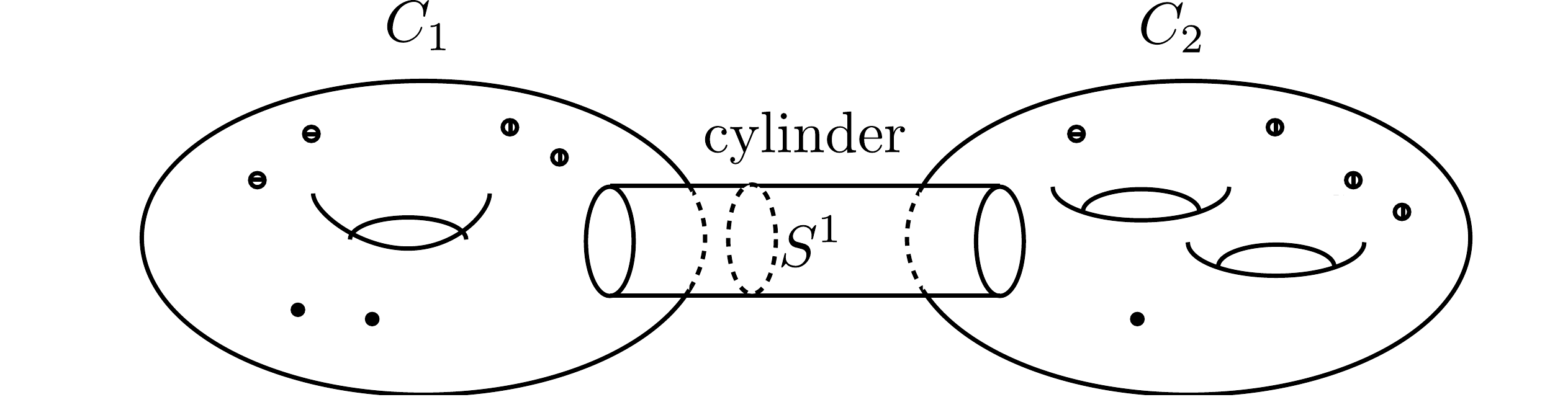}
\caption{We can glue two spectral curves $C_1$ and $C_2$ along a cylinder,
to obtain a new ``glued'' spectral curve $C$. Conversely, on the boundary of the moduli space of the 
spectral curve $C$, the curve $C$ degenerates into
two curves $C_1$ and $C_2$ connected by an infinite cylinder.
In this example we have $g_1=1, g_2=2, g=3$ and
$n_1=2, n_2=1, n=3$. }
\label{Fig_III_12}
\end{figure}

Let us first recall the ``plumbing fixture'' construction. Suppose that we have 
two punctured Riemann surface $C_1$ and $C_2$.
Let us choose points $P_{1,2}$, each from $C_1$ and $C_2$.
We assume that these points are different from the punctures of $C_{1,2}$.
Let us choose the local 
holomorphic coordinates $z_{1,2}$ at the points $P_{1,2}$, such that the points chosen correspond to $z_{1,2}=0$. 
We can then glue $C_1$ and $C_2$ by the 
identification of the two local neighborhoods via the relation
\begin{align}
z_1 z_2= e^{-t+\i \theta} \qquad (0\le t < \infty , 0\le \theta < 2\pi) \;.
\label{eq.plumbing}
\end{align}

For the discussion of integrable models 
we need to include the holomorphic one-form into this gluing discussion.
Note that in a coordinate neighborhood of the points $P_{1}$ and $P_2$  the holomorphic one-form 
can be written as 
\begin{align}
\omega_i=c_i \frac{\d z_i}{z_i} \quad (i=1,2) \;,  \quad c_i=\frac{1}{2\pi \i} \textrm{Res}_{P_i} \omega_i \;,
\end{align}
where the residue $c_i$ is a constant. This means that the 
gluing \eqn\eqref{eq.plumbing} is compatible with the choice of the one-form
if and only if  $c_1=-c_2$ or equivalently
\begin{align}
\textrm{Res}_{P_1} \omega_1 = -
\textrm{Res}_{P_2} \omega_2 \;.
\label{eq.residue}
\end{align}
When this condition is satisfied we can define the one-form $\omega$
on the glued surface $C$ by the relation
\begin{align}
\omega:= \omega_1=-\omega_2 \;.
\end{align}
Note that the minus sign in \eqn\eqref{eq.residue} originates from the 
orientation reversal of the local coordinate. 

We can of course describe the same process
in the opposite direction.

Suppose that  the spectral curve $C$ with a holomorphic one-form $\omega$
degenerates into two spectral curves $C_1, C_2$ connected by a 
cylinder.
The poles and the zeros of $\omega$ will be 
distributed into the poles and the zeros of the holomorphic one-forms $\omega_1$ and $\omega_2$ on 
$C_1$ and $C_2$. In addition, we expect that 
the degeneration will create first-ordered poles of $\omega$,
one for each curve $C_1$ and $C_2$, which have opposite residue. This means that on the cylinder the one-form
has two first order poles, and in a suitable holomorphic coordinate $z$ of $\mathbb{CP}^1$
the one-form can be written as $\omega=\d z/z$, with poles at $0$ and $\infty$.

If we give $C$ the natural metric $\omega \br{\omega}$ it becomes, as we saw in section \ref{sec:disorder}, a flat surface with conical singularities at the location of the zeroes of $\omega$. As a flat surface, $C$ can be obtained by gluing together certain regions in the plane with polygonal boundary by identifying pairs of parallel edges.  In this picture, the degeneration of $C$ into $C_1$ and $C_2$ is illustrated in \fig\ref{fig:gluing2}.

\begin{figure}
\centering
	\subfloat[Gluing two genus $1$ surfaces, each with two cylindrical boundaries, to get a genus $2$ surface.  In the center, we have a conformal WZW model. $\mc{D}$ and $\br{\mc{D}}$ indicate chiral and anti-chiral boundary conditions. ]{
		\label{fig:gluing2_a}
		
		\begin{tikzpicture}[scale=0.85]
\begin{scope}[shift={(-1,0)}]

	\node at (-1.5,1.5) {$\cdots$}; 
	\node at (-1.5,-1.5) {$\cdots$};
	\node at (-1.2,0) {$\scriptstyle \mc{D}$};
	\coordinate (A) at (3,1.5);
	\coordinate (B) at (3,-1.5);
	\coordinate (C) at (-1,1.5);
	\coordinate (D) at (-1,-1.5);
	
	\filldraw[pattern=north west lines](A)--(C)--(D)--(B)--(A);
	\draw[color=white,thick](A)--(C)--(D)--(B)--(A);
	\draw[dashed, ->-] (A) --(C);

	\draw[dashed, ->-](B)--(D);

	\begin{scope}[shift={(0,0)}]
	\coordinate (A) at (0,0);
	\coordinate (B) at (60:1);
	\coordinate (C) at ($(B) + (0:1)$);
	\coordinate (D) at ($(C) + (-60:1)$);
	\coordinate (E) at (-60:1);
	\coordinate (F) at ($(E) + (0:1)$);
	\filldraw[color=white] (A) --(B) --(C) --(D) --(F) --(E) --(A);
	\draw [->-](A) --(B) node [midway, sloped, below] {$\scriptscriptstyle a$}; 
	\draw [->-](B) --(C) node [midway, sloped, below] {$\scriptscriptstyle b$};  
	\draw [->-](C) --(D) node [midway, sloped, below] {$\scriptscriptstyle c$}; 

	\draw [->-](A) --(E) node [midway, sloped, above] {$\scriptscriptstyle c$};  
	\draw [->-](E) --(F) node [midway, sloped, above] {$\scriptscriptstyle b$};  
	\draw [->-](F) --(D) node [midway, sloped, above] {$\scriptscriptstyle a$};  
	\end{scope}
\end{scope}

	\node (n1) at (2.2,0) {$\scriptstyle \br{\mc{D}}$};
	\node (n2) at (3.8,0){$\scriptstyle \br{\mc{D}}$};
	\draw (n1) -- (n2) node[midway,below]{\tiny BRST};  

\begin{scope} 
	\coordinate (A2) at (4,1.5);
	\coordinate (B2) at (4,-1.5);
	\coordinate (C2) at (7,1.5);
	\coordinate (D2) at (7,-1.5);	
	\filldraw[pattern=north west lines](A2)--(C2)--(D2)--(B2)--(A2);
	\draw[color=white,thick](A2)--(C2)--(D2)--(B2)--(A2);
	\draw[dashed, -<-] (A2) --(C2);
	\draw[dashed, -<-](B2)--(D2);
\end{scope}

	\node(n3) at (7.2,0) {$\scriptstyle {\mc{D}}$};
	\node(n4) at (8.8,0){$\scriptstyle {\mc{D}}$};
	\draw (n3) -- (n4) node [midway, below]{\tiny BRST}; 

\begin{scope}[shift={(1,0)}] 
	\begin{scope}[shift={(9,0)}]

	\coordinate (A) at (3,1.5);
	\coordinate (B) at (3,-1.5);
	\coordinate (C) at (-1,1.5);
	\coordinate (D) at (-1,-1.5);
	
	\filldraw[pattern=north west lines](A)--(C)--(D)--(B)--(A);
	\draw[color=white,thick](A)--(C)--(D)--(B)--(A);
	\draw[dashed, ->-] (A) --(C);

	\draw[dashed, ->-](B)--(D);
	\end{scope}

	\begin{scope}[shift={(8.5,0)}]
	\coordinate (A) at (0,0);
	\coordinate (B) at (30:1);
	\coordinate (C) at ($(B) + (-10:1)$);
	\coordinate (D) at ($(C) + (-40:1)$);
		\coordinate (E) at ($(D) + (-150:1)$);
	\coordinate (F) at ($(E) + (170:1)$);
	\filldraw[color=white] (A) --(B) --(C) --(D) --(E) --(F) --(A);
	\draw [->-](A) --(B) node [midway, sloped, below] {$\scriptscriptstyle x$}; 
	\draw [->-](B) --(C) node [midway, sloped, below] {$\scriptscriptstyle y$};  
	\draw [->-](C) --(D) node [midway, sloped, below] {$\scriptscriptstyle z$}; 

	\draw [-<-](D) --(E) node [midway, sloped, above] {$\scriptscriptstyle x$};  
	\draw [-<-](E) --(F) node [midway, sloped, above] {$\scriptscriptstyle y$};  
	\draw [-<-](F) --(A) node [midway, sloped, above] {$\scriptscriptstyle z$};  
	\end{scope}
	\node at (12.5,1.5) {$\cdots$}; 
	\node at (12.5,-1.5) {$\cdots$};
	\node at (12.2,0) {$\scriptstyle \br{\mc{D}}$};
	
\end{scope}

\end{tikzpicture}}

\subfloat[As $L \to \infty$ we approach configuration (a). ]{
	\label{fig:gluing2_b}
\begin{tikzpicture}[scale=0.85]
	\node at (-1.5,1.5) {$\cdots$}; 
	\node at (-1.5,-1.5) {$\cdots$};

	\node at (-1.2,0) {$\scriptstyle \mc{D}$};
	\coordinate (A) at (12,1.5);
	\coordinate (B) at (12,-1.5);
	\coordinate (C) at (-1,1.5);
	\coordinate (D) at (-1,-1.5);
	
	\filldraw[pattern=north west lines](A)--(C)--(D)--(B)--(A);
	\draw[color=white,thick](A)--(C)--(D)--(B)--(A);
	\draw[dashed, ->-] (A) --(C);

	\draw[dashed, ->-](B)--(D);

	\begin{scope}[shift={(0,0)}]
	\coordinate (A) at (0,0);
	\coordinate (B) at (60:1);
	\coordinate (C) at ($(B) + (0:1)$);
	\coordinate (D) at ($(C) + (-60:1)$);
	\coordinate (E) at (-60:1);
	\coordinate (F) at ($(E) + (0:1)$);
	\filldraw[color=white] (A) --(B) --(C) --(D) --(F) --(E) --(A);
	\draw [->-](A) --(B) node [midway, sloped, below] {$\scriptscriptstyle a$}; 
	\draw [->-](B) --(C) node [midway, sloped, below] {$\scriptscriptstyle b$};  
	\draw [->-](C) --(D) node [midway, sloped, below] {$\scriptscriptstyle c$}; 

	\draw [->-](A) --(E) node [midway, sloped, above] {$\scriptscriptstyle c$};  
	\draw [->-](E) --(F) node [midway, sloped, above] {$\scriptscriptstyle b$};  
	\draw [->-](F) --(D) node [midway, sloped, above] {$\scriptscriptstyle a$};  
	\end{scope}

	\begin{scope}[shift={(8.5,0)}]
	\coordinate (A) at (0,0);
	\coordinate (B) at (30:1);
	\coordinate (C) at ($(B) + (-10:1)$);
	\coordinate (D) at ($(C) + (-40:1)$);
		\coordinate (E) at ($(D) + (-150:1)$);
	\coordinate (F) at ($(E) + (170:1)$);
	\filldraw[color=white] (A) --(B) --(C) --(D) --(E) --(F) --(A);
	\draw [->-](A) --(B) node [midway, sloped, below] {$\scriptscriptstyle x$}; 
	\draw [->-](B) --(C) node [midway, sloped, below] {$\scriptscriptstyle y$};  
	\draw [->-](C) --(D) node [midway, sloped, below] {$\scriptscriptstyle z$}; 

	\draw [-<-](D) --(E) node [midway, sloped, above] {$\scriptscriptstyle x$};  
	\draw [-<-](E) --(F) node [midway, sloped, above] {$\scriptscriptstyle y$};  
	\draw [-<-](F) --(A) node [midway, sloped, above] {$\scriptscriptstyle z$};  
	\end{scope}

	\node at (12.5,1.5) {$\cdots$}; 
	\node at (12.5,-1.5) {$\cdots$};

	\node at (12.2,0) {$\scriptstyle \br{\mc{D}}$};
	\draw[<->] (2,-1.7) -- (8.5,-1.7)  node [midway,below] {$\scriptstyle L$};
\end{tikzpicture}}

\caption{}
\label{fig:gluing2} 
\end{figure}

We let $z_1$, $z_2$ be the points on $C_1,C_2$ which are attached to each end of the cylinder.  As we have seen, the one-forms $\omega_1,\omega_2$ on $C_1,C_2$ have a pole at $z_1,z_2$.  To specify a theory associated to $C_1,C_2$, we need to specify the boundary conditions at these poles.  These boundary conditions can be either chiral Dirichlet $\mc{D}$, or anti-chiral Dirichlet $\br{\mc{D}}$.  

In this paper, we have focused on the case when the theory is ``balanced'', that is, has the same number of chiral and anti-chiral defects. A chiral defect can mean either a chiral disorder operator at a first-order zero of $\omega$, or a chiral Dirichlet boundary condition at a first order pole.  These are on the same footing, because a chiral Dirichlet boundary condition arises in the limit when a chiral disorder operator at a zero of $\omega$ collides with a topological Dirichlet boundary condition at a second-order pole.  

We would like to glue such balanced theories from balanced theories on $C_1,C_2$. Suppose that there are $n$ chiral and $\br{n}$ anti-chiral defects on $C$, where the balanced condition means $n = \br{n}$.  As $C$ degenerates into $C_1,C_2$ we find that $n_1,\br{n}_1$ of the chiral and anti-chiral defects  on $C$ end up on $C_1$, and $n_2,\br{n}_2$ end up on $C_2$. We have $n_1 + n_2 = n$, $\br{n}_1 + \br{n}_2 = n$.    

If we choose to have a chiral Dirichlet boundary condition at $z_1 \in C_1$, then the total number of chiral defects on $C_1$ is $n_1 + 1$.  In order for the configuration of defects on $C_1$ to be balanced, we need $n_1+1 = \br{n}_1$. This implies that $n_2 = \br{n}_2 + 1$, so that we can have a balanced configuration of defects on $C_2$ by imposing anti-chiral Dirichlet boundary conditions at $z_2$. 

We conclude that the only way to have a balanced configuration on $C_1,C_2$ is to choose Dirichlet boundary conditions of opposite chirality on $z_1,z_2$ and ask that the defects on $C$ are distributed so that $n_1 \pm 1 = \br{n}_1$, $n_2 \mp 1 = \br{n}_2$. 

When we glue two theories together, we must decide what to place on the intermediate cylinder. This amounts to choosing the boundary conditions for $(\CP^1, \d z / z$ at $z = 0$, $z = \infty$.  We can glue chiral Dirichlet boundary conditions to each other by BRST reduction, and similarly for anti-chiral Dirichlet boundary conditions.  Since we have chosen boundary conditions of opposite chirality at $z_1,z_2$ we must also choose boundary conditions of opposite chirality on $(\CP^1,\d z / z)$.  

This means that the cylinder theory we use for gluing is the conformal WZW model  discussed in section \ref{sec:WZW}. When the cylinder becomes thin,
then locally at the cylinder the four-dimensional theory can be reduced along the winding cycle of the cylinder, to obtain the three-dimensional Chern-Simons theory (recall again discussion around \eqn\eqref{eq.level}).
Moreover the monodromy integral of the one-form $\omega$ along  a
winding cycle of the cylinder ($\int_{S^1} \omega$)
is turned into the complexified level of the three-dimensional Chern-Simons theory.
This parameter also coincides (up to sign) with the 
residues $\textrm{Res}_{P_i} \omega_i$ of the one-form 
on $C_i$ ($i=1,2$) discussed above.

We will refer to the chiral action as $G_L$, and the anti-chiral action as $G_R$. 
The gluing at the level of the integrable field theory
is then written schematically as (in the notation $\mbf{IFT}(C,  \omega )$ introduced previously)
\begin{equation}	
	\begin{split}
	\lim_{L \to \infty}	\mbf{IFT}(C_L, \omega_L, n ,\br{n}  ) 
		&= \Bigg\{\mbf{IFT}\left(C_1,\omega_1, n_1,\br{n}_1 , \mc{D}(z_1)\right)  \oplus  \mbf{IFT}\left(\mathbb{CP}^1, \, \omega=\tfrac{\d z}{z}, \mc{D}(0), \br{\mc{D}}(\infty)\right)    \\ 
		&\oplus  \mbf{IFT}\left(C_2, \omega_2, n_2,\br{n}_2, \br{\mc{D}}(z_2)\right) \Bigg\} \Bigg/\!\!\!\Bigg/ (G_L\times G_R)  \;,
	\end{split}
\label{eq.T_glue}
\end{equation}
where we assumed the condition \eqref{eq.residue} for $\omega_{1,2}$.  
On the left hand side, we suppose we have a family of surfaces $(C_L,\omega_L)$ which develops a long neck, which splits the surface into $C_1$ and $C_2$ (as in \fig\ref{fig:gluing2}\subref{fig:gluing2_b}). 

The main result of this section is the proof of the equality in equation \eqref{eq.T_glue}, at the classical level. 

At the Lagrangian level the cutting/gluing construction of our theories
leads us to a rather general construction of the integrable field theories,
which often look rather non-trivial at the level of the Lagrangians for the integrable field theories. 

Moreover, given a surface data $(C, \omega, D)$ the decomposition into two surface
data $(C_1, \omega_1, D_1)$
and $(C_2, \omega_2, D_2)$ is far from unique---we can distribute the poles/zeros of $\omega$,
as well as the surface defects $D$, 
into $C_{1,2}$ in different manners, and these will give rise to different decompositions of the same theories.
In particular, this in general describes the decompositions into
chiral and non-chiral theories, all of which will be glued back into the same theory
we started with. This is the integrable-field-theory
counterpart of the ``generalized S-duality'' in class $\mathcal{S}$ theories \cite{Gaiotto:2009we}.

We have restricted ourselves to the discussion of balanced theories, with an equal number of chiral and anti-chiral boundary conditions. The story would, in many ways, be more natural if we removed this restriction. Without this restriction, however, we did not see a nice way to describe the integrable field theories associated to higher genus curves. 

Before we turn to the proof of the gluing statement, let us 
count the number of parameters of the model. Suppose that the one-form $\omega$ has $m$ double poles and $2m+2g-2$ first order zeros
on a genus $g$ curve $C$.  (First order poles, as usual, can be treated as a collision of a zero with a second-order pole). 
This has $m+(2m+2g-2)+(3g-3)=3m+5g-5$ parameters,
where we have taken into account the positions of the 
poles/zeros as well as the moduli space of the genus $g$ curve (which is of dimension $3g-3$). 

Consider the degeneration of $C$ into 
two components $C_1$ and $C_2$,
each with genus $g_1$ and $g_2$, where
$g=g_1+g_2$.
Then the poles and the zeros are
distributed such that we have
\begin{itemize}
\item $m_i$ double poles, one first-order pole, and  $2m_i+2g_i-1$ first-order zeros
\end{itemize}
for $C_i$ ($i=1,2$), with $m_1+m_2=m$.
Note that the Riemann-Roch theorem (the number of poles minus the number of zeros, counted with multiplicity, is 
$2-2g$), is satisfied on each of the curve $C_{i}$.
The complex dimension of moduli on the curve $C_i$ is $m_i+1+(2m_i+2g_i-1)+(3g_i-3)=3m_i+5 g_i-3$.

In the gluing procedure \eqn\eqref{eq.T_glue}, the right hand side
has $\sum_{i=1,2} (3m_i+5 g_i-3)=3m+5g-6$ parameters. This is one less than those of the left hand side. The remaining parameter corresponds to smoothing the node, which is a marginal deformation of the glued theory that moves us to the interior of the moduli space.  

\subsection{Proof of the Gluing Formula}

In this section we will sketch the proof of the result, stated in section \ref{sec:disorder}, that the integrable models we have constructed from a higher genus curve can be described, in the limit that the curve develops a node, in terms of the BRST reduction of models on curves of lower genus.  

Suppose that we have a family of surfaces $(C_L,\omega_L)$ which develops a long neck, as in \fig\ref{fig:gluing2}\subref{fig:gluing2_b}. In the limit the neck becomes long, we can cut the surface along the infinite cylinder to give a surface $(\til{C},\til{\omega})$, as above. It could be that $\til{C}$ is a disjoint union of $C_1,C_2$ as above; but we could also consider the situation when $\til{C}$ is connected.  

As before, we let $z_1,z_2 \in \til{C}$ be the two special points which we glue to get $\lim_{L \to \infty} C_L$.  At these points we impose chiral and anti-chiral Dirichlet boundary conditions $\mc{D}(z_1)$, $\br{\mc{D}}(z_2)$.  

In this section we will show that the limit $\mbf{IFT}(C_L, \omega_L)$ as $L \to \infty$ is obtained from $\mbf{IFT}(\til{C}, \til{\omega})$ by coupling to a conformal WZW model, and performing BRST reduction with respect to both the chiral and anti-chiral Kac-Moody algebras, as in \fig\ref{fig:gluing2}\subref{fig:gluing2_a}.  

In fact, as an intermediate step, we will suppose that we have chiral Dirichlet boundary conditions at $z_1,z_2$. Then, if we perform the diagonal BRST reduction for the Kac-Moody algebras at these boundary conditions, we will show that we find the theory associated to the curve $C$ obtained by gluing $z_1, z_2$. 

The first thing we need to understand is a description of the theory in the limit.   We will do this from the four-dimensional point of view. To do this, we need to describe the four-dimensional gauge theory on a nodal curve, in the vicinity of the node. We can describe the nodal curve locally with coordinates $z,z'$ satisfying $z z' = 0$.  

At the quantum level, one would certainly have difficulty dealing with a field theory on a singular manifold.  But at the classical level, there are no difficulties in writing down the equations of motion.  The equations of motion, on $\R^2 \times C$, where $C$ is the nodal curve, simply say that we have $G$-bundle which is flat on $\R^2$ and holomorphic on $C$, as usual.  It makes sense to talk about holomorphic (or, more precisely, complex-analytic) $G$-bundles on a Riemann surface with singularities.

If $\til{C}$ is the resolution of $C$ which separates the node, a holomorphic bundle on $C$ is the same as a holomorphic bundle on $\til{C}$ with an isomorphism between the fibres at the points $z_1,z_2$ which are the preimages of the node. 

In the same way, a field configuration on $C \times \R^2$ satisfying the equations of motion is the same as one on $\til{C} \times \R^2$, equipped with an isomorphism of flat bundles on $\R^2$ from the one obtained by restricting to $z_1 \times \R^2$, to the one obtained  by restricting to $z_2 \times \R^2$.

To match with the BRST reduction, it will be convenient to describe these field configurations in a particular gauge.  At $z_1$ or $z_2$ we have a bundle with connection on the $w$-plane.  By a gauge transformation, we can always set $A_{\wbar}(z_1)$ to zero, and similarly we set $A_{\wbar}(z_2) = 0$.  The remaining gauge transformations are independent of $\wbar$, that is, holomorphic. 

To identify the bundles at $z_1$ and $z_2$, we ask that
\begin{equation} 
	A_w(z_1) = A_w(z_2) \;.
\end{equation}
We also ask that the gauge transformations at $z_1$ and $z_2$, which are holomorphic as functions of $w,\wbar$, are the same. 

Finally we need to explain the behaviour of $A_{\zbar}$.  By using a gauge transformation proportional to $\zbar - \zbar_1$ or $\zbar - \zbar_2$, we can set $A_{\zbar}(z_1) = 0$ and $A_{\zbar}(z_2) = 0$. In this gauge, the only allowed gauge transformations are those which do not vary $A_{\zbar}$ at $z_1$ and $z_2$. This means that, when expanded in series, they do not contain a term linear in $\zbar - \zbar_1$ or $\zbar - \zbar_2$.

Now let us turn to analyzing the BRST reduction. The first thing we need to understand is the currents which generate the Kac-Moody actions at the points $z_1,z_2 \in \til{C}$.  The chiral Dirichlet boundary condition at $z_1,z_2$ sets $A_{\wbar} = 0$. We can deform this boundary condition by asking that $A_{\wbar}$ takes some non-zero value at $z_1,z_2$: say $A_{\wbar}(z_i)= A_{\wbar}^i$, $i = 0,1$, and $A_{\wbar}^i$ being a $(0,1)$ form on the $w$-plane with no $z$-dependence.

This deformation is implemented by the currents $J_1,J_2$ for the Kac-Moody actions at $z_1,z_2$, by adding a term
\begin{equation} 
	\int_{\R^2}A_{\wbar}^0 J_1 + A_{\wbar}^1 J_2 \;.  
\end{equation}
We can calculate the currents by studying the Lagrangian in the presence of background field $\what{A}_{\wbar}^i$, which depend on $z$, and where $\what{A}_{\wbar}^i = A_{\wbar}^i$ at $z = z_i$. We assume that $\what{A}_{\wbar}^i = 0$ outside of a small neighborhood of $z = z_i$.

The dependence of the Chern-Simons Lagrangian on $\what{A}_{\wbar}^i$ takes the form
\begin{multline} 
	\tfrac{1}{2} \int_{\abs{z - z_i} \le \eps} A_w \partial_{\zbar} \what{A}_{\wbar}^i \, \omega 
	-\tfrac{1}{2} \int_{\abs{z - z_i} \le \eps}(\partial_{\wbar} A_w)  \what{A}_{\wbar}^i \, \omega 
	- \int_{\abs{z - z_i} \le \eps}(\partial_w A_{\zbar})  \what{A}_{\wbar}^i \, \omega  \\
	+ \int_{\abs{z - z_i} \le \eps}[A_w, A_{\zbar}] \what{A}_{\wbar}^i \, \omega.
\end{multline}
Integrating the first line by parts, we find
\begin{equation} 
	\int_{\abs{z - z_i} \le \eps}F_{\zbar w} \what{A}_{\wbar}^i \omega + \op{Res}_{z_i}(\omega) A_w(z_i) A_{\wbar}^i \;.  \\
\end{equation}
We conclude that the current is
\begin{equation} 
	J_i = \op{Res}_{z_i}(\omega) A_w(z_i) \;. 
\end{equation}
Without loss of generality, we will assume that $\op{Res}_{z_1}(\omega) = 1$. This implies that $\op{Res}_{z_2}(\omega) = -1$. With these conventions, $J_1 = A_w(z_1)$, $J_2 = -A_w(z_2)$.

When we perform classical BRST reduction, the first step is to set the current to zero. (At the quantum level, this is implemented by the BRST operator applied to the $\sf{b}$-ghost).  We therefore find the constraint
\begin{equation} 
	A_w(z_1) = A_w(z_2)\;, 
\end{equation}
as desired.

From the chiral Dirichlet boundary conditions, we already have the constraints $A_{\wbar}(z_1) = 0$, $A_{\wbar}(z_2) = 0$ and $A_{\zbar}(z_1) = 0$, $A_{\zbar}(z_2) = 0$. Further, the chiral Dirichlet boundary conditions tell us that gauge transformations vanish at $z_1, z_2$.  

At the classical level,  the $\sf{c}$-ghost of the BRST reduction means that we introduce new gauge transformations. These are holomorphic functions of $w,\wbar$, and independent of $z$.  If we expand the gauge transformation of four-dimensional Chern-Simons theory as a function of $z-z_1$, $\zbar - \zbar_1$, before we introduce the $\sf{c}$-ghost, there is no constant term. The new gauge transformations introduced by the BRST $\sf{c}$-ghost allow us to have a constant term, which is holomorphic as a function of $w,\wbar$.

Similarly, if we expand around $z = z_2$, the $\sf{c}$-ghost allows the expansion to have a constant term, which takes the same value at $z_1$ and $z_2$.

We have found that the BRST reduction moves us from the theory associated to $\til{C}$, with chiral boundary conditions at $z_1,z_2$, to the theory on the glued surface $C$. 

We are ultimately not interested in the case when we have chiral Dirichlet boundary conditions at both $z_1$ and $z_2$.  We want to consider chiral Dirichlet at $z_1$, anti-chiral Dirichlet at $z_2$, where we glue to the conformal WZW model by both chiral and anti-chiral BRST reduction. 

The argument given above will show that the result will be the theory obtained from four-dimensional Chern-Simons theory on $\til{C}$ with a $(\CP^1,\d z / z)$ attached, where $0$ is attached to $z_1$ and $\infty$ to $z_2$.  

It turns out, though, that this is the same as the theory on the curve $C$ obtained by gluing $z_1$ to $z_2$. The point is that every stable holomorphic bundle on $\CP^1$ is trivial, so that we can choose a gauge where $A_{\zbar} = 0$ on the $\CP^1$. The equations of motion then tell us that all fields are independent of the coordinate on $\CP^1$.   Therefore, a solution to the equations of motion on $\til{C}$, with the $(\CP^1,\d z /z)$ glued in, is given by a holomorphic $G$-bundle on $\til{C}$, with a compatible flat connection on the $w$-plane, and an isomorphism between the bundles at $z_1$ and $z_2$.  This is the same as a solution to the equations of motion on $C$, as desired.

This completes the proof of the gluing statement, at the classical level. It would be very interesting to see to what extent this continues to hold at the quantum level, and further, to understand the marginal operator which moves the glued theory away from the boundary of the moduli space.  

\section*{Acknowledgments}

We would like to thank Edward Witten for collaboration in early stages of this project (as well as long-term collaboration in Part I and II of the series), and for providing many useful insights and comments which are crucial for the completion of this work. 
We also thank Benjamin Basso, Davide Gaiotto, Bogdan Stefa\'{n}ski and Benoit Vicedo for useful discussion.
K.~C.~is supported by the NSERC Discovery Grant program and by the Perimeter Institute for Theoretical Physics. Research at Perimeter Institute is supported by the Government of Canada through Industry Canada and by the Province of Ontario through the Ministry of Research and Innovation. 
M.~Y.~is partially supported by WPI program (MEXT, Japan) and by JSPS KAKENHI Grant No.\ 17KK0087, No.\ 19K03820 and No.\ 19H00689. He would like to thank the hospitality of Institute for Advanced Study, Perimeter institute, 2019 Pollica summer workshop (supposed in part by the Simons Foundation (Simons Collaboration on the Non-perturbative Bootstrap) and in part by the INFN),
and 2019 Simons Summer Workshop.

\appendix 

\section{\texorpdfstring{$\sigma$-model on K\"ahler Manifold in the Large Volume Limit}{Interlude: sigma-model on Kahler Manifold in Large Volume Limit}}\label{app:interlude}

Let $X$ be a K\"ahler manifold, with local coordinates $u^i,\br{u}^{\br{i}}$ in which the K\"ahler metric takes the form
\begin{equation} 
g = \sum_{i, \br{j}} g_{i\br{j}} \d u^i \d \br{u}^{\br{j}} \;. 
 \end{equation}
Let us consider the $\sigma$-model with target $X$.  After formally complexifying the space of fields, there are $2n$ independent complex fields $\sigma^i$, $\br{\sigma}^{\br{i}}$, which are (locally on $X$) complex valued functions on $\R^2$.  In these coordinates the standard $\sigma$-model action together with a topological term $ \int \phi^\ast \omega$ is given by
\begin{align} 
& \int g_{i \br{j}}(\sigma, \br{\sigma}) (\d \br{\sigma}^{\br{j}}) (\star \d \sigma^i)+  \int \phi^\ast \omega
= - 2 \i \int   g_{i \br{j}}(\sigma, \br{\sigma}) \partial \br{\sigma}^{\br{j}} \dbar \sigma^i   \;,
\label{eq.action_sigma}
 \end{align}
 where $\phi^{\ast}\omega=\i g_{i \br{j}} \d \sigma^i \d \br{\sigma}^{\br{j}}$ is the pull-back of the K\"ahler form on the target $X$.  

The action functional is a holomorphic function of the fields, and the path integral is performed over the contour where
\begin{equation} 
\br{\br{\sigma}^{\br{i}}} = \sigma^{i} \;. 
 \end{equation}

We can rewrite the action \eqref{eq.action_sigma} in the first order formalism as
\begin{equation}
  \int \beta_i \dbar \gamma^i + \int \br{\beta}_{\br{i}} \partial \br{\gamma}^{\br{j}}  - \int g^{i \br{j}} \beta_i \br{\beta}_{\br{j}} \;.    
  \label{eq.action_sigma_2}
\end{equation}
Here we renamed $\sigma, \br{\sigma}$ by $\gamma, \br{\gamma}$, and 
we introduced auxiliary fields 
\begin{equation} 
\begin{split}
\beta_i &\in \Omega^{1,0}(\C)\;,\\
\br{\beta}_{\br{i}} & \in \Omega^{0,1}(\C) \;.
\end{split}
 \end{equation}
Geometrically, $\beta$ is a $(1,0)$ form on $\C$ with  values in the pull-back under $\gamma$ of the $(1,0)$ cotangent bundle of $X$, and $\br{\beta}$ is a $(0,1)$ form valued in the pull-back of the $(0,1)$ cotangent bundle of $X$.  $g^{i \br{j}}$ is the inverse to the K\"ahler metric $g_{i \br{j}}$ on $X$.  It is straightforward to integrate out auxiliary fields $\beta, \br{\beta}$ from \eqn \eqref{eq.action_sigma_2}, to reproduce the 
original expression \eqn \eqref{eq.action_sigma} (up to an overall constant factor and $\sigma$ replaced by $\gamma$).

The action \eqref{eq.action_sigma_2} is the action for the $\beta-\gamma$ system and its complex conjugate, together with a deformation by a $\beta-\br{\beta}$ coupling given by the inverse of the K\"ahler metric. 
The latter vanishes in the large volume limit, as claimed above.

\section{Fateev Three-Dimensional Sausage Solution}
\label{app:3d_sausage}

Let us briefly summarize the Fateev three-dimensional sausage solution \cite{Fateev:1996ea}.
The Fateev solution is a two-parameter family of 
deformation of the metric. Here we choose a one-parameter locus, whose metric reads\footnote{In
the notation of \cite{Fateev:1996ea}, the two parameters are given by $\nu$ and $k$ with $\nu>0$ and $k^2<1$.
We choose the limit $k\to 1$ with $\lambda=\nu/(2(1-k^2))>0$ kept finite. The parameter $\lambda$ is an overall constant factor of the metric.}
\begin{equation}
ds^2=ds^2_{\rm std} - (\tanh^2 \zeta) (\phi_1+\phi_2)^2 \;,
\end{equation}
where  we used the coordinates $x_1, x_2, x_3, x_4$ satisfying the condition $x_1^2 + \dots +x_4^2=1$,
so that the canonical metric on the three-sphere is induced from the embedding into $\mathbb{R}^4$:
$ds^2_{\rm std} =dx_1^2+ \dots +dx_4^2$.
We also defined 
\begin{equation}
\begin{split}
\phi_1&=x_1 dx_2-x_2 dx_1 \;,\\
\phi_2&=x_3 dx_4-x_4 dx_3 \;.
\end{split}
\end{equation}

Let us switch to complex coordinates as defined by $u_1=x_1+\i x_2, u_2=x_3+\i x_4$,
so that we have $ds^2_{\rm std}=du_1 d\bar{u}_1+du_2 d\bar{u}_2$ and
\begin{equation}
\begin{split}
\phi_1=\frac{1}{2\i} (\br{u}_1 du_1-u_1 d\br{u}_1) \;,\\
\phi_2=\frac{1}{2\i} (\br{u}_2 du_2-u_2 d\br{u}_2) \;.
\end{split}
\end{equation}
The sausage metric is then
\begin{equation}
ds^2=du_1 d\bar{u}_1+du_2 d\bar{u}_2 + \frac{1}{4}(\tanh^2 \zeta) (\br{u}_1 du_1-u_1 d\br{u}_1+\br{u}_2 du_2-u_2 d\br{u}_2)^2 \;,
\end{equation}
or, using the constraint $\d (\abs{u_1}^2+\abs{u_2}^2)= 0$, 
\begin{equation}
ds^2=du_1 d\bar{u}_1+du_2 d\bar{u}_2 + (\tanh^2 \zeta) (\br{u}_1 du_1+\br{u}_2 du_2)^2 \;.
\end{equation}
This metric coincides with that in \eqn \eqref{eq.S3_sausage} in the main text, up to an exchange of $u_2$ and $\br{u}_2$.


\clearpage


\bibliography{bib_draft3}
\bibliographystyle{nb}

\end{document}